\RequirePackage{fix-cm}

\documentclass[smallextended]{svjour3}       
\smartqed  
\usepackage{graphicx}

 \journalname{Bulletin of Mathematical Biology}

\usepackage{hyperref}

\usepackage{amsmath} \usepackage{amsfonts,euscript,amssymb}
\usepackage{graphicx} \usepackage{subcaption} \usepackage{xcolor}

\usepackage{mathrsfs}

\usepackage{verbatim}

\makeatletter
\newcommand{\vast}{\bBigg@{4}}
\makeatother
\newcommand\cross{\!\times\!}

\newcommand\partt[1]{\frac{\partial{#1}}{\partial t}}

\newcommand\odiff[2]{\frac{d{#1}}{d{#2}}}

\renewcommand\th{$^\text{th}$\space}
\newcommand\ah[1]{{\color{black}{#1}}}
\newcommand\ahh[1]{{\color{black}{#1}}}

\usepackage{xcolor}

\newcommand{\corr}[1]{{ \color{black} #1}}

\def\R{\mathbb{R}}
\def\I{\mathcal{I}}
\def\P{\mathcal{P}}
\def\D{\mathcal{D}}
\def\Y{\Upsilon}

\newcounter{myappendixctr}

\begin{document}

%
%

\title{Signal propagation in sensing and reciprocating cellular systems with spatial and structural heterogeneity}

\author{Arran Hodgkinson   \and	Giles Uz\'e	\and	Ovidiu Radulescu	\and	Dumitru Trucu}

\institute{
	Arran Hodgkinson \at
    DIMNP - UMR 5235, Universit\'e de Montpellier,
           Pl. E. Bataillon, 34095 Montpellier, France \\
	\email{arran.hodgkinson@etu.umontpellier.fr}
\and
	Giles Uz\'e \at
    DIMNP - UMR 5235, Universit\'e de Montpellier
           Pl. E. Bataillon, 34095 Montpellier, France \\
	\email{giles.uze@umontpellier.fr}
\and
	Ovidiu Radulescu \at
    DIMNP - UMR 5235, Universit\'e de Montpellier
           Pl. E. Bataillon, 34095 Montpellier, France \\
	\email{ovidiu.radulescu@umontpellier.fr}
\and
	Dumitru Trucu \at
	School of Science and Engineering, University of Dundee, Dundee, DD1 4HN, Scotland \\
	\email{trucu@maths.dundee.ac.uk}
}
\date{Received: 26th October 2017 / Accepted: date}

\maketitle

\begin{abstract}
Sensing and reciprocating cellular systems (SARs) are important for the operation of many biological systems. Production in interferon (IFN) SARs is achieved through activation of the Jak-Stat pathway, and downstream upregulation of IFN regulatory factor (IRF)-3 and IFN transcription, but the role that high and low affinity IFNs play in this process remains unclear. We present a comparative between a minimal spatio-temporal partial differential equation (PDE) model and a novel spatio-structural-temporal (SST) model for the consideration of receptor, binding, and metabolic aspects of SAR behaviour. Using the SST framework, we simulate single- and multi-cluster paradigms of IFN communication. Simulations reveal a cyclic process between the binding of IFN to the receptor, and the consequent increase in metabolism, decreasing the propensity for binding due to the internal feedback mechanism. One observes the effect of heterogeneity between cellular clusters, allowing them to individualise and increase local production, and within clusters, where we observe `subpopular quiescence'; a process whereby intra-cluster subpopulations reduce their binding and metabolism such that other such subpopulations may augment their production. Finally, we observe the ability for low affinity IFN to communicate a long range signal, where high affinity cannot, and the breakdown of this relationship through the introduction of cell motility. Biological systems may utilise cell motility where environments are unrestrictive and may use fixed system, with low affinity communication, where a localised response is desirable.

\end{abstract}


\section{Introduction}


\subsection{\corr{Sensing and Reciprocating Systems and their Mathematical Treatment}}

In order for biological \corr{systems} to initiate \corr{changes in behaviour at the scale of a group of cells or of a tissue in response to a localized event}, it is necessary for small signals to be transformed into large signals and sequentially communicated to other cells. This is no more apparent than in the human immune response where T-cells are actively recruited to the site of infection through the amplification and dispersion of the precursor signal \corr{\cite{mackay1996chemokine}.} The intermediate signal must be received and amplified, in order that distant cells may receive the signal with sufficient veracity as to respond.

In the case of the immune system, the cell-to-cell communication can be at least partially orchestrated by dynamic changes of the cell membrane receptors and by secretion of communication proteins such as chemokines \cite{mackay1996chemokine} and cytokines \cite{OylerYaniv2017}. Other cell-to-cell communication and amplification mechanisms are used by bacteria in a phenomenon known as ``quorum sensing'' \cite{Ng2009} and by yeast to optimize mating efficiency \cite{barkai1998protease}. In order to synchronise the phenotypes expressed by a local group of cells, bacteria and yeast posses internal feedback loops that amplify incoming diffusible chemical signals. Similar examples where local behavior spreads by cell-to-cell communication can be found in animal development, when blocks of tissues can be developed from sheets of cells by a phenomenon called ``community effect'' \cite{gurdon1988community} or when cell fate is specified by ``sequential patterning''  such as in the spatial
regulation of Delta-Notch signalling \cite{hoyos2011quantitative,henrique1997maintenance}. Collective synchronous behaviour of cells is also needed in insulin secretion by pancreatic islets but, in this case, the possible cell-to-cell communication mechanisms are still under debate \cite{pedersen2005intra}. We call such systems sensing and reciprocating systems (SARs), on the basis that the initial chemical signals are replicated and amplified, which is similar to the concept of secrete and sensing cells \cite{Maire2015,Olimpio2017}.

SARs are ubiquitous in biology and some mathematical models dealing with properties of such systems exist. The versatility of collective properties of secrete and sensing cells was studied using phenomenological, compartment based models and ordinary differential equations (ODEs) \cite{Youk2014}. The same type of
formalism was used for metabolic synchronisation of insulin secretion in islets
\cite{pedersen2005intra} and for studying cell-to-cell communication in the immune system \cite{Francois2016}. ODE based models allow rather detailed descriptions of intracellular signalling and metabolic dynamics but do not cope accurately with cell proliferation, migration, and cell-to-cell interactions.

Although not yet used for SARs, frameworks based on partial differential equations (PDEs) could integrate many of these processes and explain aspects related to spatial heterogeneity such as the role played of spatial arrangement of cells in determining the conveyance of these signals \cite{OylerYaniv2017}. However, in PDE models, non-spatial heterogeneity, resulting from the fact that cells in close spatial proximity do not necessarily respond synchronously to stimuli, is lost by averaging. This ``structural heterogeneity'' can be an essential part of a complex cell dynamics, in which cell sub-populations behave differently to the average, and may be essential to understanding the complicated dynamics of biological systems. As an example, such models have predicted that below a certain threshold value, interferon (IFN) signalling allows the activity of the cellular population to decay entirely \cite{Hart2014}.

\corr{A paradigm which seems appropriate to exploring the possible structural dimensions}  of biological problems, in a mathematical context, is that of the continuous structural approach \cite{Brikci2008,Lorz2012,Lorz2015}. This approach encompasses the \corr{genetic or epigenetic state of a cell}, under temporal conditions which are consistent with the continuous nature of dynamic biological problems by employing the application of PDEs \corr{in structure, rather than in spatial position}. On the other hand, these approaches neglect the spatial dimensions associated with chemical communication between cells and, thusly, do not provide the descriptive breadth necessary to analyse these situations.

One recent ``spatio-structural-temporal'' (SST) framework, which demonstrates the potential to represent greater details of dynamical processes in dimensions of both structure and space, was developed in order to model the urokinase plasminogen activator system in breast cancer \cite{Domschke2016,Trucu2016}. Herein, we present a similar derivation in order to augment the generality of this framework and present a modelling form capable of capturing the intricacies, and important heterogeneous features of SARs. Compared to \cite{Domschke2016,Trucu2016} we introduce new metabolic structural variables and conjugated advection fluxes that are derived from the continuity equation and Liouville's theorem. These variables are needed for modelling stimulated amplification in SARs. The use of Liouville theorem is a major advance in the SST framework as it can relate any single cell ODE dynamics to population dynamics
in structure space.

\subsection{An example of a SAR system: Cellular Interferon (IFN) System}

We look, here, specifically at a detailed model for the IFN binding process of a given cell and the concurrent metabolic processes that result from this binding process. This SAR shall serve as an exemplar biological system on which to base models that will explore the efficacy of the framework to be proposed.

There are 13 forms of IFN$\alpha$ and 1 of IFN$\beta$, which we subcategorise as low and high affinity  and denote as IFN$\alpha$ and IFN$\beta$ respectively. Their ability to activate a cell's internal infrastructure is dependent on their ability to concurrently bind the IFN-$\alpha/\beta$ receptors 1 (IFNAR1) and 2 (IFNAR2) on the surface of the cell. The association rate of IFN with IFNAR2 is approximately $10\times$ that of IFNAR1, therefore the primary interaction is with the jak1 signalling complex of IFNAR2 \cite{Gavutis2005,Gavutis2006}. It is also essential, however, that IFN bind the lower affinity IFNAR1 and so IFNAR1 is recruited to the location of the bound IFN/IFNAR2 complex \cite{Gavutis2005,Gavutis2006}. These tyk2 and jak1 protein phosphorylate one another to initiate what is known as the Jak-Stat pathway \cite{Stark1998}.

The Jak-Stat pathway is predicated on the fact that the phosphorylated Jak1-tyk2 complex is capable of phosphorylating the transcription factors Stat1 and Stat2. These two factors are then able to bind the IFN regulatory (transcription) factor (IRF)-9 in order to form the IFN stimulated gene factor (ISGF)-3 complex \cite{Stark1998,Samuel2001}, which is capable of entering the nucleus \cite{Lin1998}. Having achieved this step, this complex can bind to the promoter region of IFN stimulated genes (\textbf{ISG}s) and effectively initiate their transcription \cite{Stark1998,Samuel2001}.

One particularly significant ISG is the IRF-7 protein who is capable of the downstream binding of and IRF-3. This IRF-7-3 complex is directly responsible for the promotion of \textbf{IFN-$\alpha$} and \textbf{IFN-$\beta$} genes \cite{Haller2006}. Another effect of transcribing ISGs is the transcription of USP18, which will compete with jak1 for binding of the intracellular domain of IFNAR2 \cite{FrancoisNewton2011}. IFNAR2s bound by USP18 have also been shown to be ineffective at affecting the transcription of IRF-7 \cite{Randall2006,Wilmes2015,Arimoto2017}.

\begin{figure}[t!] \centering
	\includegraphics[width=.6\textwidth]{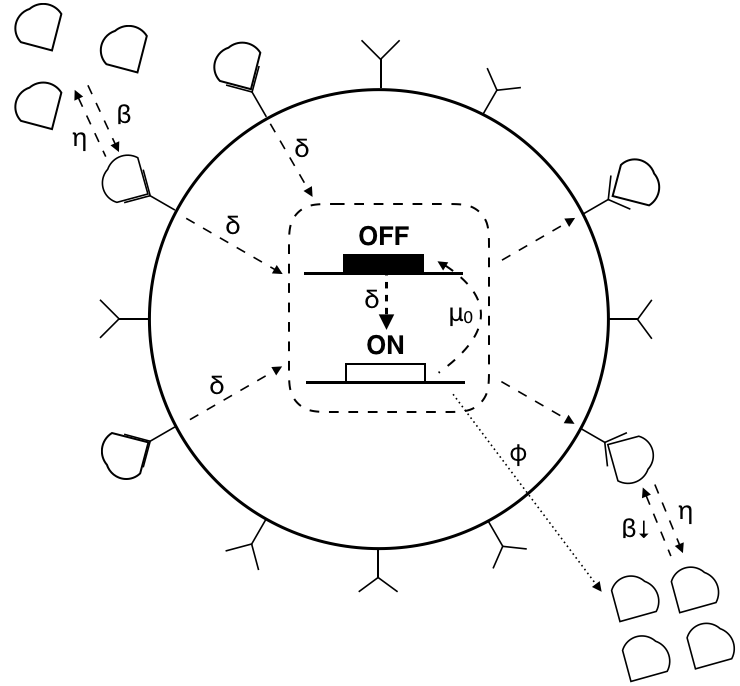}
	\caption{Diagram describing the simplified IFN cell-regulatory system. Unbound IFN (\textit{top left}) will attempt to bind unbound receptors \ah{(\textit{`Y's})} on the surface of the cell (\textit{circle}), in accordance with its affinity for these receptors, with binding rate $\beta$. Likewise, these bound IFN-receptor complexes unbind with some rate $\eta$. The other way in which the proportion of surface bound molecules may decrease in through the internalisation of IFN receptor complexes with rate $\delta$. The internalisation of IFN, through a complicated biological pathway, leads to a metabolic switching of the IFN-producing cell infrastructure from the default state of dormant (OFF) to active (ON), in which state the cell produces greater levels of IFN with rate $\phi$ (\textit{centre}). The cell infrastructure attempts to return to the default (OFF) state with a constant rate $\mu_0$. In the active (ON) state the conformation of receptors, in the presence of IFN, is reduced which can be modelled through the reduction of the ability of IFN molecules to bind their receptors (i.e. $\beta\downarrow$, \textit{bottom right}).
} \label{fig:Interferon_Cell_Diag}
\end{figure}

Therefore, this system can be looked at through the simplified lens of two major and important processes:
\begin{itemize}
	\item[(a)]	the binding of IFN to the surface of the cell, and
	\item[(b)]	the activation of the metabolic pathway which eventually leads to the creation of new IFN molecules.
\end{itemize}
We use the phrase `metabolic activation' in order to characterise the state of the cell in terms of the chemical activity levels of those proteins involved in the Jak/Stat pathway and, ultimately, the transcription of the genes necessary for the synthesis of IFN. Thus, when one describes the metabolic activation of the cell, with regards to the IFN pathway, one is actually describing, in some way, the spatially differentiated presence of IRF-7-3 within the cell (Figure \ref{fig:Interferon_Cell_Diag}).

Moreover, one review of experimental data plotted the relationship over time between the activation of genes within the cell and the fractional levels of bound and unbound surface receptors, for both IFN$\alpha$2 and IFN$\beta$ \cite{Schreiber2015}. This graph importantly showed that, for low levels of IFN$\alpha$2, as the number of surface receptors decreased, the metabolic activation level rose concurrently. Further, as genetic activation levels decreased, one could observe a corresponding normalisation of the fractional surface receptor levels \cite{Lavoie2011}. Comparably, for high levels of IFN$\beta$, one finds that the cells genetic mechanism is activated in a locally irreversible process and that the fraction of IFNAR1 receptors is maintained at approximately 40\% \cite{Schreiber2015}.


\medskip
In order to demonstrate the descriptive power within the existing modelling frameworks, we choose the biological IFN system in T-cells as an illustrative example of such a system of SARs. This will serve as a comparative case for the development of a framework, which is capable of significantly improving upon one's existing capacity.

\section{A Simple, Continuous Mathematical Model for the IFN System}

If one were to create the simplest possible system of SARs, one would begin with only the population of SARs, themselves, and the molecular population of SAR \corr{diffusing ligands}. In reality, however, these systems are rarely as simplistic as this and often require consideration of spatially intermediate cells which may mediate the levels of the SAR \corr{ligands}, by consuming these proteins without reciprocally producing them. This is the case in the biological IFN system and, as such, we call such intermediate cells `consumers' and the SAR cells as `producers', within a system that considers only such a responsive protein.

Therefore, begin by defining a temporal domain, given by $\I=[0,T]$ with $t\in\I$, and a \corr{two-dimensional} spatial domain, given by \corr{$\D\subseteq\R^2$} with $x\in\D$. We then write cellular population functions such that $c_1:\I\cross\D\rightarrow\R$ gives the population of IFN producing cells and $c_2:\I\cross\D\rightarrow\R$ gives the global population of consumer cells, whilst $m:\I\cross\D\rightarrow\R$ gives the non-dimensionalised concentration of IFN molecules.

In order to write as simple a model as is possible, we begin by ignoring all dynamics in the cellular populations are given simply by $c_1(t,x):=c_1(0,x)$ and $c_2(t,x):=1$, respectively. This is so that one might analyse only the communicative capabilities of the IFN itself.

We then write the dynamics of the system as a whole as an ordinary differential equation in $m(t,x)$, such that the spatial dynamics are given entirely by the diffusion of this molecule in the solution. Interferon is then systematically consumed by $c_2$, at a rate $\lambda$, and is autoreplicated within $c_1$ cells, at a rate $\phi_2$, and where this autoreplication is further \corr{stabilized by negative self-regulation}, with the rate constant $\phi_3$. Therefore, we have that
\begin{equation} \label{eq:simple_sys}
\frac{dm}{dt} = D_m\nabla_x^2m - \lambda mc_2 + (\phi_2m^2 - \phi_3m^3)c_1
\end{equation}
where $D_m$ is the coefficient for diffusion of IFN.

\begin{figure} \begin{tabular}{cccccc}
	$\lambda$ \hspace{-2em} & $t=0$ & $t=25$ & $t=50$ & $t=75$ & $t=100$ \\ $\begin{array}{c}\\[-4.5em] 0.01\end{array}$ \hspace{-1em} &
		\includegraphics[width=.16\textwidth]{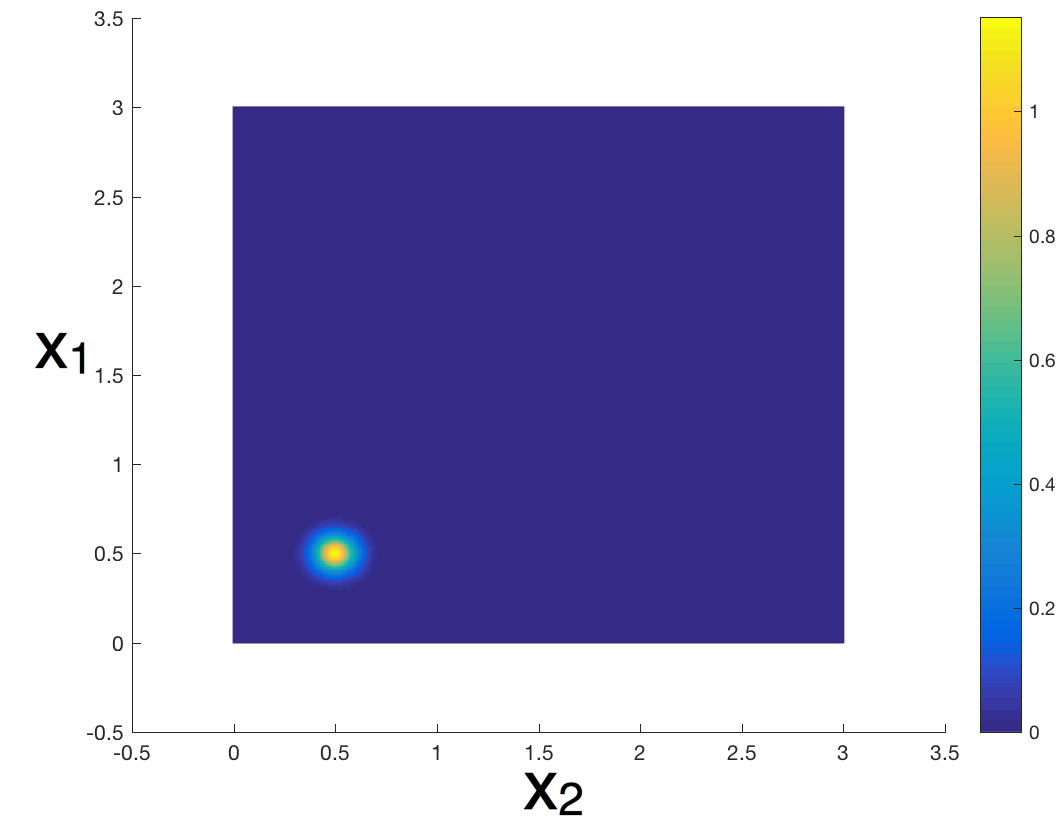} &
		\includegraphics[width=.16\textwidth]{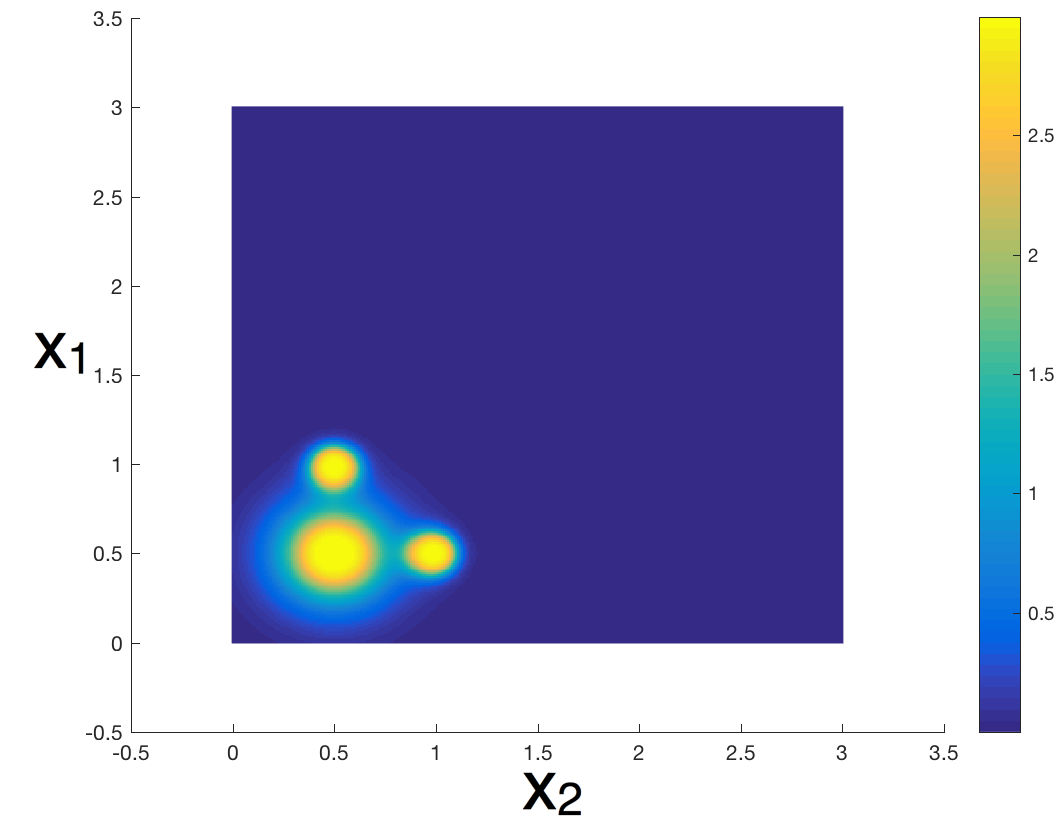} &
		\includegraphics[width=.16\textwidth]{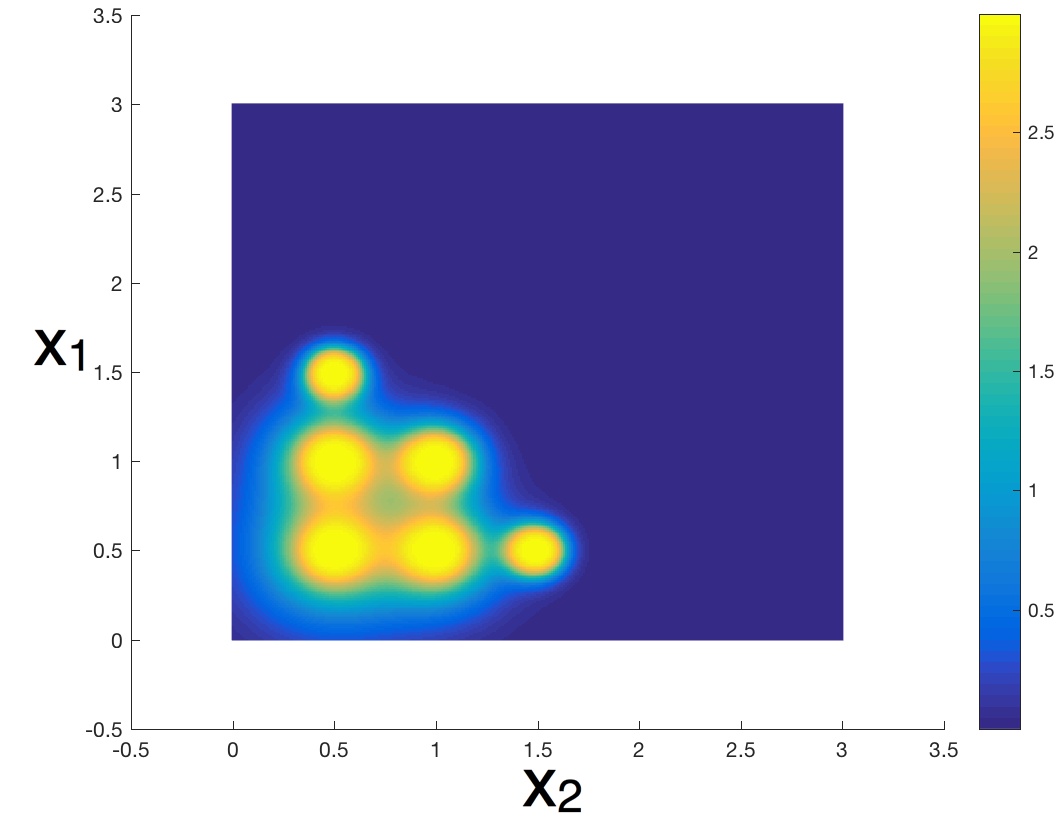} &
		\includegraphics[width=.16\textwidth]{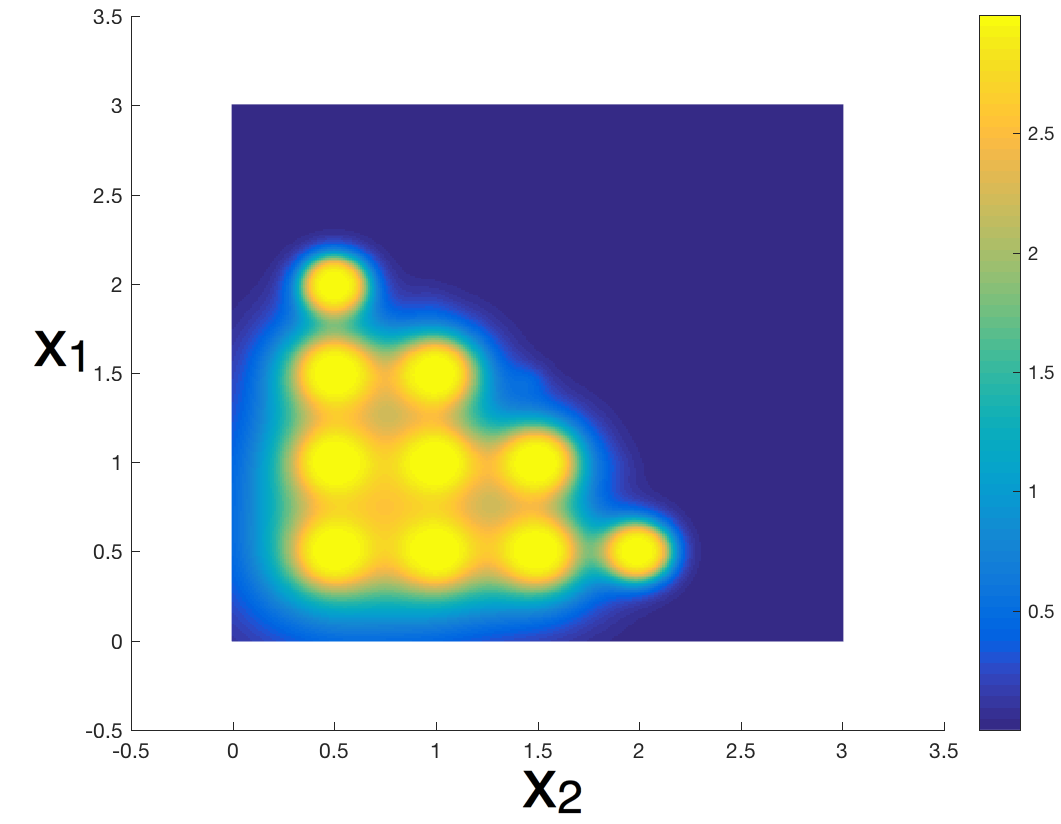} &
		\includegraphics[width=.16\textwidth]{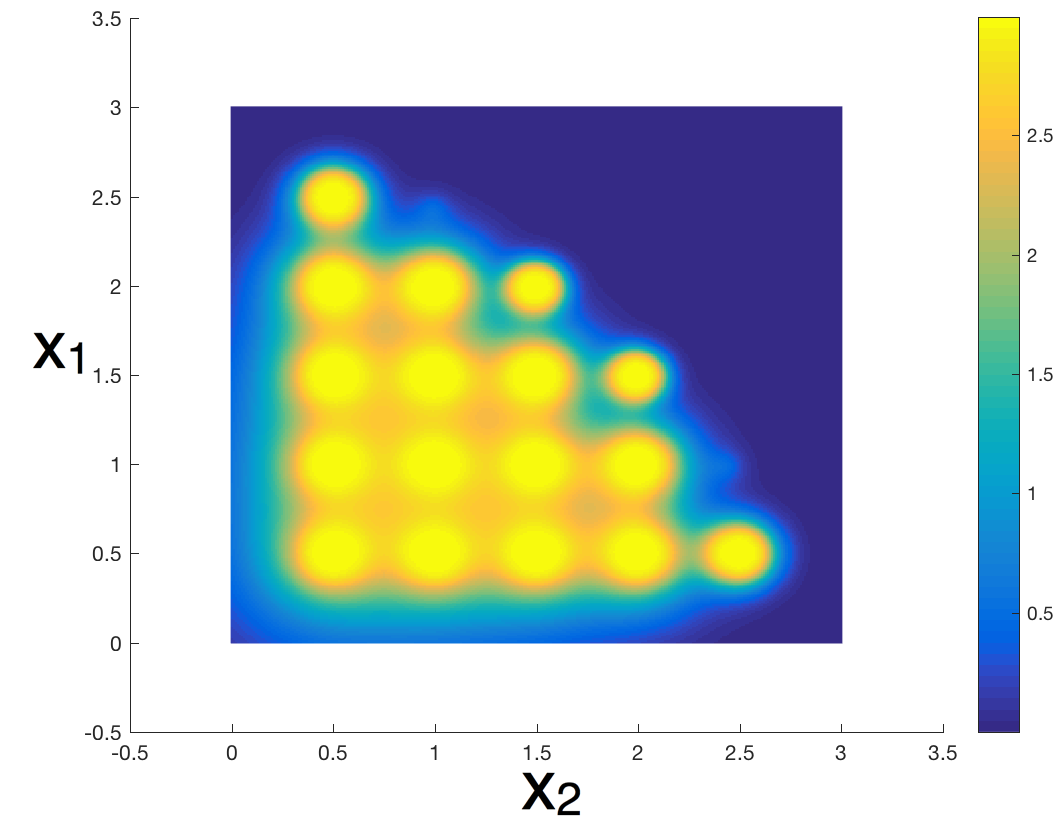} \\ $\begin{array}{c}\\[-4.5em] 0.1\end{array}$ \hspace{-1em} &
		\includegraphics[width=.16\textwidth]{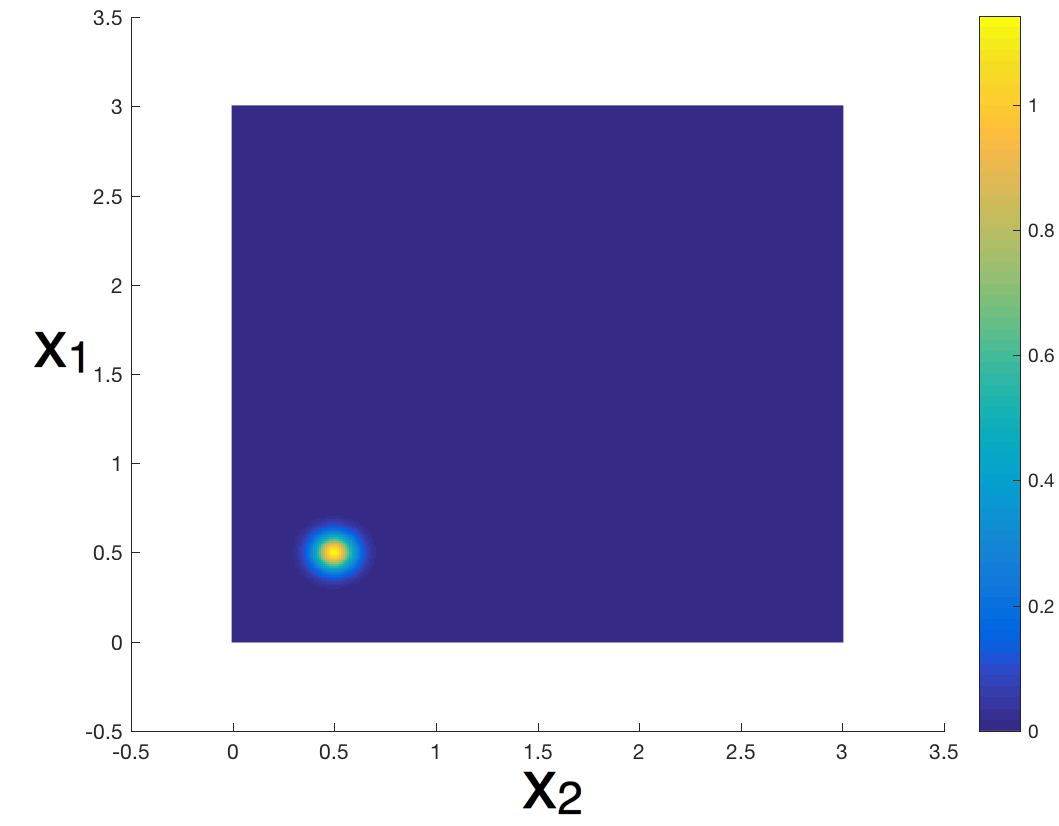} &
		\includegraphics[width=.16\textwidth]{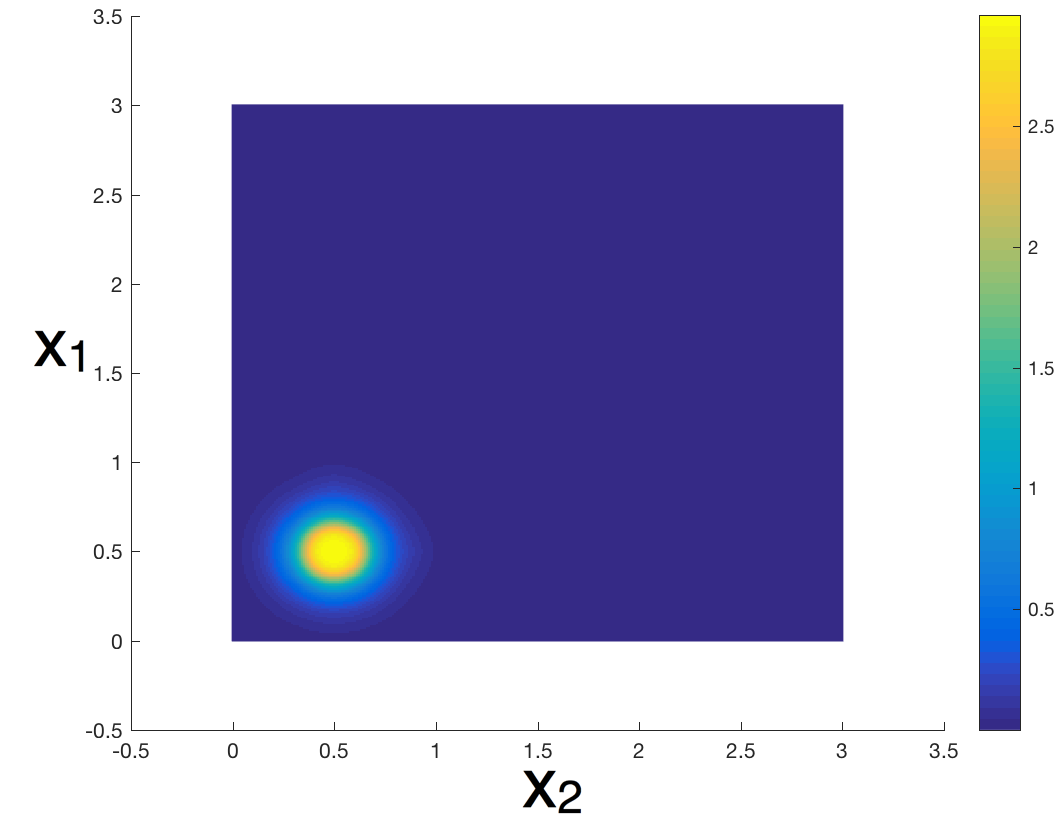} &
		\includegraphics[width=.16\textwidth]{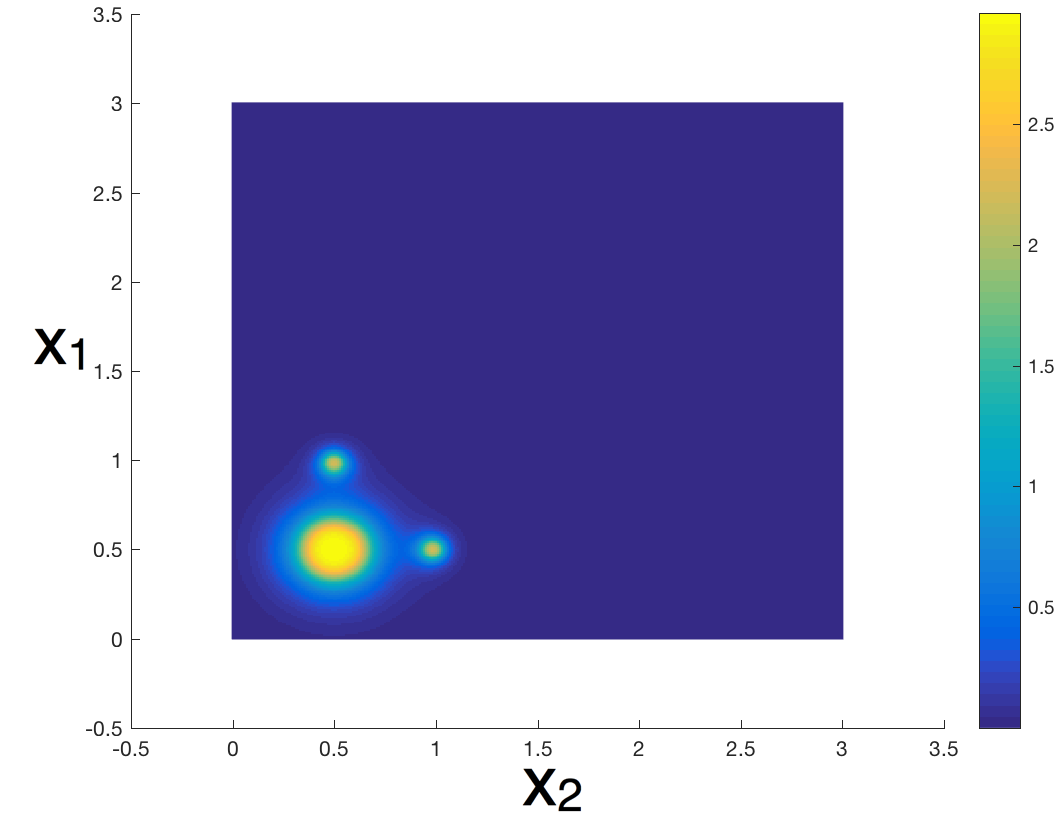} &
		\includegraphics[width=.16\textwidth]{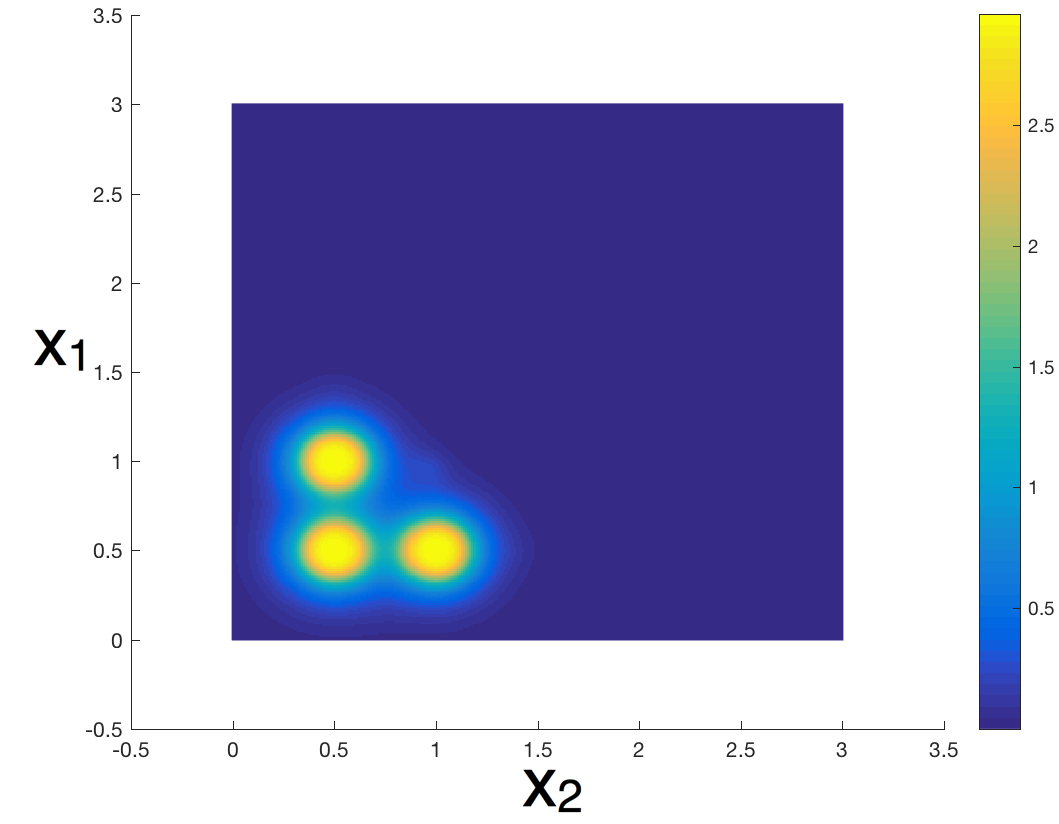} &
		\includegraphics[width=.16\textwidth]{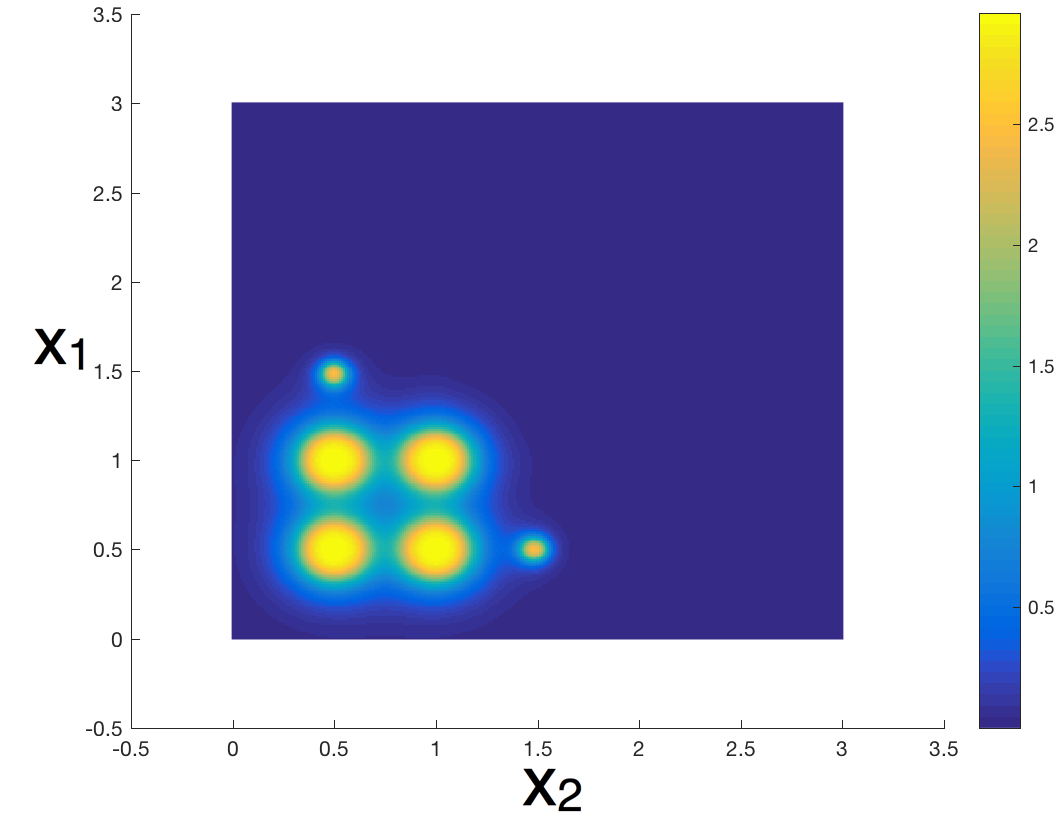} \\ $\begin{array}{c}\\[-4.5em] 0.15\end{array}$ \hspace{-1em} &
		\includegraphics[width=.16\textwidth]{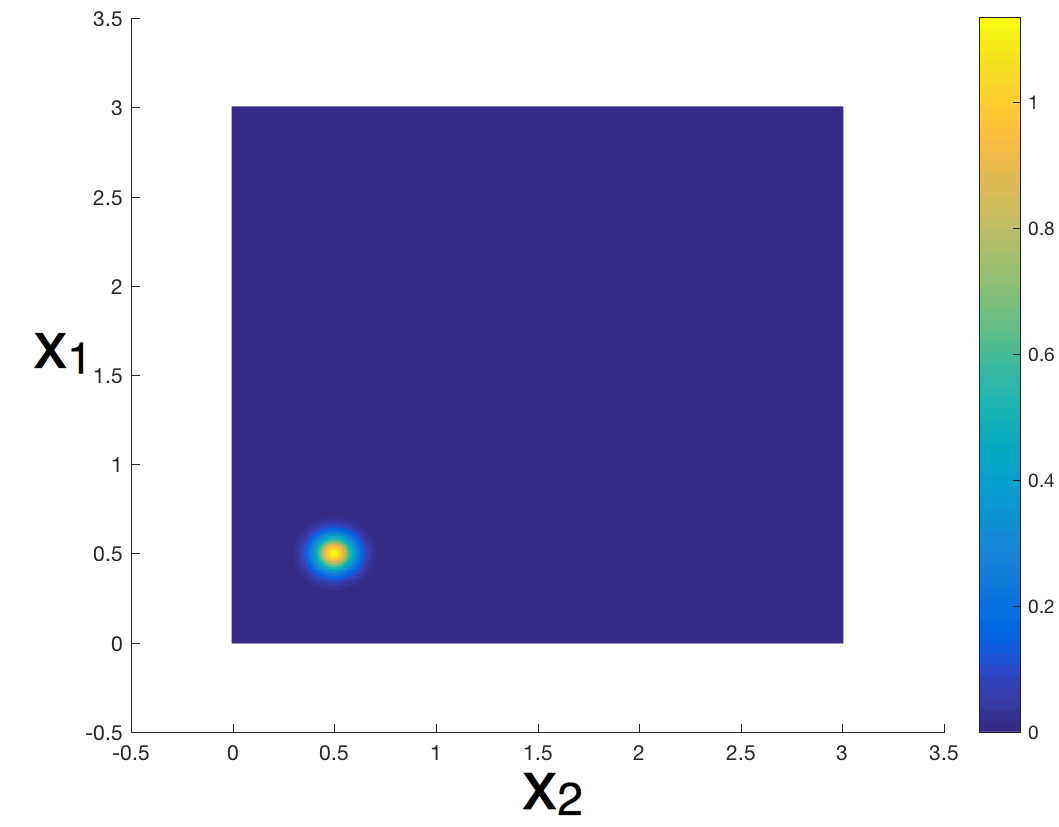} &
		\includegraphics[width=.16\textwidth]{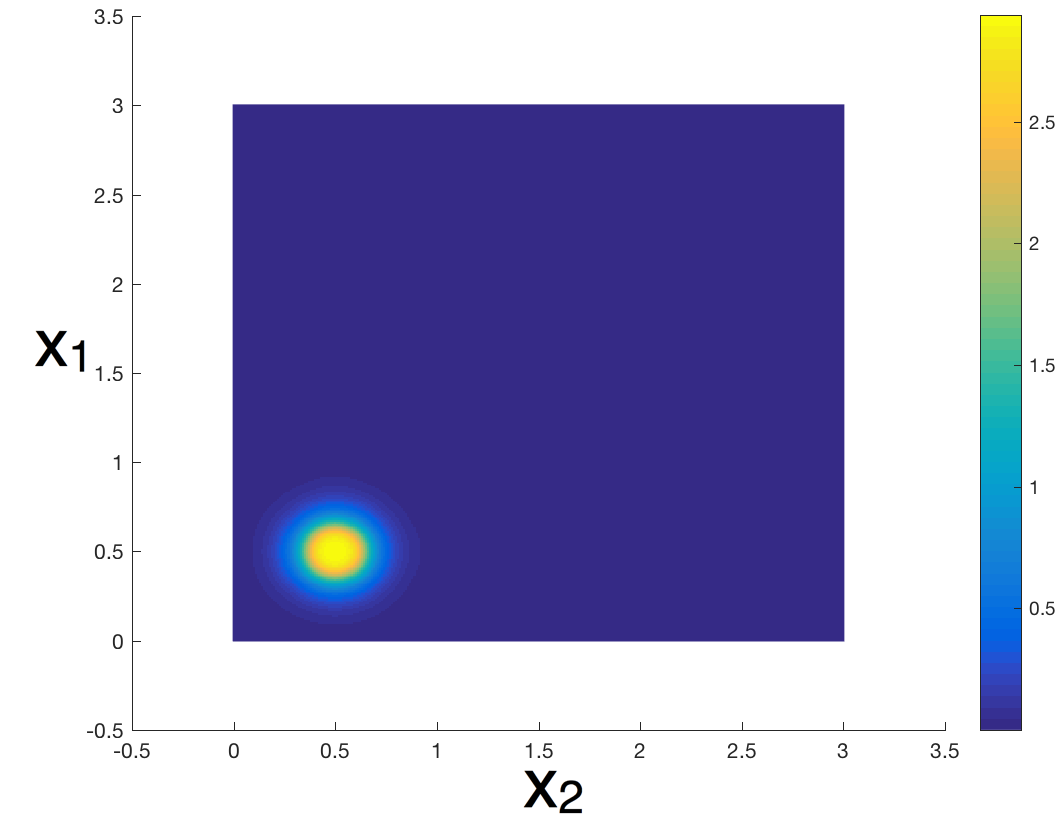} &
		\includegraphics[width=.16\textwidth]{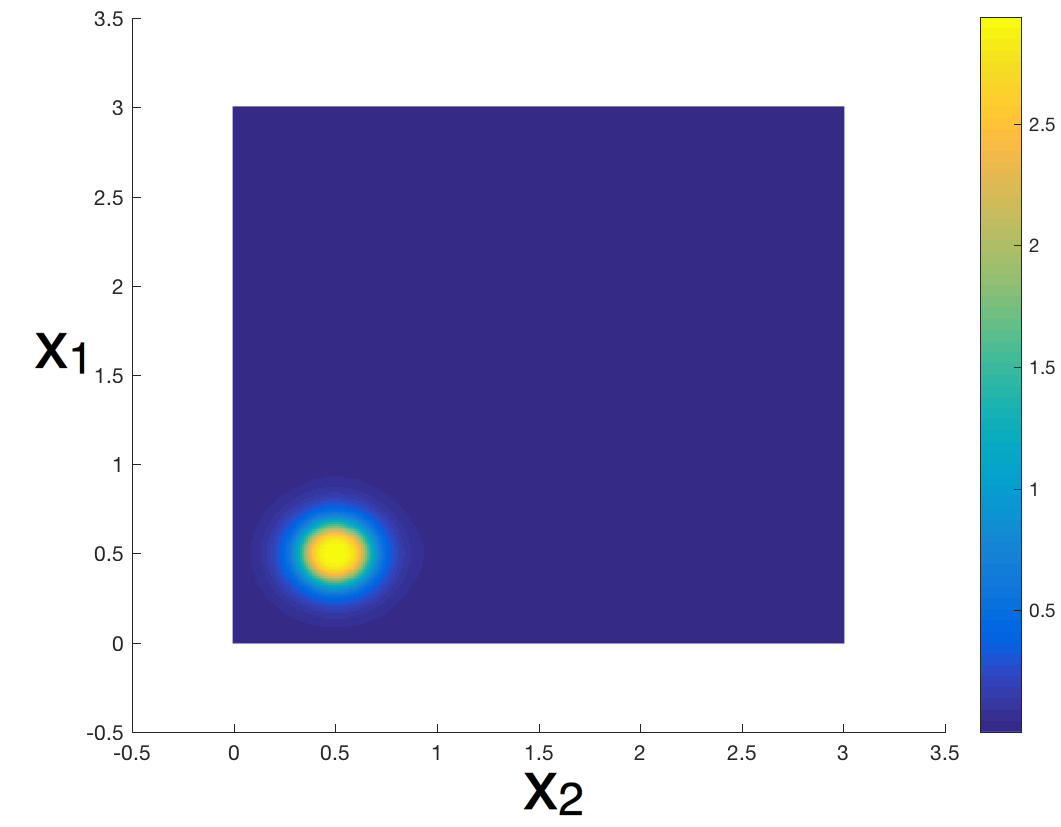} &
		\includegraphics[width=.16\textwidth]{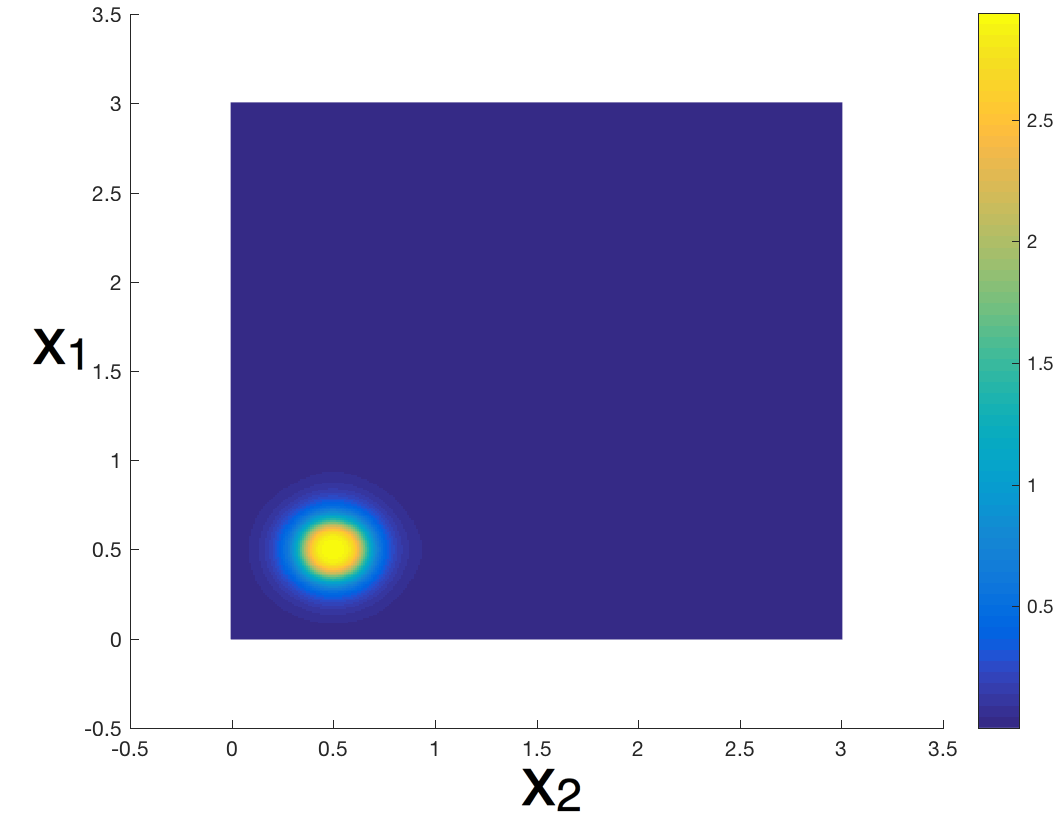} &
		\includegraphics[width=.16\textwidth]{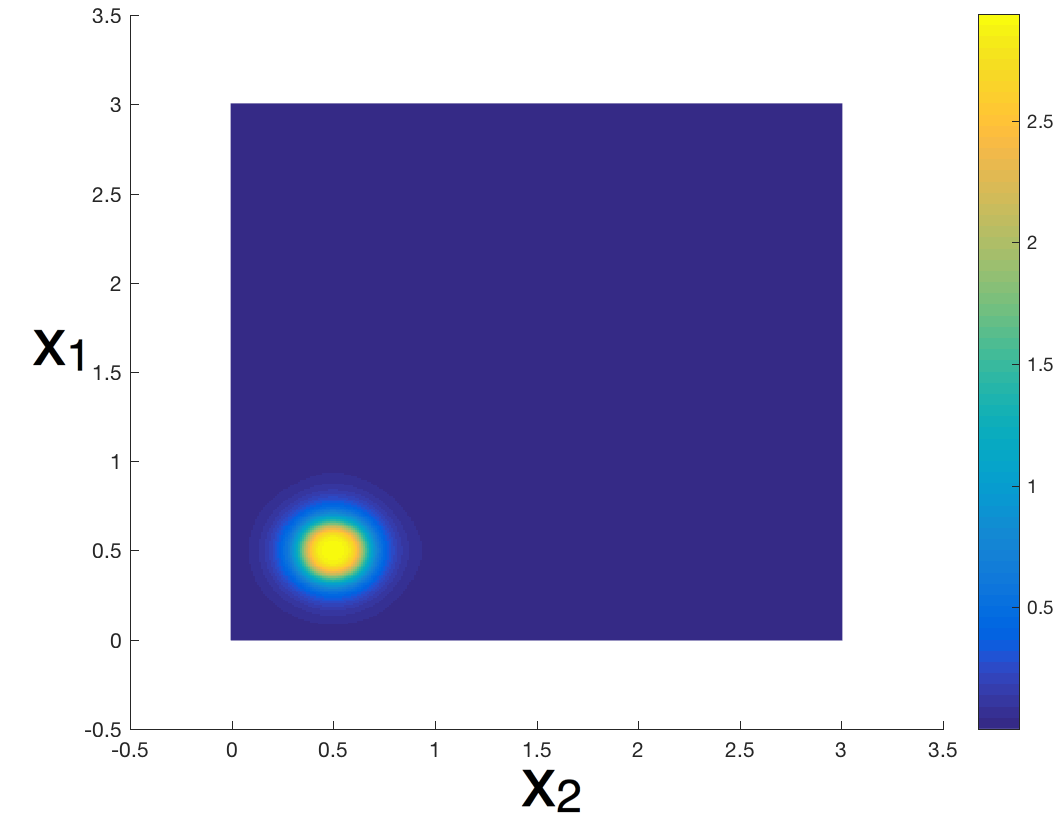} 
	\end{tabular}
	\caption{Multi-cluster results in the concentration of molecular species $m(t,x)$ from simulation of the simple model (\ref{eq:simple_sys}) for varying affinities, $\lambda\in\{0.01,0.1,0.15\}$, and for $t\in\{0,25,50,75,100\}$ respectively.} \label{fig:Results_noStruc}
\end{figure}

Simulations were performed for this system using a 4\th order Runge-Kutta predictor and MacCormack corrector, with a central difference formula used for the calculation of diffusion terms. Initial conditions for the producer cells are given by
\[ c_1(0,x) = \sum\limits_{j=1}^{5}\sum\limits_{i=1}^{5}\exp\left[ -(x_1-\frac{1}{2}i)^2-(x_2-\frac{1}{2}j)^2 \right]\,, \]
and for the IFN concentration is given by the \corr{Gaussian} distribution
\[ m_0:=m(0,x) = \exp\left[ -(\corr{x_1}-\frac{1}{2})^2 - (\corr{x_2}-\frac{1}{2})^2 \right]\,, \]
with the rate constants given by $D_m=10^{-3}$, $\phi_2=\frac{3}{4}$, and $\phi_3=\frac{1}{8}$ and $\lambda$ being variable between simulations.

The results for the simulation of system (\ref{eq:simple_sys}) show, most simply, that communicative capability increases with decreasing values for affinity of IFN for its consumer cells (Fig. \ref{fig:Results_noStruc}). The approximate threshold value for which this is true falls in the interval $\lambda\in(0.1,0.15)$ (Fig. \ref{fig:Results_noStruc}), given the values chosen for $\D_m,\,\lambda,\,\phi_2,\,\phi_3$.

This may, to some extent, give a mathematical explanation for why it may be biologically advantageous to maximise the utilisation of lower affinity IFN in a system where one wishes to stop the spread of \corr{the infection}. It could be that cells employ this methodology in order to spread a panic signal upon the initial detection of a virus and initialisation of a local IFN signal.

The explanation given by this simple model, however, does not explain the nature of the interaction between molecules and cells that allows this system to proffer communicative capabilities as it does. For example, we artificially introduce the notion that increasing the affinity of IFN molecules will increase their consumption but must still question what effect this alteration should have on the interaction with producer cells. It is difficult to intuit, also, how this increase in affinity should change the interactions that impact the metabolism of IFN within the cell. One might expect that affinity would increase production but would it also increase feedback sufficiently to dampen that response? Alterations to equation \ref{eq:simple_sys}, however, require suppositions on the desired final behaviour of the system, rather than \textit{a priori} biological assumptions.

Therefore, in response to this fundamental issue, we aim to create a more biologically descriptive model that will serve to quantify dynamics in the cell-surface receptors; the binding of these receptors by free molecules; and the consequential alterations in metabolism in a spatial context. We will also realise the interactions between these various dynamical behaviours, in order that one might better understand how the biological reality is affected by changing individual characteristics.

\section{\corr{General SAR model within the SST framework}}

\corr{We introduce here a general SST model for SAR systems. Various instances of this model can serve to study different biological problems.}

\corr{In this framework we consider that cells of the same type can differ in their states.
The cell state is described by three variables $\xi\in\Y\subset\R^\upsilon$, $y\in\P\subset\R^p$ and $\alpha \in \Gamma \subset\R^\gamma$, where $\xi$, $y$, $\alpha$ represent the total density of receptors on the cell membrane; the part of receptors that have bound ligands; and the metabolic variables, respectively.
We consider that there are $q$ different diffusible ligands  of concentrations  $m_l(t,x), 1 \leq l \leq q$. As a simplifying assumption we consider that ligands $m_k$ bind with no competition to their cognate receptors $\xi_i,\, 1 \leq i \leq \upsilon$. Competition could be easily introduced by considering that the same receptor can bind several ligands, but in this case the $y$ space has to be supplemented with extra dimensions corresponding to the simple and double charge of the receptors. The binding event can trigger the signalling and activation of metabolic variables $\alpha_k, 1 \leq k \leq \gamma$ that are responsible of the production of the ligands $m_j, 1 \leq j \leq q$.}

\corr{A spatially and structurally heterogenous cell population is described by a structured cell density, namely by a positive, integrable function $\hat c(t,x,\xi,y,\alpha)$, with $t \in (0,T]$, $x \in \D \subset \R^d$, $\xi\in\Y$, $y\in\P$, and $\alpha\in\Gamma$.

The spatial cell density $c(t,x)$ can be obtained as the marginal distribution of the structured
cell density
\begin{equation}
c(t,x) = \int\limits_{\Y\cross\P\cross\Gamma} \hat c (t,x,\xi,y,\alpha) d\xi\, dy\, d\alpha.
\end{equation}
The dynamics of the structured cell density is described by
\ahh{\begin{equation} \label{eq:GenFlux} \begin{array}{rll}
	\partt{}\hat c(t,x,\xi,y,\alpha)	 &=& \hat S(t,x,\xi,y,\alpha) - \nabla_x \cdot \hat F(t,x,\xi,y,\alpha) - \nabla_{\xi}\cdot \hat G(t,x,\xi,y,\alpha)
\\[.7em]	&&	- \nabla_y\cdot \hat H(t,x,\xi,y,\alpha)
	\ahh{- \nabla_{\alpha}\cdot \hat K(t,x,\xi,y,\alpha)}
\end{array} \end{equation}}
whose full derivation is based upon work by Domschke \textit{et al.}\cite{Domschke2016} and is given in \ref{app:Formalism}, along with a novel derivation of a structural source term, where $\hat S$ is a source term and where \ahh{$\hat F, \hat G, \hat H, \hat K$} are space-structure fluxes conjugated to the variables $x,\xi,y,\alpha$, respectively.

We then proceed to more clearly define each of the flux terms in (\ref{eq:GenFlux}) as follows.
}

\subsection{Spatial Flux}

\corr{The general form of the spatial flux equation is commonly obtained from Fick's law and is given by
\begin{equation} \begin{array}{rll} \label{eq:Flux}
	\hat F(t,x,\xi,y,\alpha)	&=	& -D_c\nabla_x \hat c(t,x,\xi,y,\alpha) \\[1em] &&
	+ \hat c(t,x,\xi,y,\alpha)\chi_{v}\nabla_xv(t,x) \\[.6em]&&
	+ \hat c(t,x,\xi,y,\alpha) \sum\limits_{i=1}^q \chi_{i}(y) \nabla_x m_i(t,x),
\end{array} \end{equation}
where the first term represents the spatial undirected diffusion of cells, the second term and third terms correspond to directed haptotactic and chemotactic cell migration, respectively.  }

\subsection{Structural Fluxes}

\corr{
We consider here the dynamics of a cell population in structure space.
Each cell of the population is characterised by its structure state vector $s=(\xi, y, \alpha)$ and by its location $x \in \D$. We consider that cells in the same location follow a dynamics defined by the vector field $\Psi$ on $s \in \Y\cross\P\cross\Gamma$, with $c(t,x)$, $m(t,x)$, $v(t,x)$ as parameters defining the local environment
\begin{equation}\label{eqdiff}
\odiff{s}{t} = \Psi (s; c(t,x), m(t,x), v(t,x)).
\end{equation}

Different cells have different initial conditions at $t=t_0$, whose distribution is given by $\hat{c}(t_0,x,s)$. Let  $s(t) = \Phi_{t,t_0}(s_0)$ be the unique solution of (\ref{eqdiff})
starting from $s_0$ at $t_0$.

Let us consider the cell sub-population located in bounded spatial $V \subset \D$ and structural $U  \subset \Y\cross\P\cross\Gamma$ boxes. A population in which each cell follows (\ref{eqdiff}) fulfils the continuity equation, namely
\begin{equation} \label{continuity} \begin{aligned}
	\int\limits_{V} \!\int\limits_{\Phi_{t,t_0}(U )} \hspace{-1.2em} \hat c (t,x,s) \,ds\,dx	=&
		\!\int\limits_{V} \!\int\limits_{U } \hspace{-.2em} \hat c (t_0,x,s) \,ds\,dx
		- \!\int\limits_{t_0}^t \!\int\limits_{\partial V} \!\int\limits_{\Phi_{t',t_0} (U)} \hspace{-1.2em} \hat{F}(t',x,s) \cdot \mathfrak{n}(x) \,ds\,d\sigma(x)\,dt' \\&
		+ \!\int\limits_{t_0}^t \!\int\limits_{V} \!\int\limits_{\Phi_{t',t_0} (U)} \hspace{-1.2em} \hat{S}(t',x,s) \,ds\,dx\,dt'\,,
\end{aligned} \end{equation}
where $\Phi_{t,t_0}(U)$ is the image of $U$ by $\Phi_{t,t_0}$, $\partial V$ is the boundary of $V$, $\mathfrak{n}(x)$ is the normal vector and $d \sigma(x)$ is the surface measure on this boundary. Performing a change of variables in the left hand side of (\ref{continuity}) we get
\begin{equation}\label{continuityp}
\int\limits_{\Phi_{t,t_0}(U)} \hat{c}(t,x,s) \,ds = \int\limits_{U}
\hat{c}(t,x,\Phi_{t,t_0}(s)) J_{t,t_0}  \,ds\,,
\end{equation}
where $J_{t,t_0} = |\det\odiff{\Phi_{t,t_0}}{s} |$ is the Jacobian determinant.

Using Stokes theorem and the first fundamental theorem of calculus in (\ref{continuity}) and further using (\ref{continuityp}) it follows
\begin{equation}\label{continuityppp}
\begin{aligned}
\int\limits_{V} \int\limits_{U}  \odiff{}{t}[\hat c (t,x,\Phi_{t,t_0}(s)) J_{t,t_0}] \,ds\,dx  &=& -\int\limits_{V} \int\limits_{ \Phi_{t,t_0} (U) } \nabla_x \cdot \hat F(t,x,s)  \,ds\,dx  \\
&+&  \int\limits_{V} \int\limits_{\Phi_{t,t_0} (U) } \hat S(t,x,s) \,ds\,dx.
\end{aligned}
\end{equation}
After changing the structure variables in the two integrals in the right hand side of
\eqref{continuityppp} we get
\begin{equation}\label{continuity4p}
\odiff{}{t}[\hat c (t,x,\Phi_{t,t_0}(s)) J_{t,t_0}] = -\nabla_x \cdot \hat F(t,x,s) J_{t,t_0} + \hat S(t,x,s) J_{t,t_0}.
\end{equation}

Using $\frac{1}{J}\odiff{J}{t} = \nabla_s \cdot \Psi (s, c(t,x), m(t,x), v(t,x))$, from (\ref{continuity4p}) we obtain the Liouville equation
\begin{equation}\label{liouville}
\partt{\hat c(t,x,s)} = - \nabla_s \cdot (\hat c(t,x,s) \Psi (s, c(t,x), m(t,x), v(t,x)))
-\nabla_x \cdot \hat{F}(t,x,s) + \hat{S}(t,x,s).
\end{equation}
Comparing this result to (\ref{eq:GenFlux}) it follows that the structural fluxes
$\hat G$,  $\hat H$, $\hat I$ are advection fluxes
\begin{eqnarray}
\hat G &=&   \hat c \Psi_{\xi} (\xi,y,\alpha; c(t,x), m(t,x), v(t,x)), \\
\hat H &=&   \hat c \Psi_{y} (\xi,y,\alpha; c(t,x), m(t,x), v(t,x))), \\
\hat I &=&   \hat c \Psi_{\alpha} (\xi,y,\alpha; c(t,x), m(t,x), v(t,x)),
\end{eqnarray}
where $\Psi_{\xi}$, $\Psi_{y}$, $\Psi_{\alpha}$ are the components of the vector $\Psi$
on the directions $\xi$, $y$, $\alpha$, respectively.
}

\subsection{Dynamics in receptoro-binding space}

\corr{Notice that each ligand binds to the available cognate receptors. Thus, the binding rate depends on the free receptor amount $\xi_i - y_i$ and is proportional to the ligand concentration $m_i$
\begin{equation} \label{binding}
b_i(\xi,y,m) = \beta_i \vartheta(\xi_i - y_i) m_i,
\end{equation}
where $\vartheta$ is a function allowing to cope with the situation when binding is thresholded in the concentration of free receptors. The unbinding rate is simply proportional to the fraction of the carrying capacity of bound receptors
\begin{equation} \label{unbinding}
u_i(y) = \eta_i y_i.
\end{equation}
Bound receptors are internalised with a rate
\begin{equation} \label{internalisation}
\iota_i (y) = k_i y_i.
\end{equation}
A subset of these internalised receptors are recycled. The timescale $\zeta_i^{-1}$ of this process results from complex interactions between receptors and scaffolds inside the endosome \cite{Grant2009} and depends nonlinearly on $y$. Therefore, the recycling rate reads
\begin{equation} \label{recycling}
r_i (y) = \zeta_i (y) y_i,\quad 0 \leq \zeta_i (y) \leq k_i.
\end{equation}
Receptors are synthesised by the cell with a rate $p_i(\alpha, \xi)$ that depends on the metabolic variables $\alpha$ and also on actual concentration of receptors $\xi$ and are lost by various mechanisms with a rate proportional to $\xi$
\begin{equation}
d_i(\xi) = d_i \xi_i.
\end{equation}
In summary, the receptoro-binding variables of a single cell follow the differential equations
\begin{eqnarray}\label{odesreceptors}
\odiff{\xi}{t} &=& \Psi_{\xi}(\xi,y,\alpha)  = {\mathcal P}(\alpha,\xi) - {\mathcal D} \xi + ({\mathcal R} (y)-{\mathcal I}) y \\
\odiff{y}{t} &=& \Psi_{y}(\xi,y)  = \beta {\mathcal B}(\xi - y) m - ({\mathcal U}+ {\mathcal I}) y,
\end{eqnarray}
where $\mathcal P$, $\mathcal D$, $\mathcal R$, $\mathcal I$,$\beta$, $\mathcal B$, $\mathcal U$ are diagonal matrices with diagonal entries $p_i$, $d_i$, $\zeta_i$, $k_i$, $\beta_i$, $\vartheta(\xi_i-y_i)$, $\eta_i$, respectively.

It follows that the advection fluxes in receptor and binding spaces are
\begin{eqnarray}\label{fluxesreceptors}
\hat G(t,x,\xi,y,\alpha) &=& \hat c(t,x,\xi,y,\alpha) [{\mathcal P}(\alpha,\xi) -
{\mathcal D} \xi + ({\mathcal R} (y)- {\mathcal I}) y ]\\
\hat H(t,x,\xi,y,\alpha) &=&  \hat c(t,x,\xi,y,\alpha) [\beta {\mathcal B}(\xi - y) m - ({\mathcal U}+ {\mathcal I})y].
\end{eqnarray}
 }

\corr{\subsection{Dynamics in metabolic space}

The part of internalisation flux that is not recycled and that escapes lysosome degradation triggers signaling and induces changes of the metabolic variables $\alpha$.
We use a flux-based description of these variables that considers that there are $\gamma$ irreversible metabolic fluxes each one producing a different molecule. The reversible case can be simply obtained by doubling the number of variables for each reversible flux. To each one of these fluxes we associate a scalar variable $0 \leq \alpha_i \leq 1$, meaning no production activity and maximum production activity for $\alpha_i=0$ and $\alpha_i =1$,
respectively.
 In order to represent competition between fluxes  we impose the condition $\sum_{i=1}^r \alpha_i \leq 1$. Thus $\alpha \in \Gamma$, where $\Gamma =  \{(\alpha_1,\ldots,\alpha_r) \mid  0 \leq \alpha_i \leq 1, \sum_{i=1}^r \alpha_i \leq 1 \}$ is a simplex. This description is equivalent to the space of admissible fluxes in stoichiometric and flux balance analysis of metabolic networks where $\alpha_i, \, 1 \leq i \le r$ represent activities of extreme pathways or currents \cite{clarke1988stoichiometric,schilling2000theory}. The dynamics in the metabolic space is described phenomenologically imposing the invariance of the simplex $\Gamma$ as fundamental property. A possible such choice is
\begin{equation}\label{odemetab}
\odiff{\alpha_i}{t} = \Psi_{\alpha_i}(y,\alpha) = f_i(y) ( 1 - \alpha_i) - \mu_i \alpha_i \,,
\end{equation}
where $\mu_i \geq \mu_0 > 0$, $f_i \leq f_0,\, f_0 > 0$, $r f_0 < (f_0 + 1)$. The corresponding advection
flux in the metabolic structure space is $\hat I = (I_1,\ldots,I_\gamma)$ with the components
\begin{equation}\label{fluxmetab}
I_i = \hat c [f_i(y) (1 - \alpha_i) - \mu_i \alpha_i]\,.
\end{equation} }

\corr{\subsection{Spatial dynamics of diffusible ligands}}

Begin by denoting $\bar{m}:=[m_1,\dots,m_p,m_{p+1},\dots,m_q]^T$, with $m_j:=m_j(t,x)$, as the total vector of molecular species, where there exist $q$ molecular species of which the first $p\leq q$ species are binding ligands.

Then, the spatial dynamics of all molecular species are defined by a diffusive process, and with a species specific diffusion coefficient \corr{$D_{m_j}$} for $m_j(t,x)$. The binding ligands, within the molecular species, are removed from the population of free molecules through binding. All molecules are produced by the cellular population, in a metabolic-activity-dependent manner, and are either contributed to or detracted from by a situation specific sink or source function $\bar\Theta(t,x)$. Therefore, denoting the $q$-dimensional vectors of parameters $\bar{\cdot}:=[\cdot_1,\dots,\cdot_q]^T$, we obtain the relations for molecular species as
\corr{
\begin{equation} \begin{aligned}
	\partt{\bar{m}} =& \nabla_x\cdot\text{diag}( D_{\bar{m}} )\nabla_x\bar{m} - \varepsilon\int\limits_\Gamma \int\limits_\P \int\limits_\Y (\text{diag}(\bar{\beta} \vartheta(\xi - y)) \bar{m} - \text{diag}(\bar{\eta})\bar y) \hat{c} \,d\xi\,dy\,d\alpha \\&
	+ \int\limits_\Gamma \int\limits_\P \int\limits_\Y \bar{\phi}_\alpha(\alpha)(1 - \bar{m}) \hat{c} \,d\xi\,dy\,d\alpha + \bar\Theta(t,x)\,,
\end{aligned} \end{equation}
}
with $\bar{\phi}_\alpha(\cdot):\Y\to\R^q$ defining a vector of production values for each molecular species given the cellular metabolic activity level, $\alpha$; $\bar\beta=[\beta_1,\dots,\beta_p,0,\dots,0]^T$; and $\bar\eta=[\eta_1,\dots,\eta_p,0,\dots,0]^T$;\corr{ $\varepsilon$ is a constant converting surface to volume binding/unbinding rates.}

\section{Particularised IFN-Based Model}

It is necessary to first have a discussion about the context into which we shall place this model, with respect to the generalised SST framework for SARs. First of all, and for simplicity, we neglect the receptor space and source terms in the IFN case. This is due to the fact that we do not consider the creation of IFN SARs, but rather their behaviour and spatial recruitment, and the change in binding in the IFN case appears to be related to affinity rather than flux of the binding proteins themselves.

It should be clear, \corr{that a main concern in modelling the IFN system} is the numerical simplification involved in the reduction of the number of necessary dimensions under consideration. This has succeeding consequences in terms of our ability to intuit the results of the system and better understand both the SST framework, and the internal processes for communicative SARs. For this reason, we also neglect, initially, the spatial dynamics of the SAR cells and concentrate first on cell-cell communication mediated by the diffusible ligand.

Now, we contextually define the binding variable, $y\in\P$ with \corr{$v=p=1$}, $0 \leq y \leq \xi=1$, such that increasing values of $y$ correspond to the increasing concentration of bound IFN-IFNAR1-IFNAR2 complexes for some given \ahh{$(t,x,\alpha)\in\I\cross\D\cross\Gamma$}.

The metabolic variable, $\alpha\in\Gamma$, is somewhat more complicated in biological terms since we wish to encapsulate a state of the cell under which a certain reaction is more likely to take place. In the particular case of IFN, for example, we understand the metabolic variable as describing a state of the cell wherein \ah{ISGs implicated in the production of or response to IFN (such as IRF-7, implicated in production, or USP18, a key regulator of the cellular response to IFN)} are more frequently transcribed. Therefore, begin by describing $\alpha=0$ as a state in which ISGs are not transcribed and $\alpha=1$ as some state where ISGs are transcribed at their physiologically maximal rates. Then we understand $\alpha$, itself, as encapsulating the propensity for the cell to proactively transcribe ISGs through the activity of the Jak-Stat pathway.

Within this paradigm, then, these two variables will interact in the following way. Begin by considering a scenario in which one cluster of IFN SARs are stimulated by a single initial dosage of IFN. The cell will bind these IFN molecules and increase in binding state of the cell, $y$, will form the IFN-IFNAR1-IFNAR2 complex and initiate the reactions of the Jak-Stat pathway. This will subsequently increase the cells metabolic state, $\alpha$, of the cell and cause the increased production of IFN. \ah{The increase in transcription of ISGs, specifically USP18, will also cause a decrease in the efficacy of the ternary complex (IFNAR1-IFN-IFNAR2) assembly  \cite{Wilmes2015} or maximal effective binding, $y$.} This, in turn, will subsequently lead to a decrease the physiological concentration of the Jak-Stat reactions and reduce the metabolic state, $\alpha$, of the cell.

\subsection{Unthresholded Binding Model}
\ah{Throughout this model, we assume a homogeneous and constant concentration of biological pathogen, such that IFN response is consistently encouraged.} We have chosen illustrative values for the binding rates, consistently with previous models \cite{Domschke2016,Trucu2016}, but with the difference that we consider here the negative feedback loop of the IFN system between the metabolic state of the cell and the binding of molecular species to the surface. In this respect, we consider binding to be non-dimensionalised and that feedback causes the maximal binding rate to decrease linearly with the metabolic state of the cell such that the range of values of $y$ for which positive binding exists is given by $y<1-\alpha$. Thus, we consider that the binding dynamics of molecular species to the surface, \ahh{$b : \I\cross\D\cross\P\cross\Gamma\rightarrow\mathbb{R}$, can be given by
\[ b(m,y,\alpha) := \beta(1-y-\alpha)m \]}
where $\beta$ is the binding rate constant for IFN.

When we say 'binding' in this context, we actually make a generalisation of the concept of 'meaningful binding' which is to say that binding is sufficient to allow for recruitment of the secondary complex (IFNAR2) and subsequent co-phosphorylation of their protein tails.

\corr{The rate of removal of bound molecular species from the surface of the cell
has a first component corresponding to unbinding and a second component corresponding
to internalisation and degradation of bound receptors.}
Therefore, we consider that the removal of species from the cell surface, $d : \mathcal{P}\rightarrow\mathbb{R}$, can be given by
\[ d(y) := (\eta + \delta)y\,, \]
where $\eta$ gives the unbinding rate of molecules from the surface of $c_1(t,x,y,\alpha)$ and $\delta$ gives the rate of cellular degradation of bound IFN.

Further, we make the assumption that the gene responsible for regulating the production of IFN has a default transcriptional state of `off', such that the gene is not transcribed unless appropriately upregulated. Therefore we arrive at a relation for the advective rate for change in metabolic profile, \ahh{$\mu : \mathcal{P}\times\Gamma \rightarrow \mathbb{R}$, of the cells  which is given by
\[ \mu(y,\alpha) := \delta y(1-\alpha) - \mu_0\alpha\, \]}
where $\delta y$ is the internalisation-degradation rate (as above) and $\mu_0$ is the intrinsic metabolic restoration rate, the purpose of which is to restore the default metabolic position of the cell $\alpha = 0$. The term \corr{$(1-\alpha)$ is chosen such that the metabolic state of the cell might never exceeds a maximum value
normalized to one}.

Production of $m$ with respect to the metabolic state of the cell is given by the production rate function \ahh{$\phi : \Gamma\rightarrow\mathbb{R}$} and is assumed to be of the form
\[ \phi(\alpha) := \phi_\alpha\alpha(\alpha - \theta_\alpha)\,, \]
where $\phi_\alpha$ is the rate constant for metabolic production of $m$ and $\theta_\alpha$ is some thresholding value above which the cell become metabolically active with respect to the production of IFN, $m$.

\medskip

For reasons that will become clear in the following subsection we call this model the {\em unthresholded binding} model, which is then written \ahh{
\begin{equation} \left\{ \begin{array}{rll}
	\frac{\partial c_1}{\partial t}	& = &	-\nabla_y\cdot\left[ \beta(1-y-\alpha)m - (\eta + \delta)y \right]c_1 - \nabla_\alpha\cdot\left[ \delta(1-\alpha)y - \mu_0\alpha \right]c_1 	\\[.3cm]
	\frac{\partial c_2}{\partial t} 	& = & 	0	\\[.3cm]
	\frac{\partial m}{\partial t}		& = &	\nabla_x D_m \nabla_x m - \int\limits_{\Gamma} \int\limits_{\mathcal{P}} (\beta(1-y-\alpha)m-\eta y)\varepsilon c_1(t,x,y,\alpha) \,dy\,d\alpha \\[.3cm] &&
	\int\limits_{\mathcal{P}} \int\limits_{\Gamma} \phi_\alpha\alpha(\alpha - \theta_\alpha)(1-m)c_1(t,x,y,\alpha) \,d\alpha\,dy - \lambda m c_2 \,.
\end{array} \right. \label{eq:IFNmodel} \end{equation} }

\subsection{Thresholded Binding Model}

\corr{There are several alternative interpretations of potency of a ligand
for meaningful binding and signaling triggering \cite{Kersh1998}. One interpretation
associate this potency to the product between concentration and affinity of the ligand,
suggests that ligands are detected irrespective to their quality as long as their
concentration is above a threshold. Thresholds in the number of triggered receptors have been  observed for immune T cells \cite{Viola1996}.
The second interpretation is based on kinetic proof-reading  and suggests that  a minimal binding
time is needed for a given ligand to trigger signaling \cite{Francois2016}. 
The correlation between binding time characteristics and immune cell activation
is confirmed by several studies \cite{Kersh1998,Gascoigne2001}.
Furthermore, recent dynamical studies demonstrated the phosphorylation of STAT2 to follow the formation of the complex (which is more or less instantaneous, $<1$ second) by approximately 8 seconds \cite{Lochte2014} for complete activation. All these studies
suggest the intrinsic assumption that meaningful binding requires that receptor-ligand complex to be bound for at least a minimal time $\tau_{\min}$.
In general, depending on the comparison between the timescales of meaningful complex
formation and dissolution and those of activation of the signaling processes it is possible that
both concentration and temporal thresholds apply to the ligand recognition.  
We do not aim to resolve this issue here. Because our model does not account for
binding time heterogeneity, we simply replace the temporal threshold by a concentration one, considering that there is a function $\tau_b(m)$ relating the concentration of
ligands to the binding time.}   
Then, for some concentration $m(t,x)=\theta_m$ we have that
\[ \tau_b(\theta_m) = \tau_{\min} \]
such that $\theta_m$ gives the concentration of $m$ sufficient for effective binding of the IFNAR2 protein and IFNAR1-IFNAR2 complex. \corr{In order to cope with this threshold effect,} we rewrite the binding flux term as
\[ b(y,\alpha,m) := \beta(1-y-\alpha)(m-\theta_m). \]

Substituting this new relation back into our model, we obtain \corr{the {\em thresholded binding model}} \ahh{
\begin{equation} \left\{ \begin{array}{rll}
	\frac{\partial c_1}{\partial t}	& = &	\!-\!\nabla_y\!\cdot\!\left[ \beta(1-y-\alpha)(m-\theta_m) - (\eta + \delta)y \right]c_1 \!-\!	\nabla_\alpha\!\cdot\!\left[ \delta(1-\alpha)y - \mu_0\alpha \right]c_1 	\\[.3cm]
	\frac{\partial c_2}{\partial t} 	& = & 	0	\\[.3cm]
	\frac{\partial m}{\partial t}		& = &	\nabla_x D_m \nabla_x m - \int\limits_{\Gamma} \int\limits_{\mathcal{P}} (\beta(1-y-\alpha)m-\eta y)\varepsilon c_1(t,x,y,\alpha) \,dy\,d\alpha \\[.3cm] &&
	\int\limits_{\mathcal{P}} \int\limits_{\Gamma} \phi_\alpha\alpha(\alpha - \theta_\alpha)(1-m)c_1(t,x,y,\alpha) \,d\alpha\,dy - \lambda m c_2 \,,
\end{array} \right. \label{eq:IFNmodel_th} \end{equation} }

\subsubsection{Spatially Dynamic, Thresholded Binding Model}

Finally, we consider a spatially dynamic system wherein cells are able to move through the spatial domain. We choose to endow this system with 2 primary functions of migration:
\begin{itemize}
	\item[(i)]	diffusion, by virtue of immune cells' natural inclination to motility, and
	\item[(ii)]	chemotaxis, by virtue of immune cells' ability to actively respond to an immune-response signal as a recruitment signal.
\end{itemize}
In stating this, we therefore assume that the immune cell will interpret the presence of IFN as a response to, for example, a viral threat to the body and respond to this signal by migrating towards its origin. We further assume that even in the absence of an IFN gradient, cell Brownian motion
will generate spatial fluxes leaving regions of highest cell concentration.

We thusly rewrite the system as
\begin{equation} \left\{ \begin{array}{rll}
	\frac{\partial c_1}{\partial t}	& = &	\nabla_x\cdot D_{c_1}\nabla_xc_1 - \nabla_y\cdot\left[ \beta(1-y-\alpha)(m-\theta_m) - (\eta + \delta)y \right]c_1 \\[.2cm]&&
	- \chi_m\nabla_x\cdot c_1\nabla_xm - \nabla_\alpha\cdot\left[ \breve\mu_+(1-\alpha)y - \mu_- \right]c_1 	\\[.3cm]
	\frac{\partial c_2}{\partial t} 	& = & 	0	\\[.3cm]
	\frac{\partial m}{\partial t}		& = &	\nabla_x D_m \nabla_x m - \int\limits_{\Upsilon} \int\limits_{\mathcal{P}} (\beta(1-y-\alpha)-(\eta + \nu_r\delta)y)\varepsilon c_1(t,x,y,\alpha) \,dy\,d\alpha \\[.3cm] &&
	\int\limits_{\mathcal{P}} \int\limits_{\Upsilon} \phi_\alpha\alpha(\alpha - \theta_\alpha)(1-m)c_1(t,x,y,\alpha) \,d\alpha\,dy - \lambda m c_2 \,,
\end{array} \right. \label{eq:IFNmodel_ch} \end{equation}

\section{Results from Numerical Simulations} \label{sec:Non_Spatial_Results}

\begin{figure}[t!] \centering \begin{tabular}{ccccc}
	$\begin{array}{c} \\[-2.5cm] c_\alpha \end{array}$ &\hspace{-2em}
		\includegraphics[width=.22\textwidth]{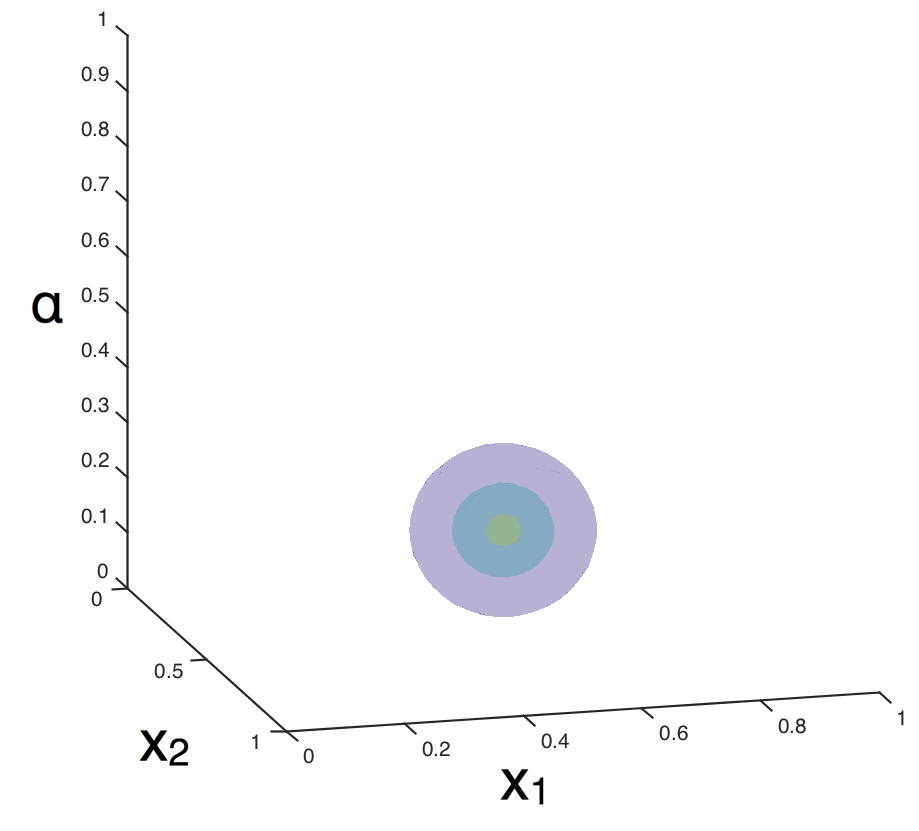} &\hspace{-2em}
		\includegraphics[width=.22\textwidth]{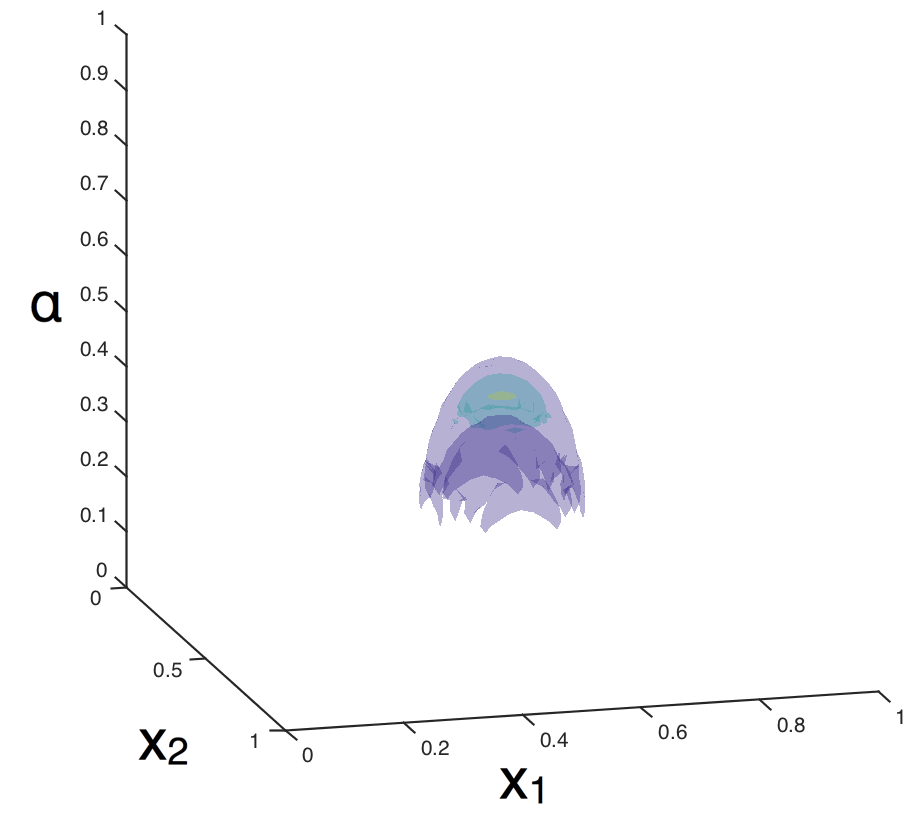} &\hspace{-2em}
		\includegraphics[width=.22\textwidth]{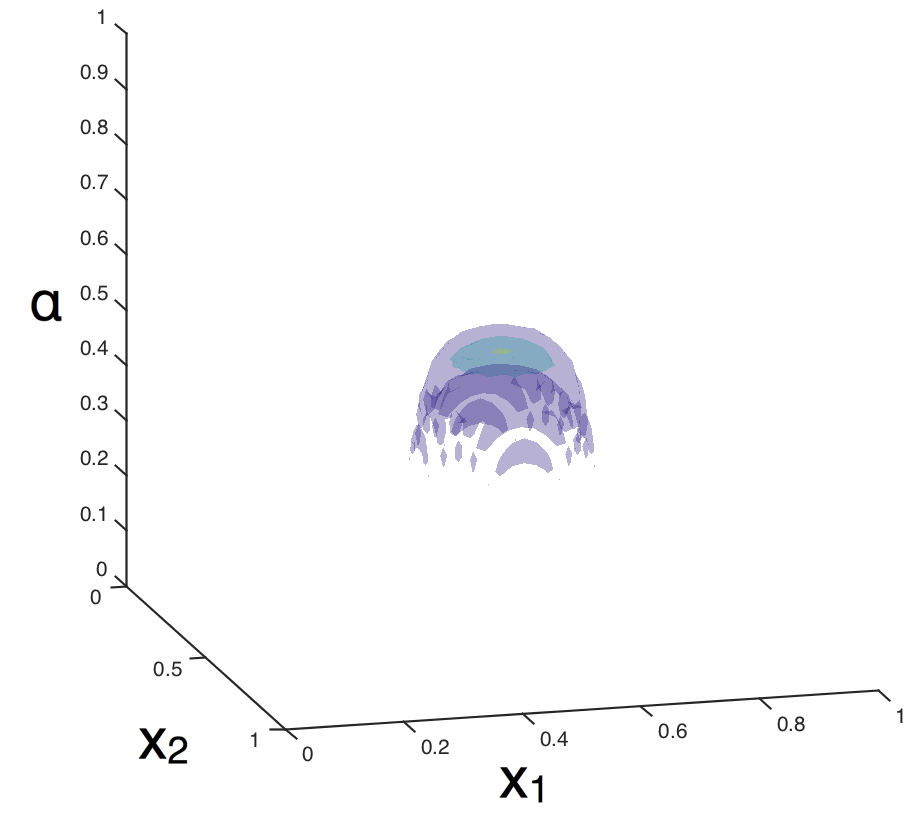} &\hspace{-2em}
		\includegraphics[width=.22\textwidth]{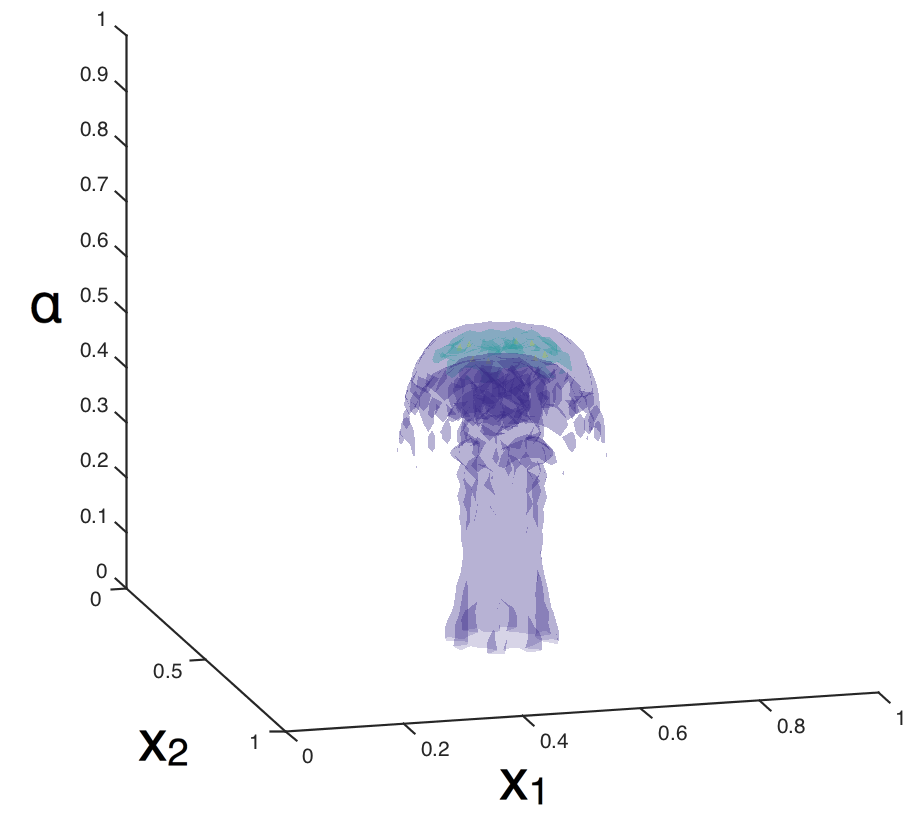}
	\\ $\begin{array}{c} \\[-2.5cm] c_y \end{array}$ &\hspace{-2em}
		\includegraphics[width=.22\textwidth]{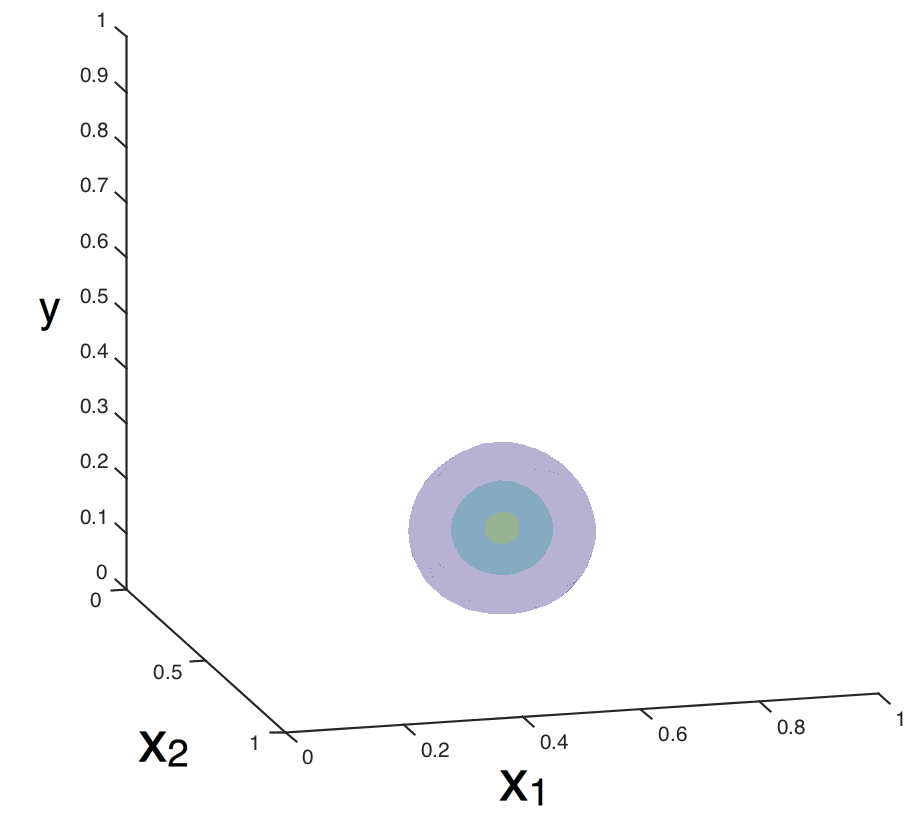} &\hspace{-2em}
		\includegraphics[width=.22\textwidth]{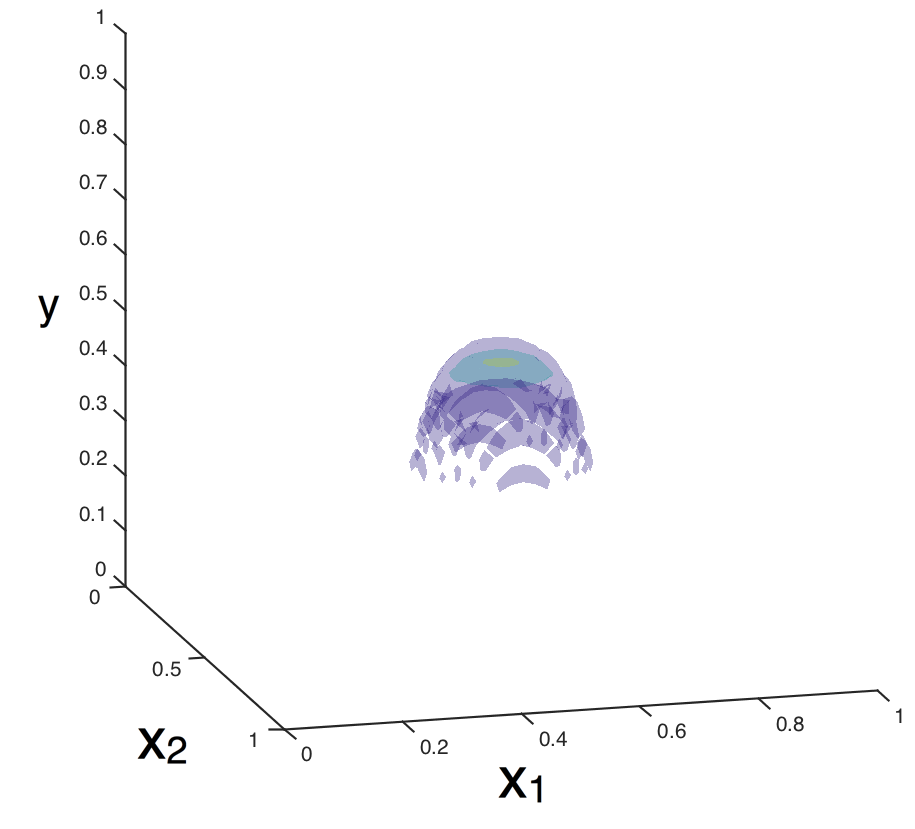} &\hspace{-2em}
		\includegraphics[width=.22\textwidth]{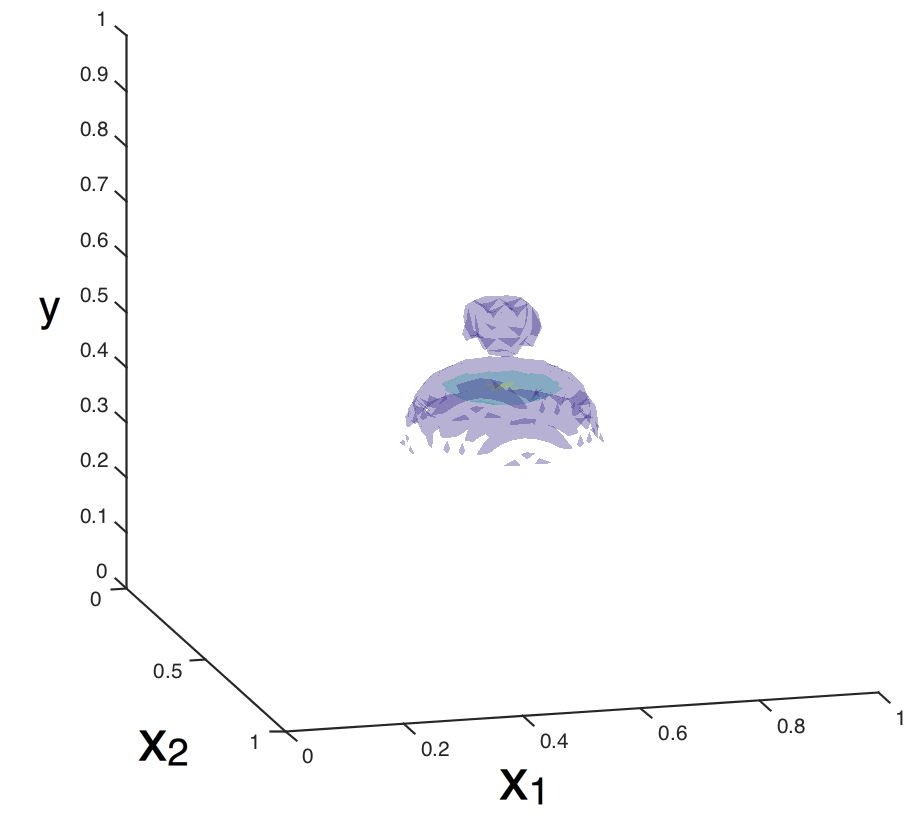} &\hspace{-2em}
		\includegraphics[width=.22\textwidth]{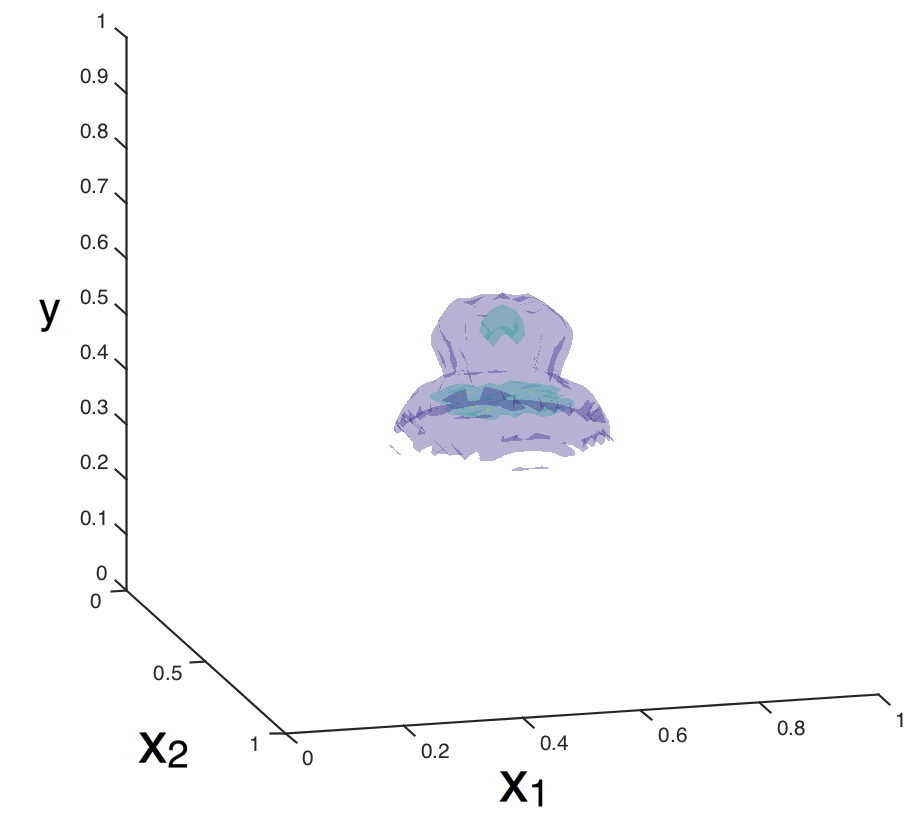}
	\\ $\begin{array}{c} \\[-2.5cm] \breve{c} \end{array}$ &\hspace{-2em}
		\includegraphics[width=.22\textwidth]{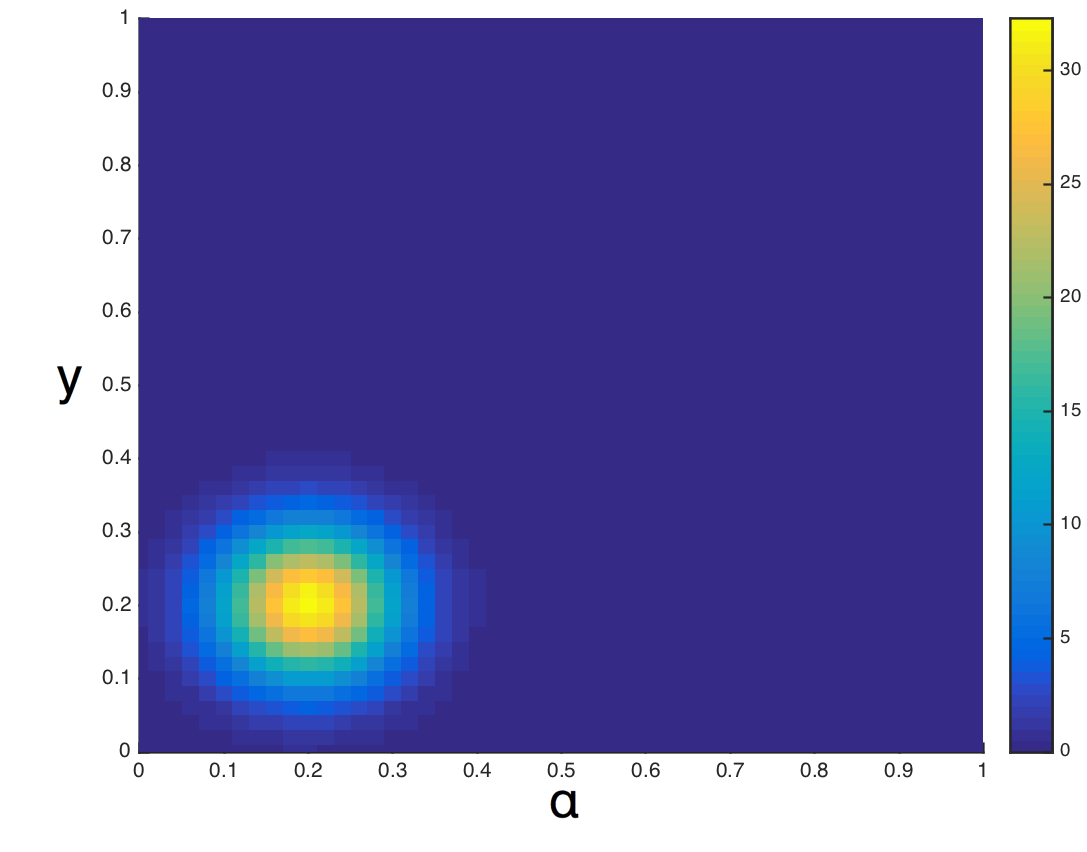} &\hspace{-2em}
		\includegraphics[width=.22\textwidth]{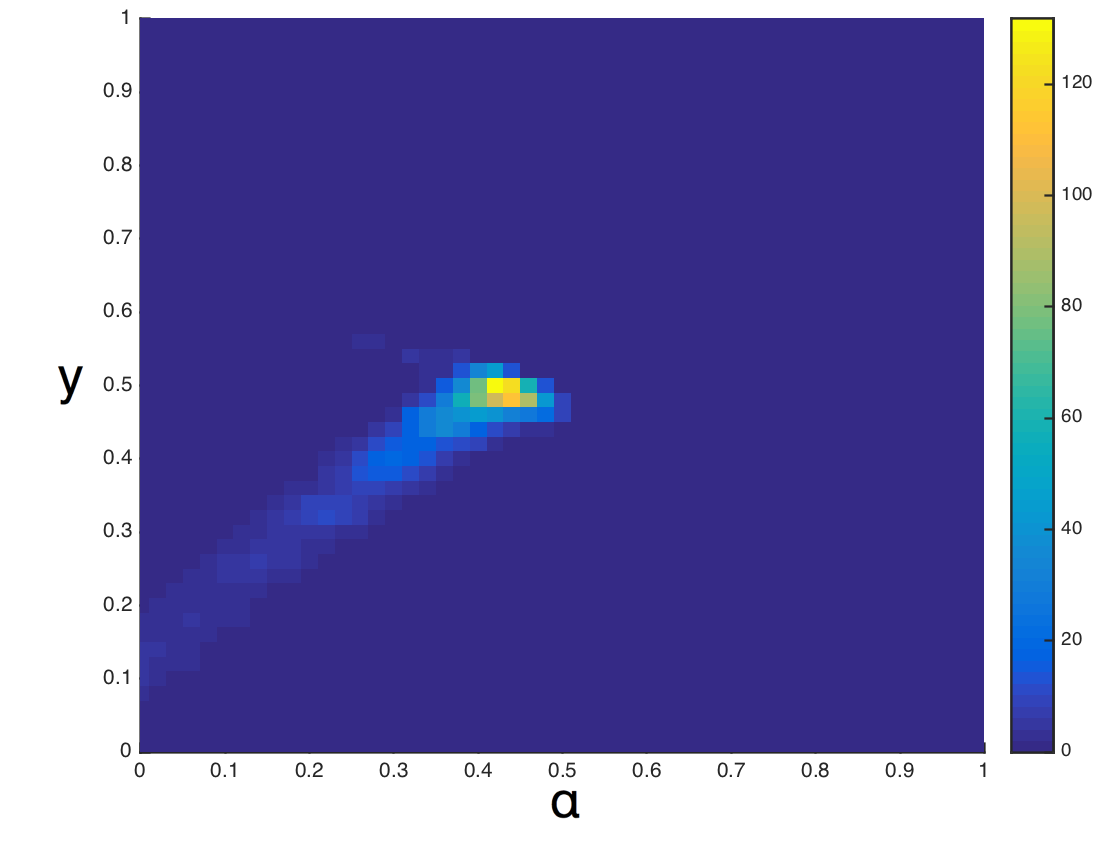} &\hspace{-2em}
		\includegraphics[width=.22\textwidth]{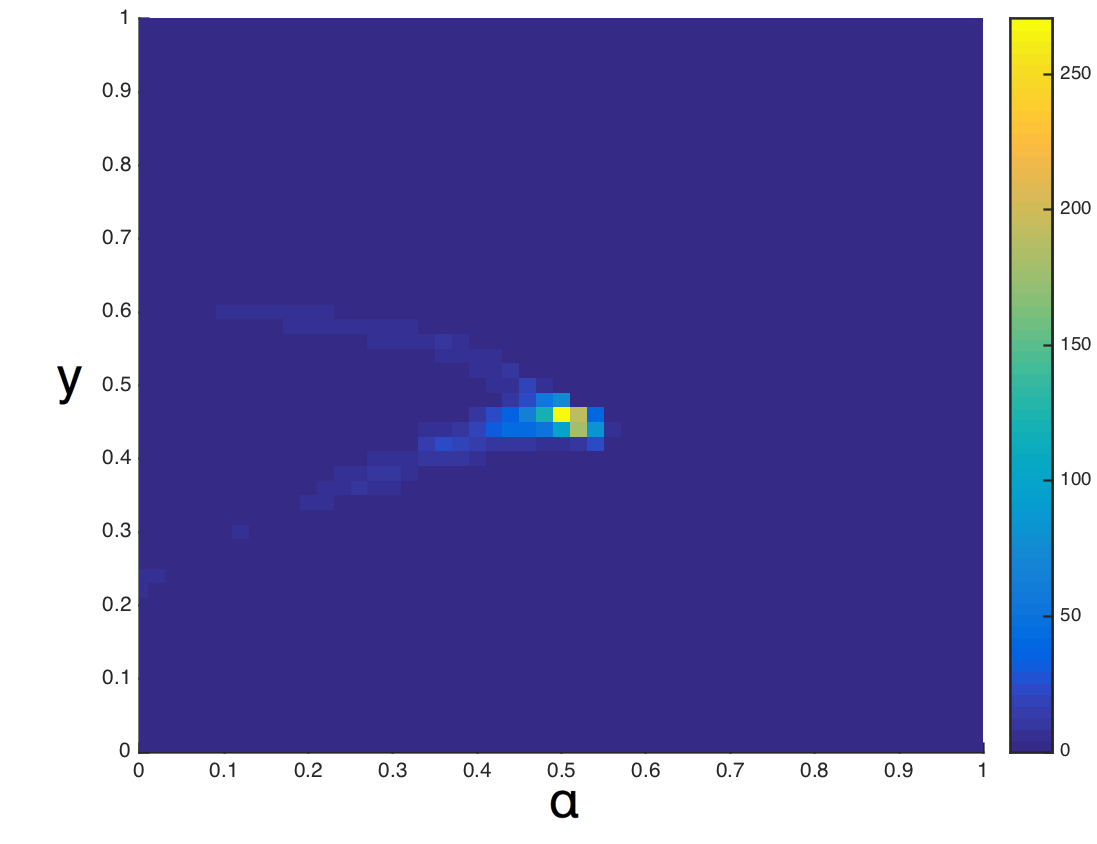} &\hspace{-2em}
		\includegraphics[width=.22\textwidth]{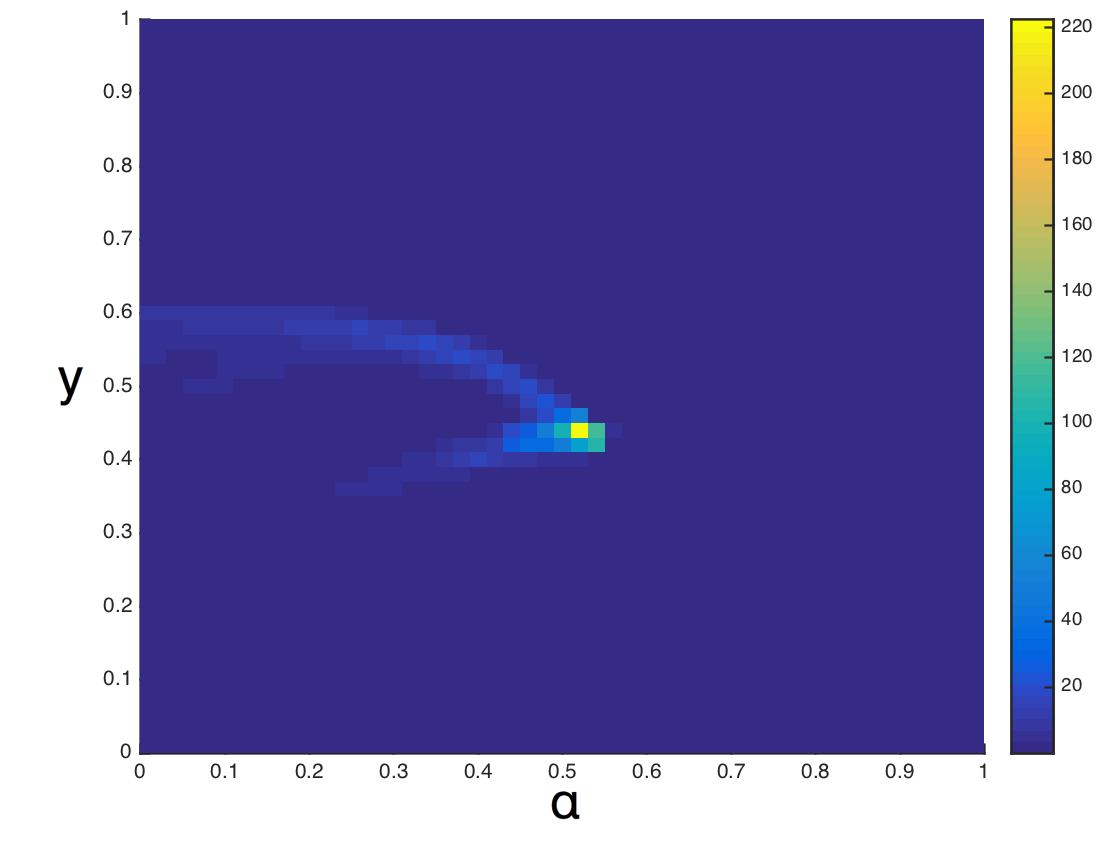}
	\\ $\begin{array}{c} \\[-2.5cm] m_1 \end{array}$ &\hspace{-2em}
		\includegraphics[width=.22\textwidth]{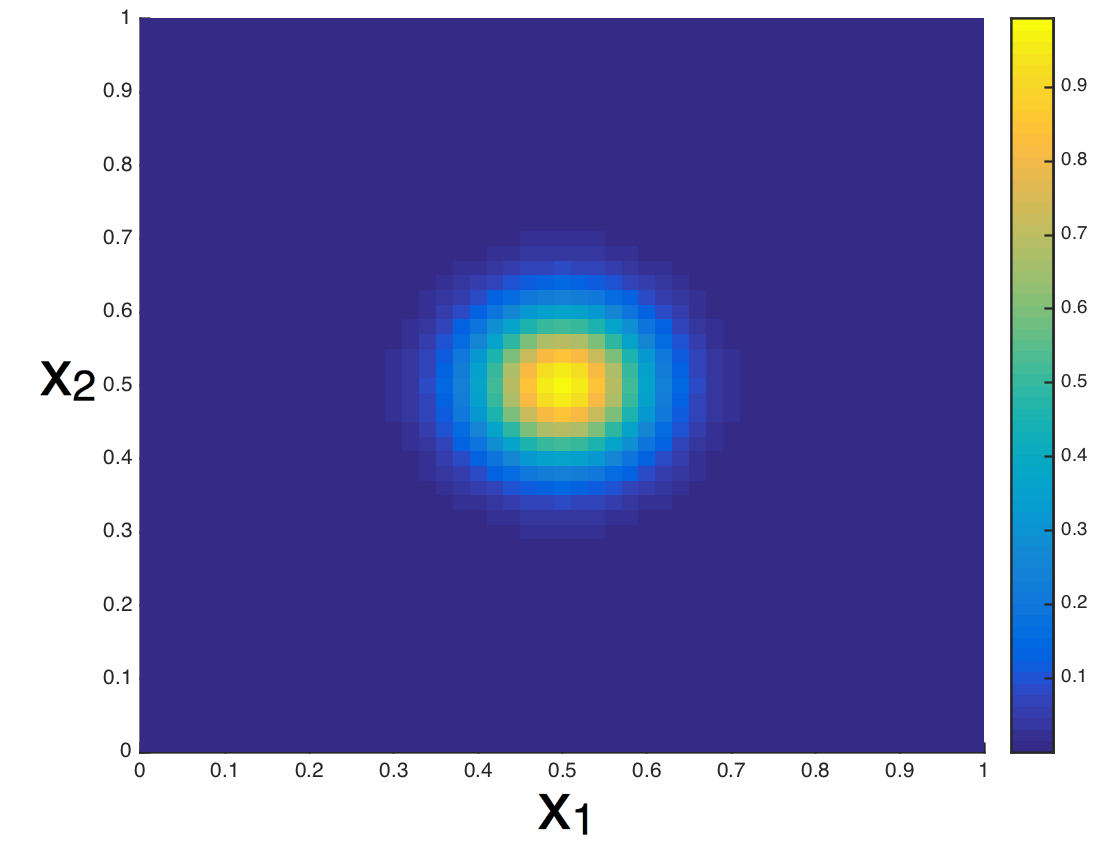} &\hspace{-2em}
		\includegraphics[width=.22\textwidth]{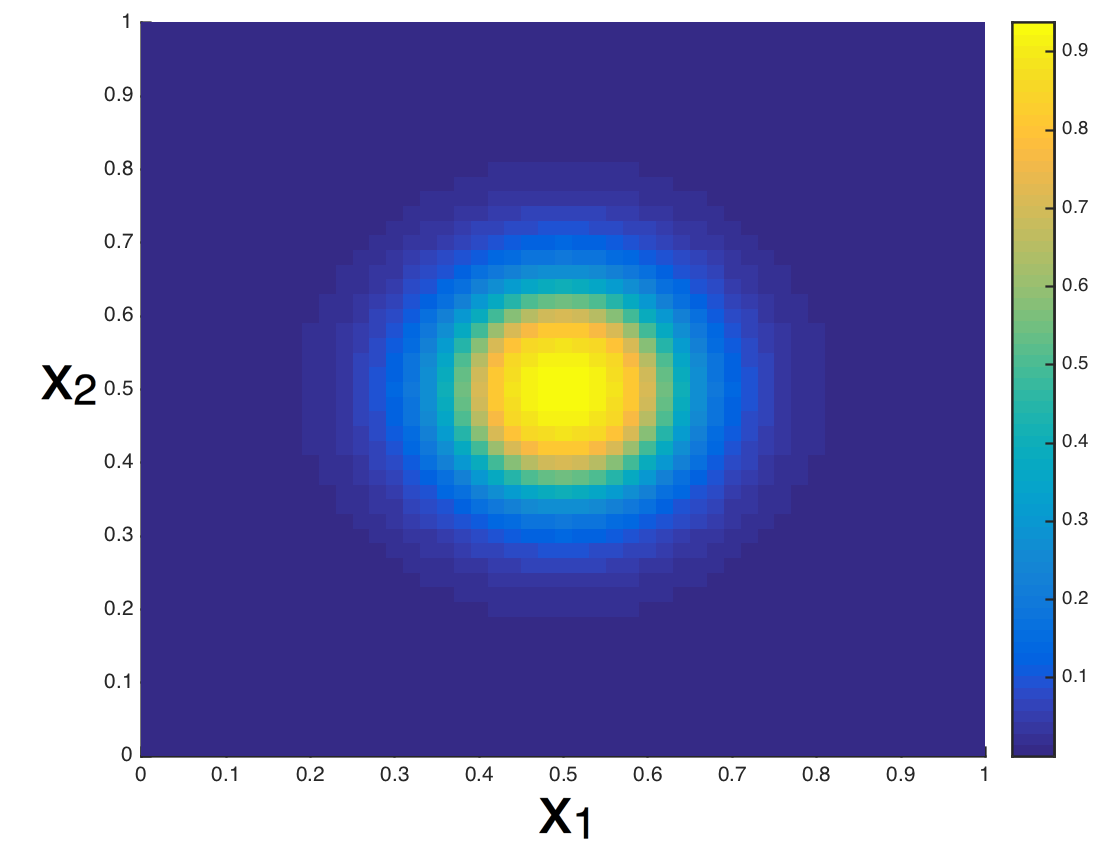} &\hspace{-2em}
		\includegraphics[width=.22\textwidth]{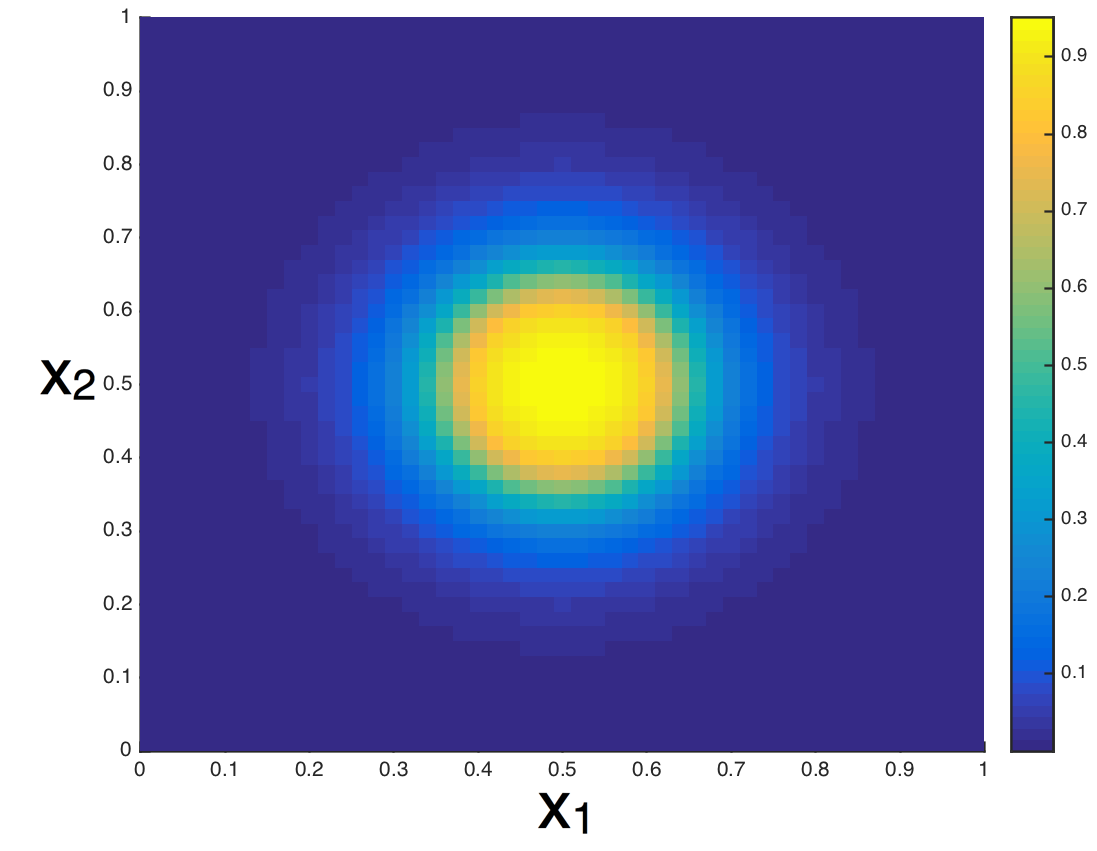} &\hspace{-2em}
		\includegraphics[width=.22\textwidth]{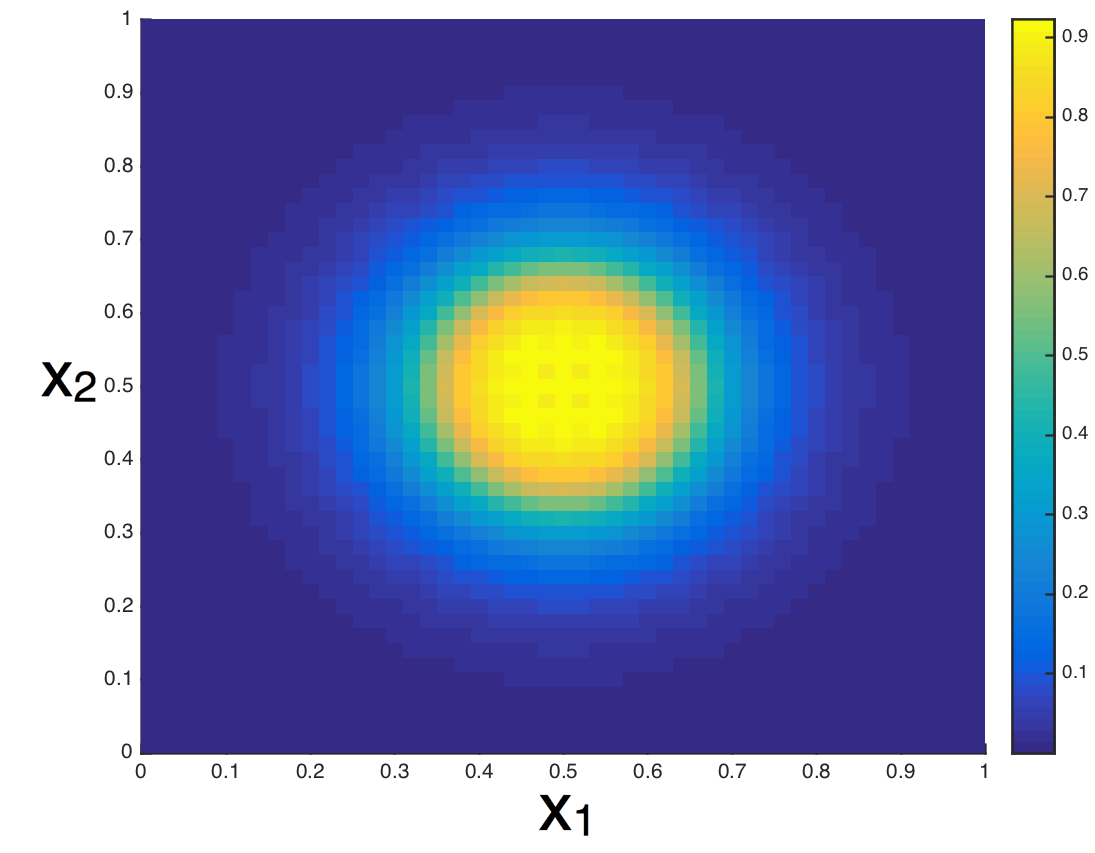}
	\\	&\hspace{-2em} $t=0$ &\hspace{-2em} $t=5$ &\hspace{-2em} $t=10$ &\hspace{-2em} $t=15$
	\end{tabular}
	\caption{Single-cluster results from simulation of model (\ref{eq:IFNmodel}) for low affinity ($\lambda=0.1$) are given for $c(t,x,y,\alpha)$ in the spatio-metabolic domain (\textit{$1^{st}$ row}, $c_\alpha$), with $x$ on the horizontal plane and $\alpha$ on the vertical axis; in the spatio-binding domain (\textit{$2^{nd}$ row}, $c_y$), with $x$ on the horizontal plane and $y$ on the vertical axis; in the metabolo-binding domain (\textit{$3^{rd}$ row}, $\breve{c}$), with $\alpha$ on the horizontal axis and $y$ on the vertical axis; and for $m(t,x)$ in space (\textit{$4^{th}$ row}), for $t\in\{0,5,10,15\}$ respectively.
} \label{fig:Single_cell,1}
\end{figure}
\begin{figure}[t!] \centering \begin{tabular}{ccccc}
	$\begin{array}{c} \\[-2.5cm] c_\alpha \end{array}$ &\hspace{-2em}
		\includegraphics[width=.22\textwidth]{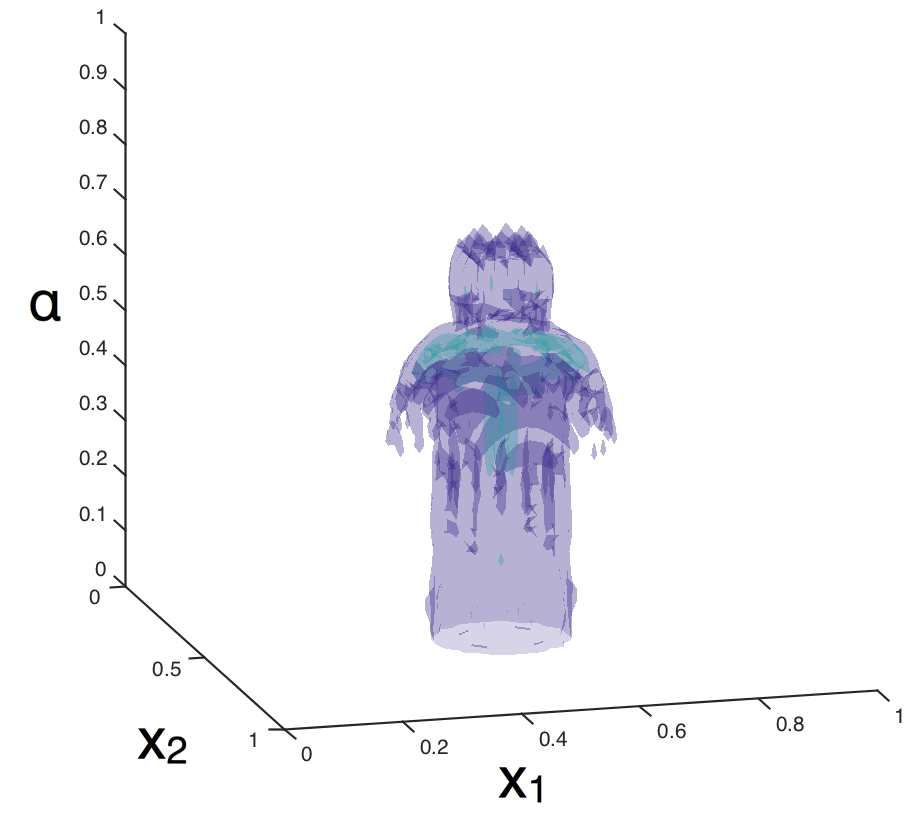} &\hspace{-2em}
		\includegraphics[width=.22\textwidth]{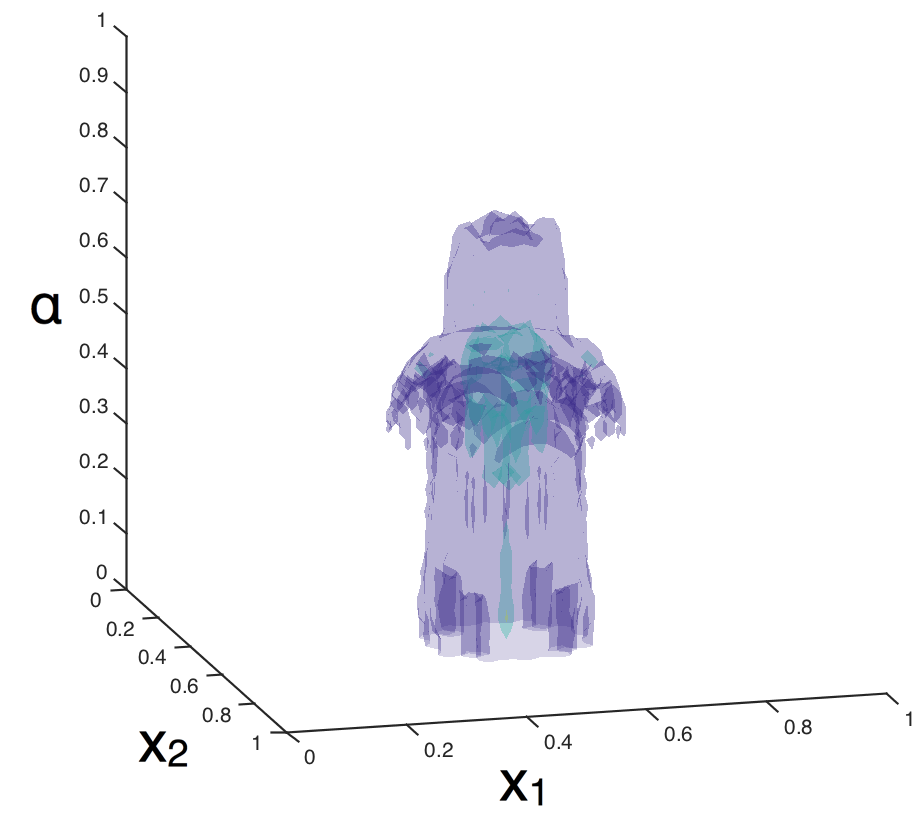} &\hspace{-2em}
		\includegraphics[width=.22\textwidth]{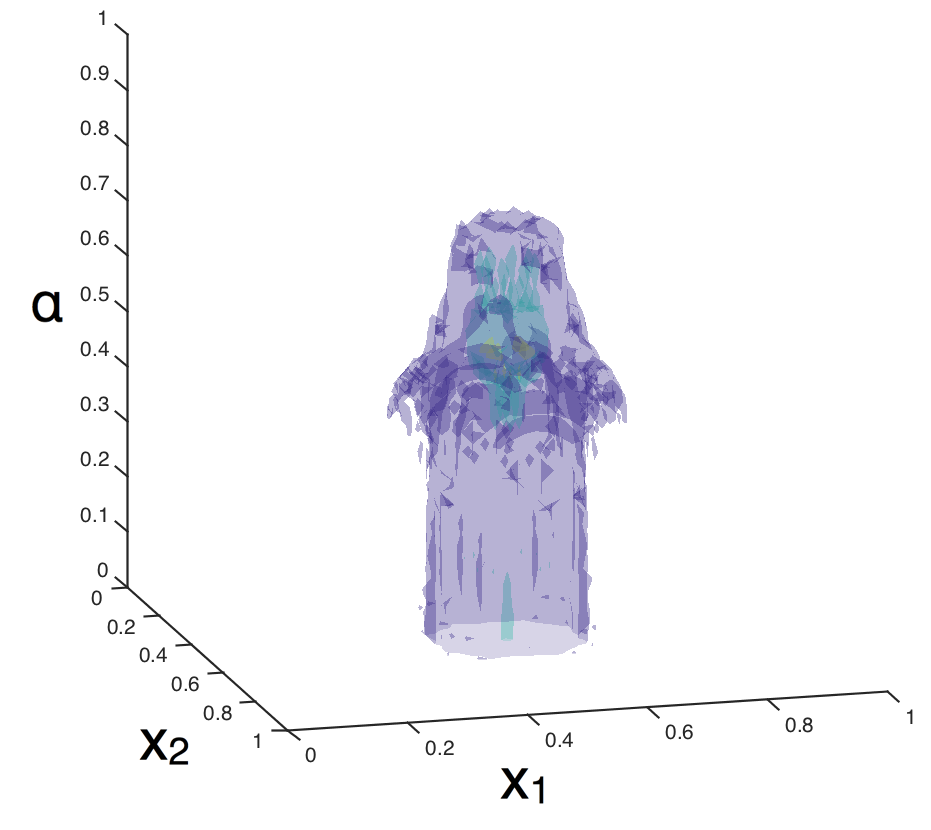} &\hspace{-2em}
		\includegraphics[width=.22\textwidth]{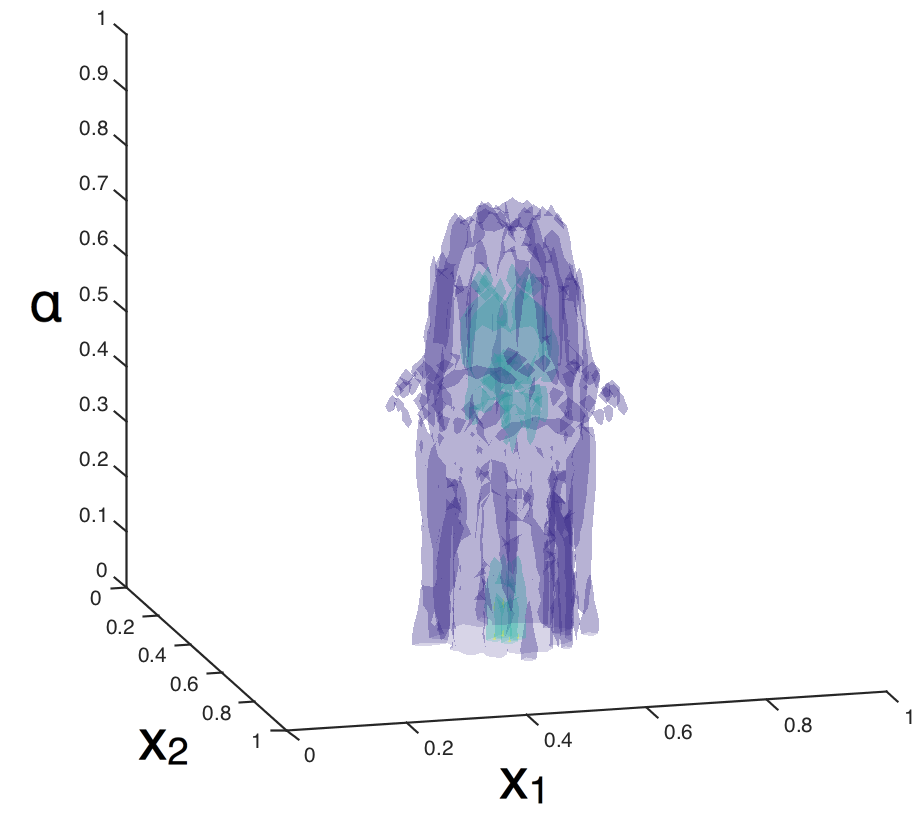}
	\\ $\begin{array}{c} \\[-2.5cm] c_y \end{array}$ &\hspace{-2em}
		\includegraphics[width=.22\textwidth]{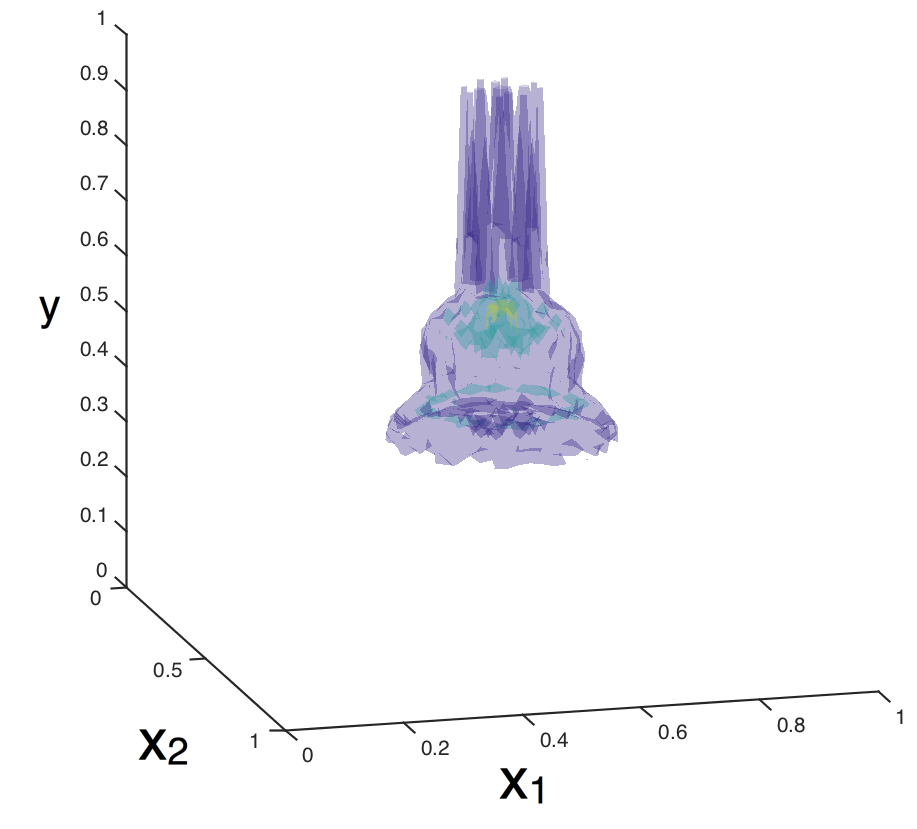} &\hspace{-2em}
		\includegraphics[width=.22\textwidth]{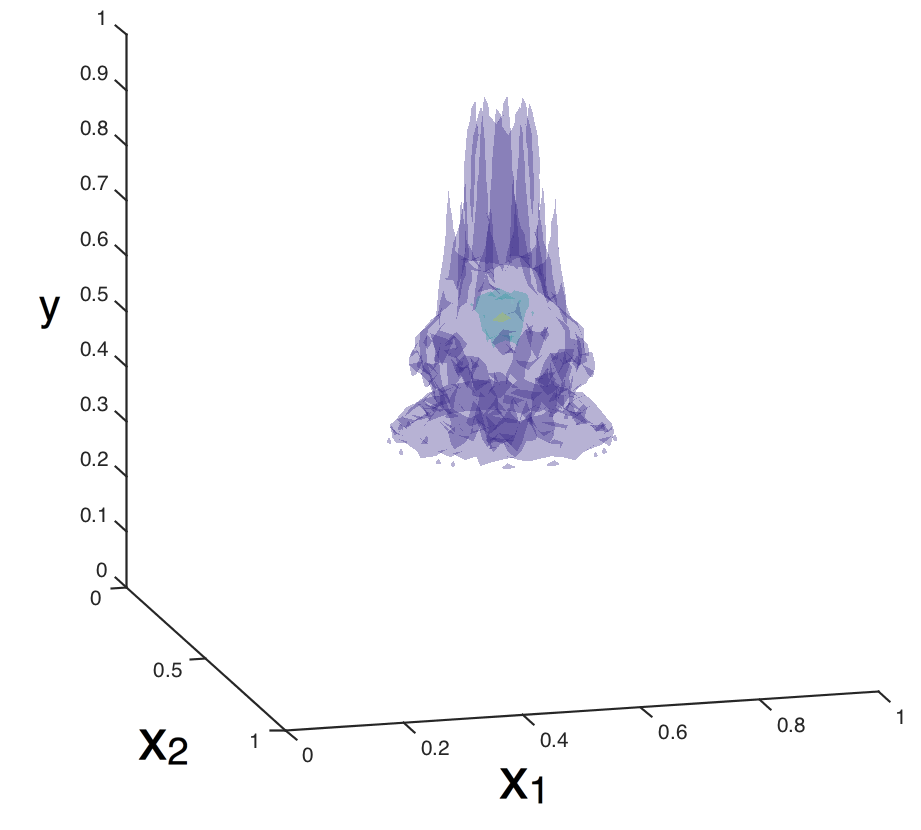} &\hspace{-2em}
		\includegraphics[width=.22\textwidth]{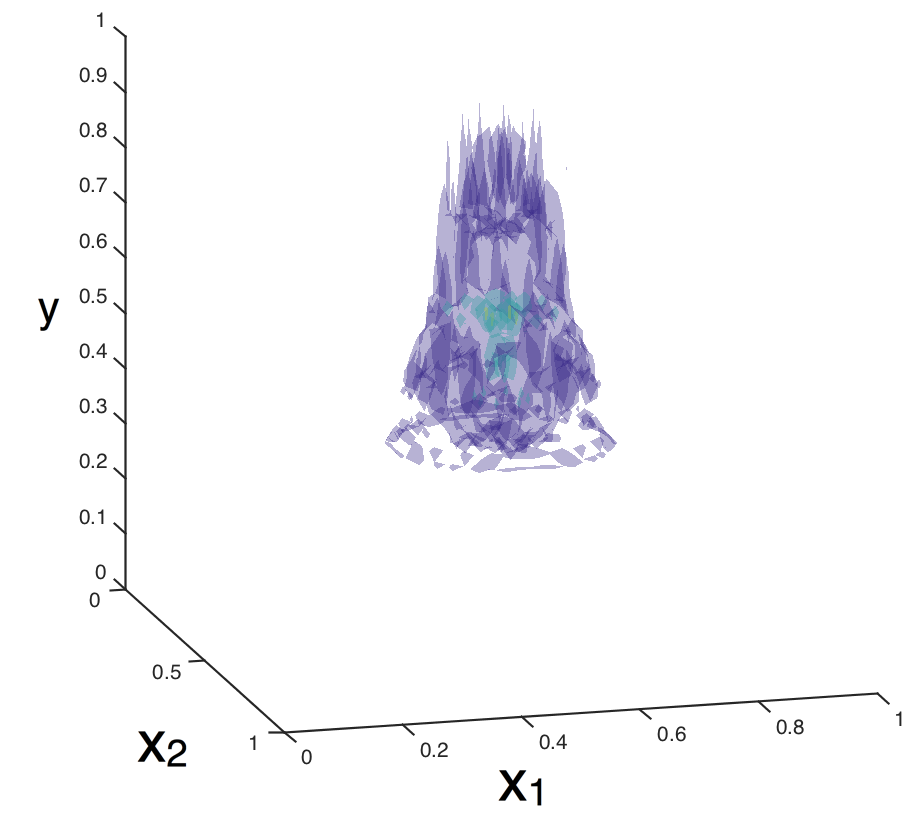} &\hspace{-2em}
		\includegraphics[width=.22\textwidth]{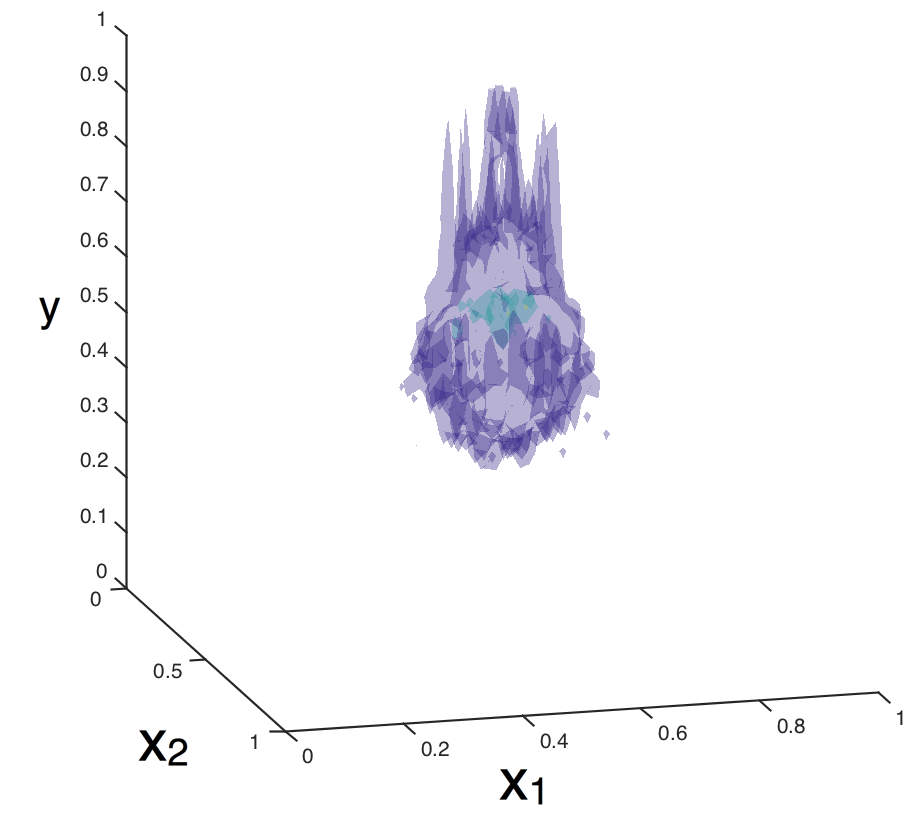}
	\\ $\begin{array}{c} \\[-2.5cm] \breve{c} \end{array}$ &\hspace{-2em}
		\includegraphics[width=.22\textwidth]{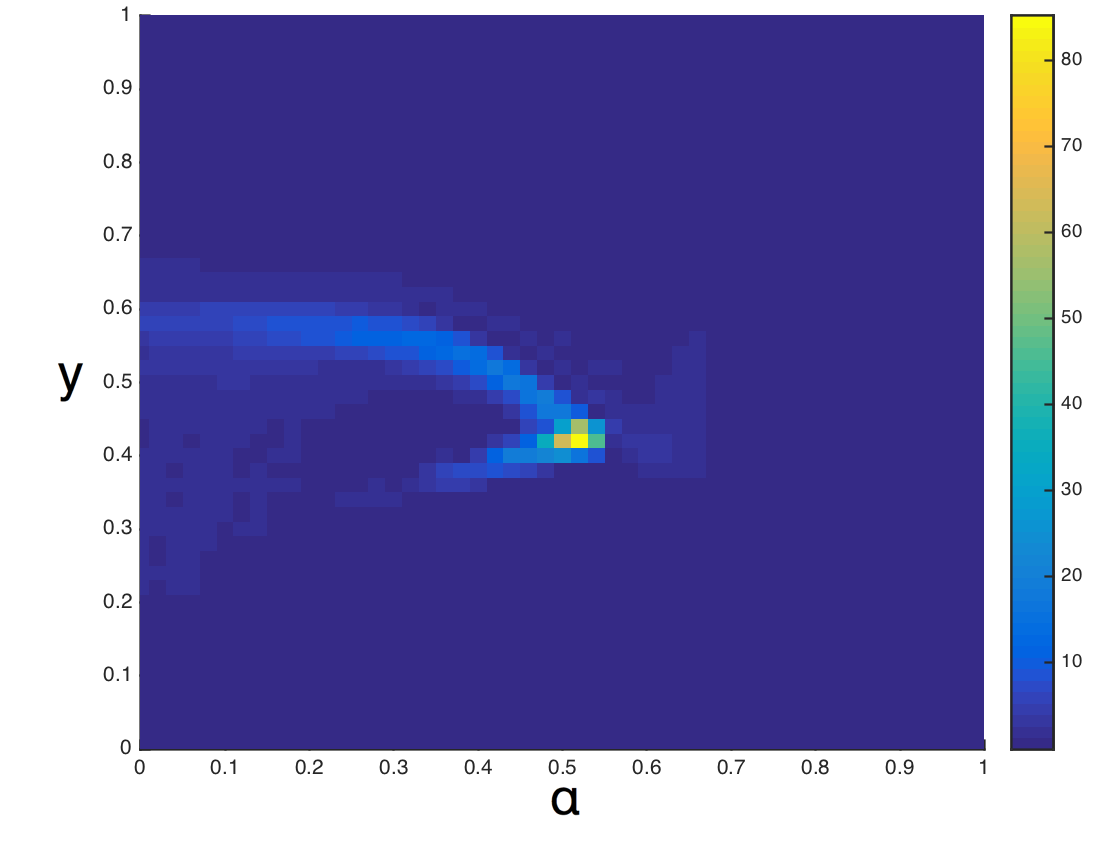} &\hspace{-2em}
		\includegraphics[width=.22\textwidth]{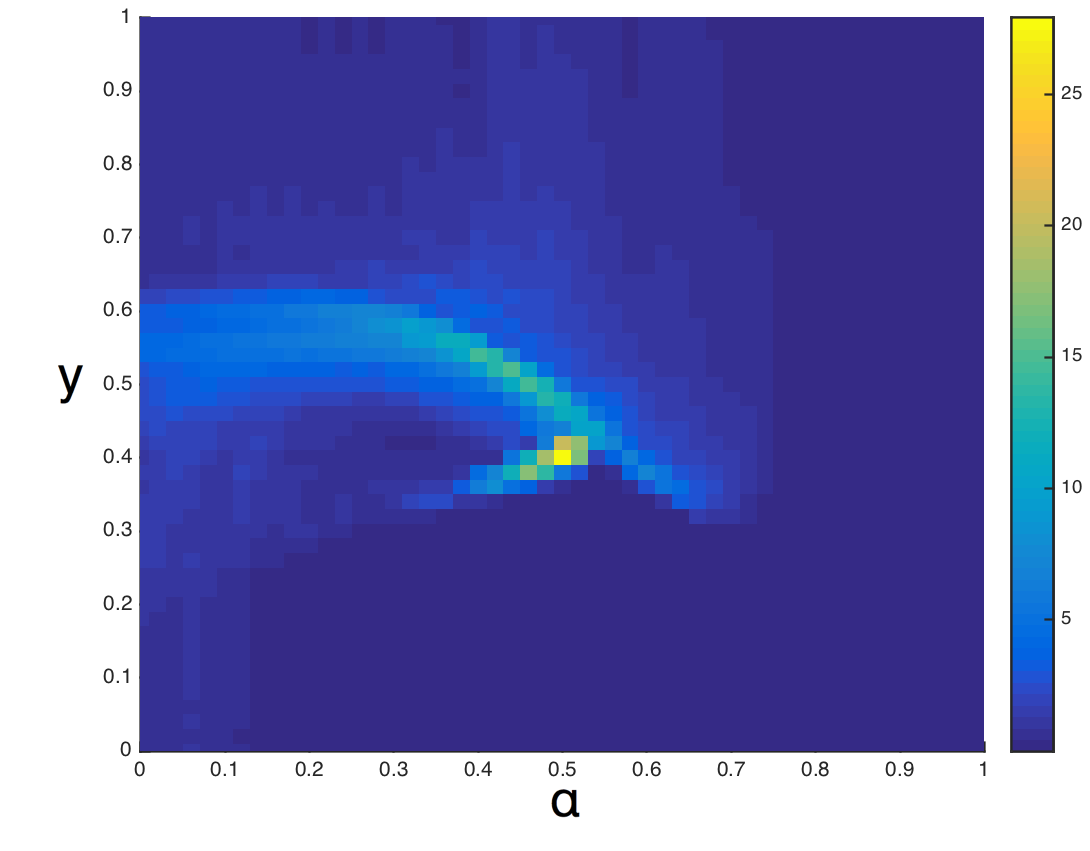} &\hspace{-2em}
		\includegraphics[width=.22\textwidth]{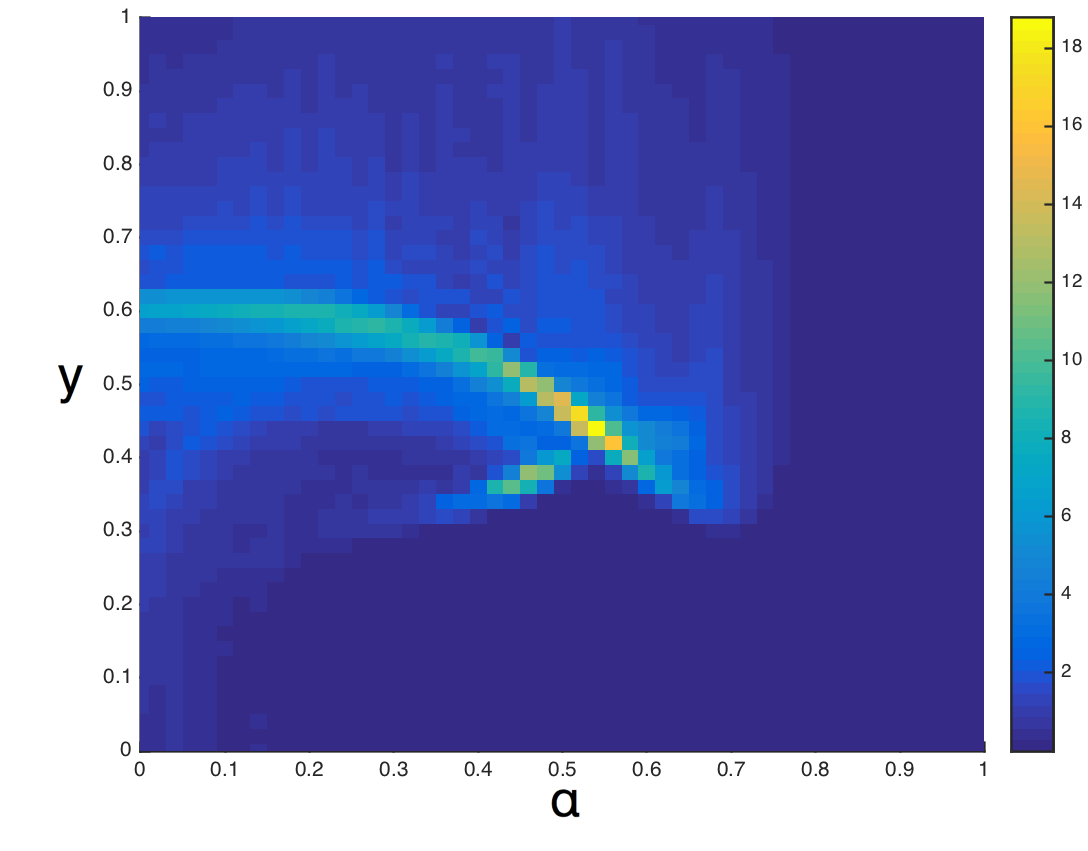} &\hspace{-2em}
		\includegraphics[width=.22\textwidth]{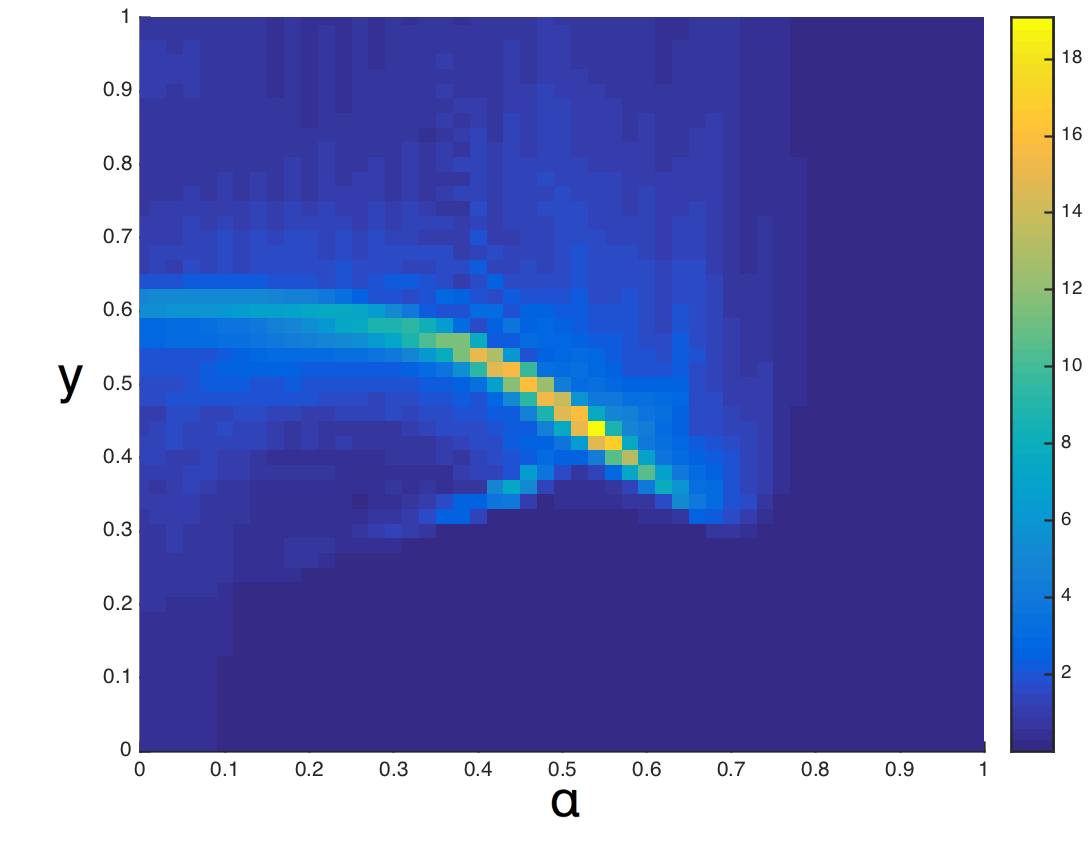}
	\\ $\begin{array}{c} \\[-2.5cm] m_1 \end{array}$ &\hspace{-2em}
		\includegraphics[width=.22\textwidth]{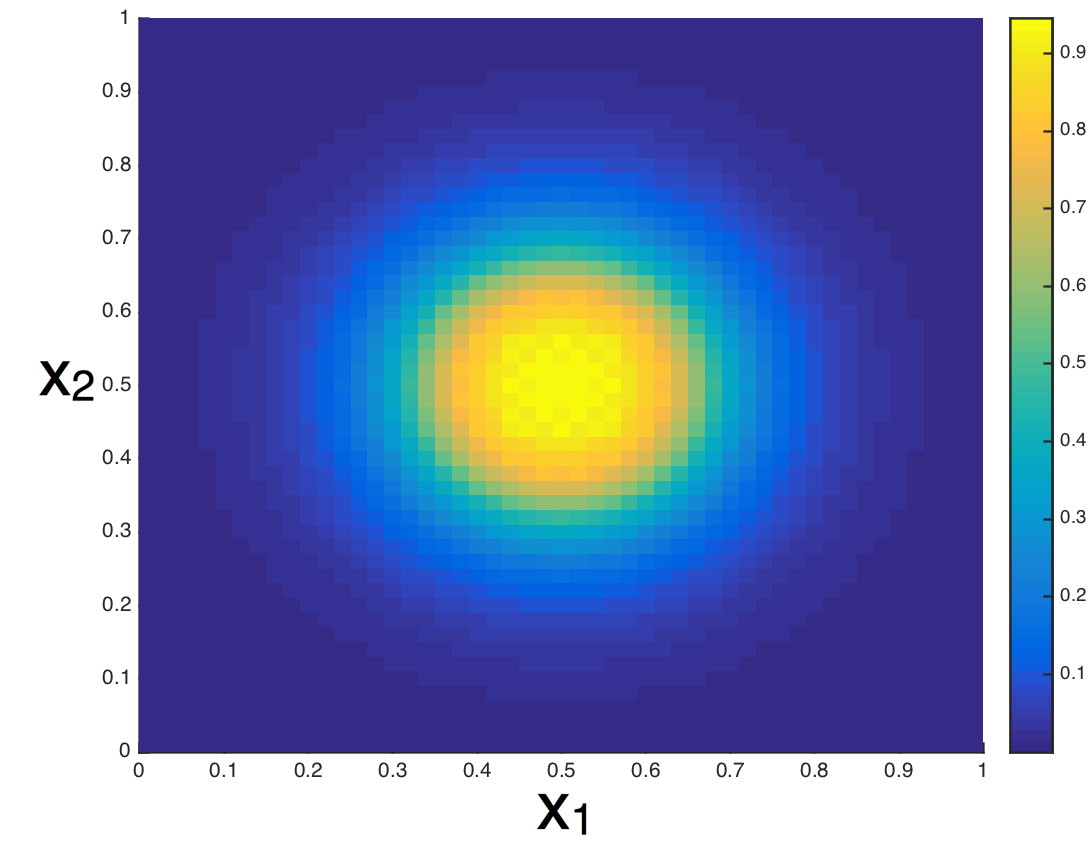} &\hspace{-2em}
		\includegraphics[width=.22\textwidth]{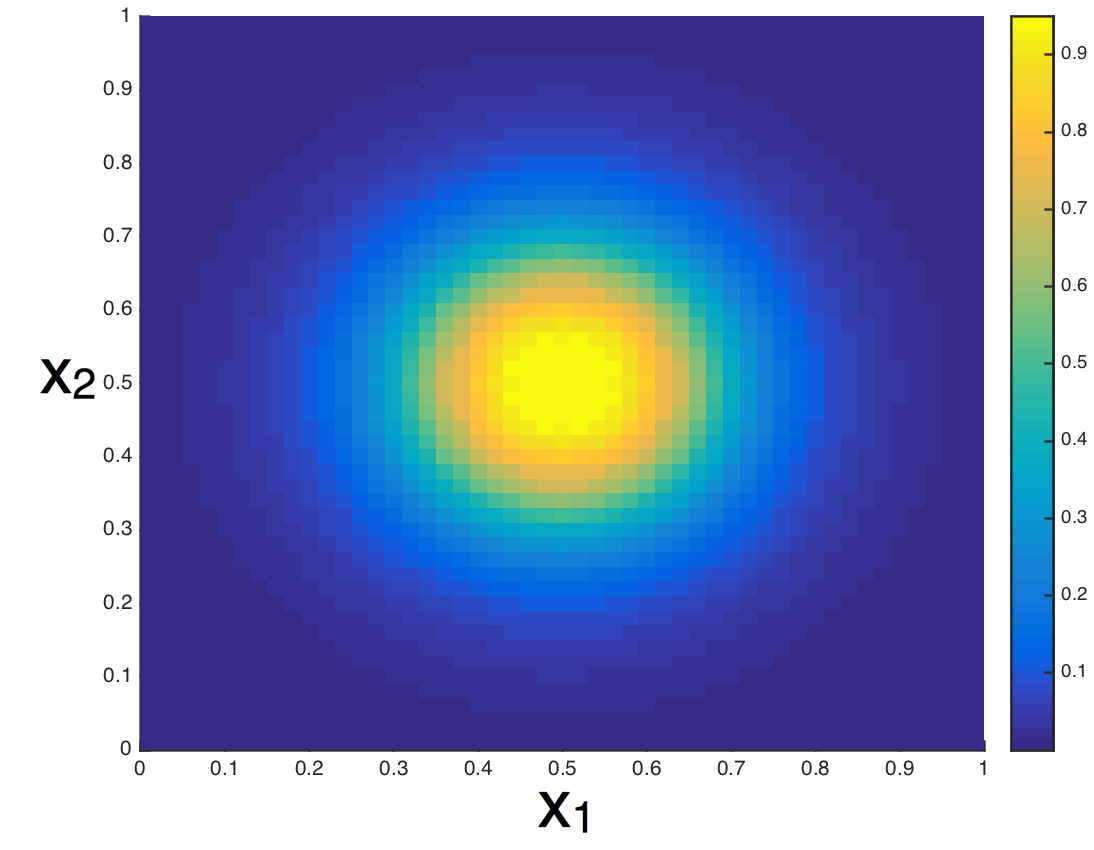} &\hspace{-2em}
		\includegraphics[width=.22\textwidth]{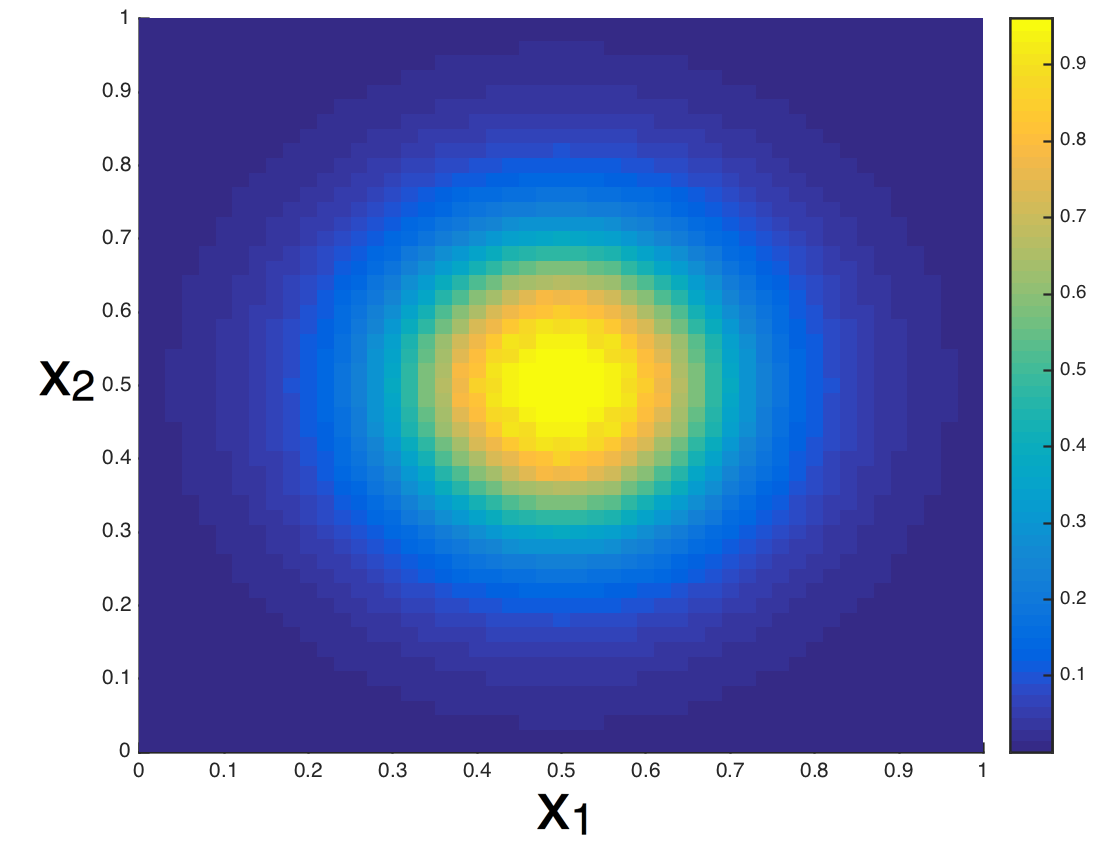} &\hspace{-2em}
		\includegraphics[width=.22\textwidth]{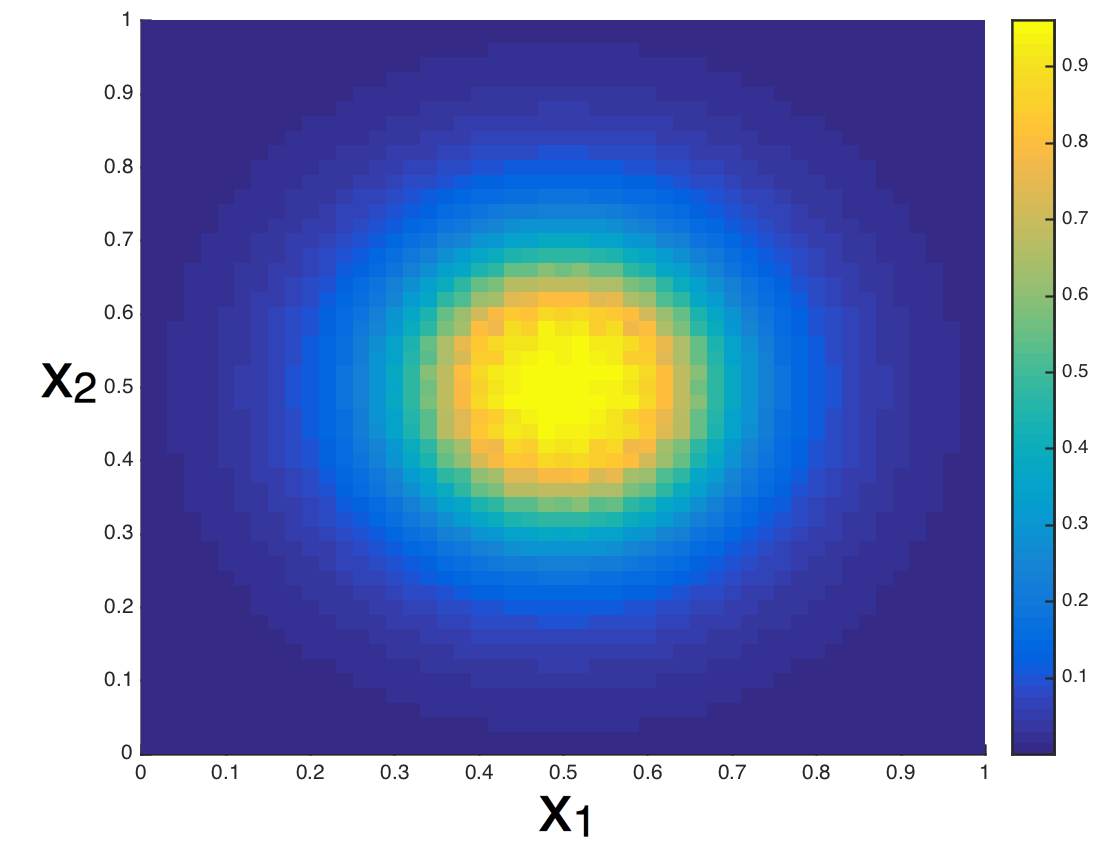}
	\\	&\hspace{-2em} $t=20$ &\hspace{-2em} $t=25$ &\hspace{-2em} $t=30$ &\hspace{-2em} $t=35$
	\end{tabular}
	\caption{Single-cluster results from simulation of model (\ref{eq:IFNmodel}) for low affinity ($\lambda=0.1$) are given for $c(t,x,y,\alpha)$ in the spatio-metabolic domain (\textit{$1^{st}$ row}, $c_\alpha$), with $x$ on the horizontal plane and $\alpha$ on the vertical axis; in the spatio-binding domain (\textit{$2^{nd}$ row}, $c_y$), with $x$ on the horizontal plane and $y$ on the vertical axis; in the metabolo-binding domain (\textit{$3^{rd}$ row}, $\breve{c}$), with $\alpha$ on the horizontal axis and $y$ on the vertical axis; and for $m(t,x)$ in space (\textit{$4^{th}$ row}), for $t\in\{20,25,30,35\}$ respectively.} \label{fig:Single_cell,2}
\end{figure}

\begin{figure}[t!] \centering \begin{tabular}{cccc}
	$\begin{array}{c} \\[-2.5cm] c_\alpha \end{array}$ &\hspace{-2em}
		\includegraphics[width=.3\textwidth]{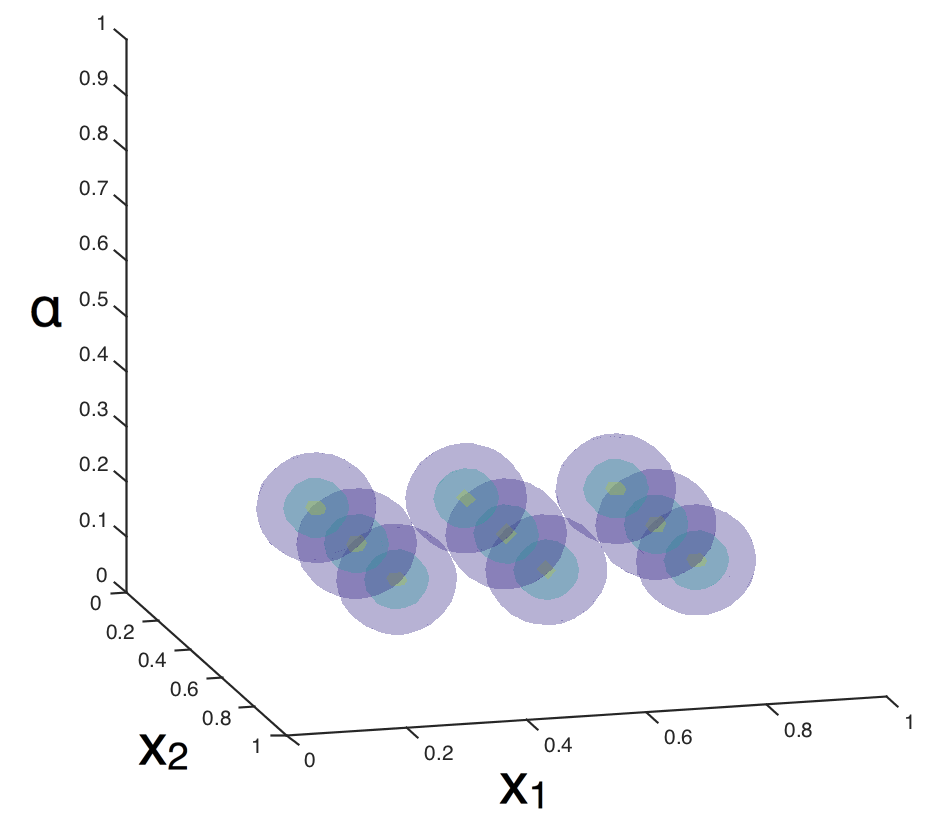} &\hspace{-2em}
		\includegraphics[width=.3\textwidth]{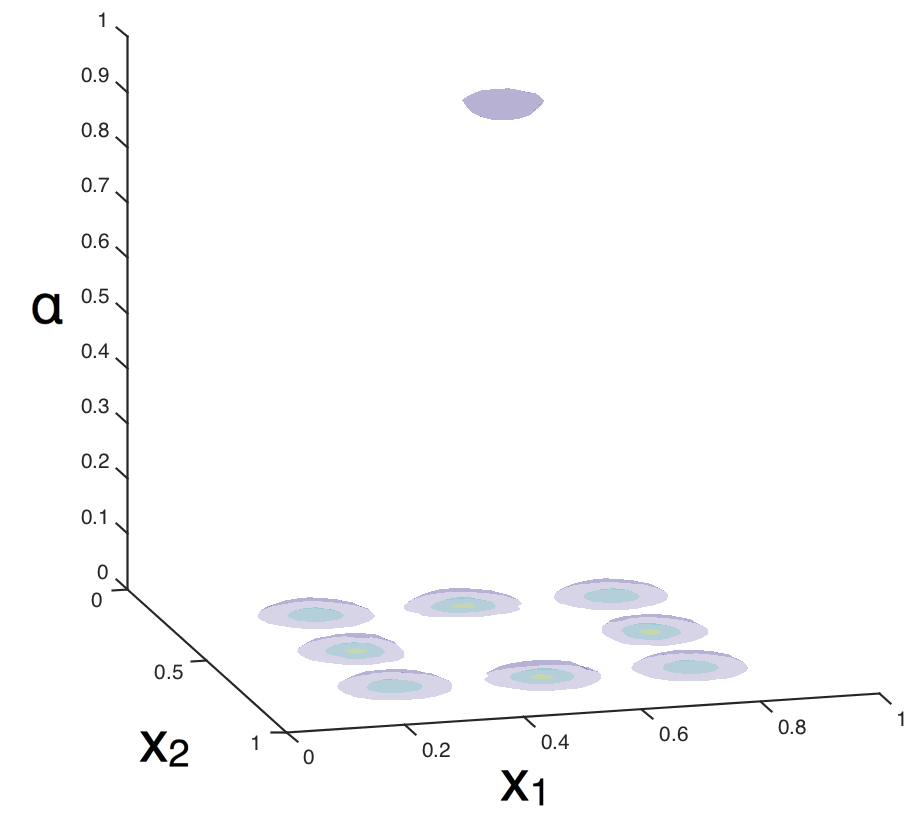} &\hspace{-2em}
		\includegraphics[width=.3\textwidth]{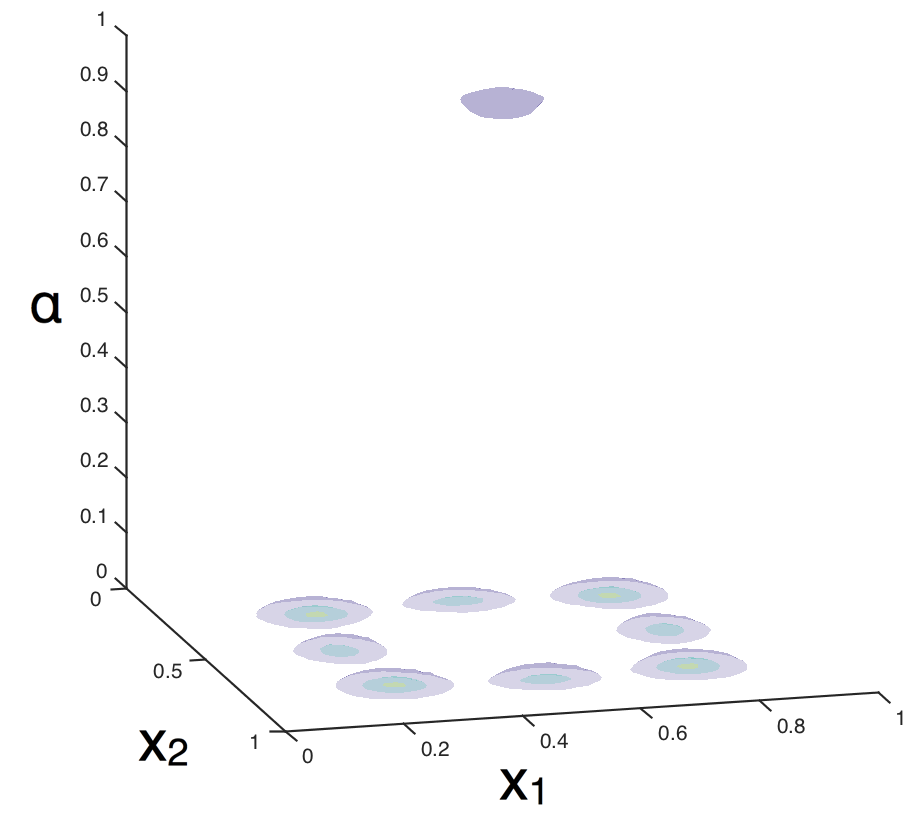}
	\\ $\begin{array}{c} \\[-2.5cm] c_y \end{array}$ &\hspace{-2em}
		\includegraphics[width=.3\textwidth]{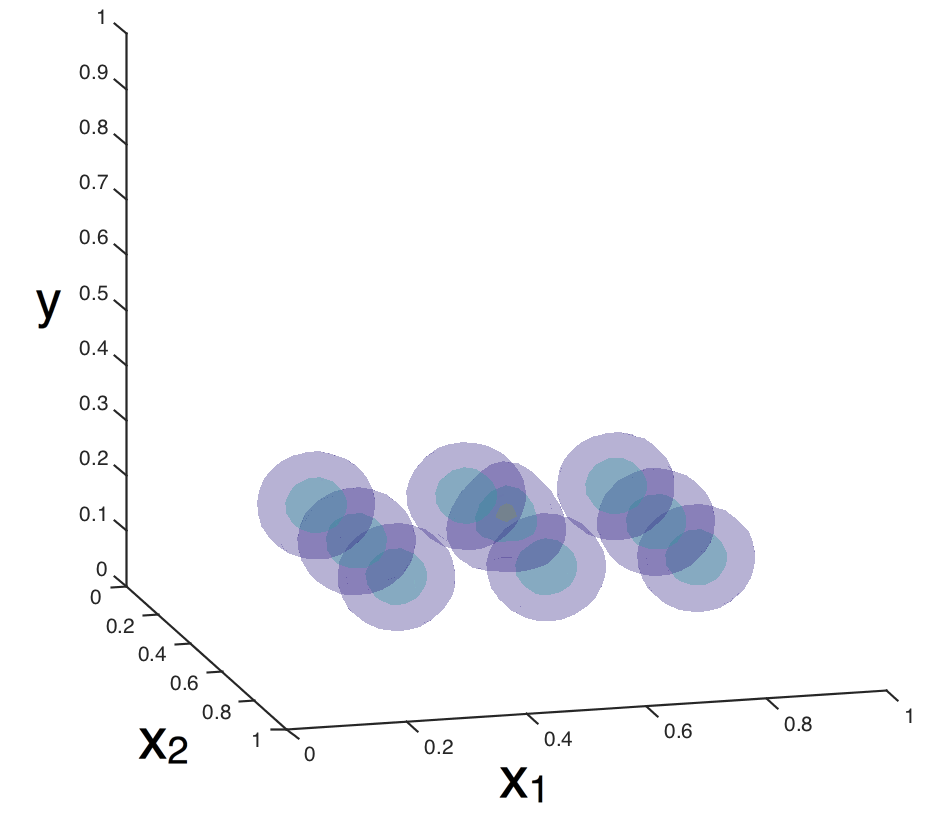} &\hspace{-2em}
		\includegraphics[width=.3\textwidth]{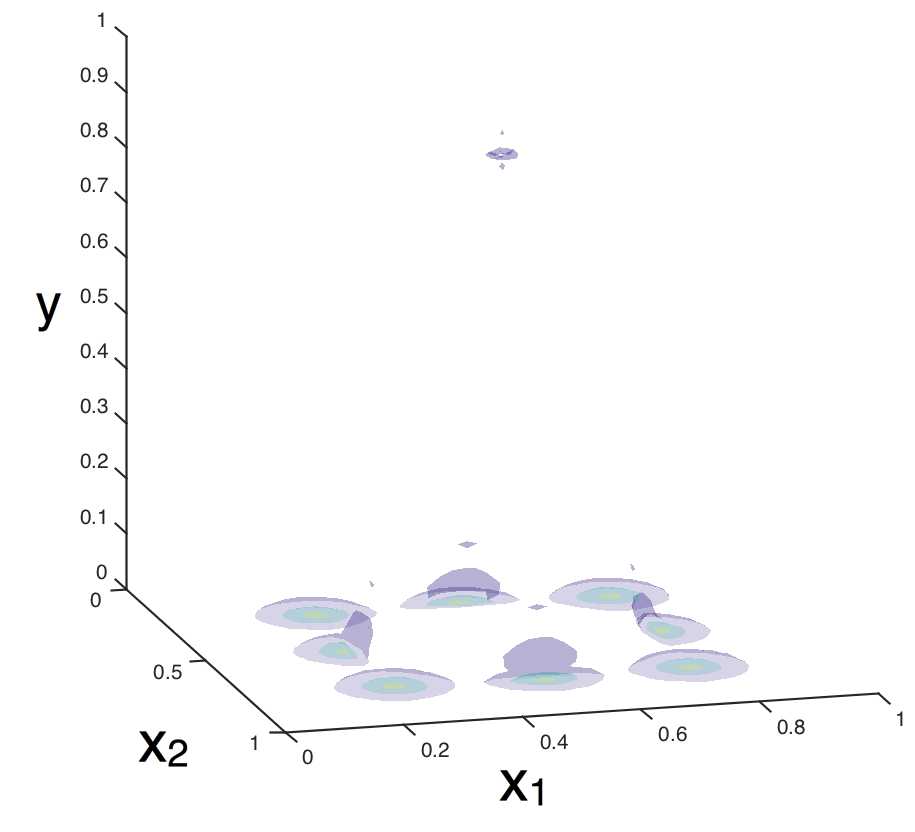} &\hspace{-2em}
		\includegraphics[width=.3\textwidth]{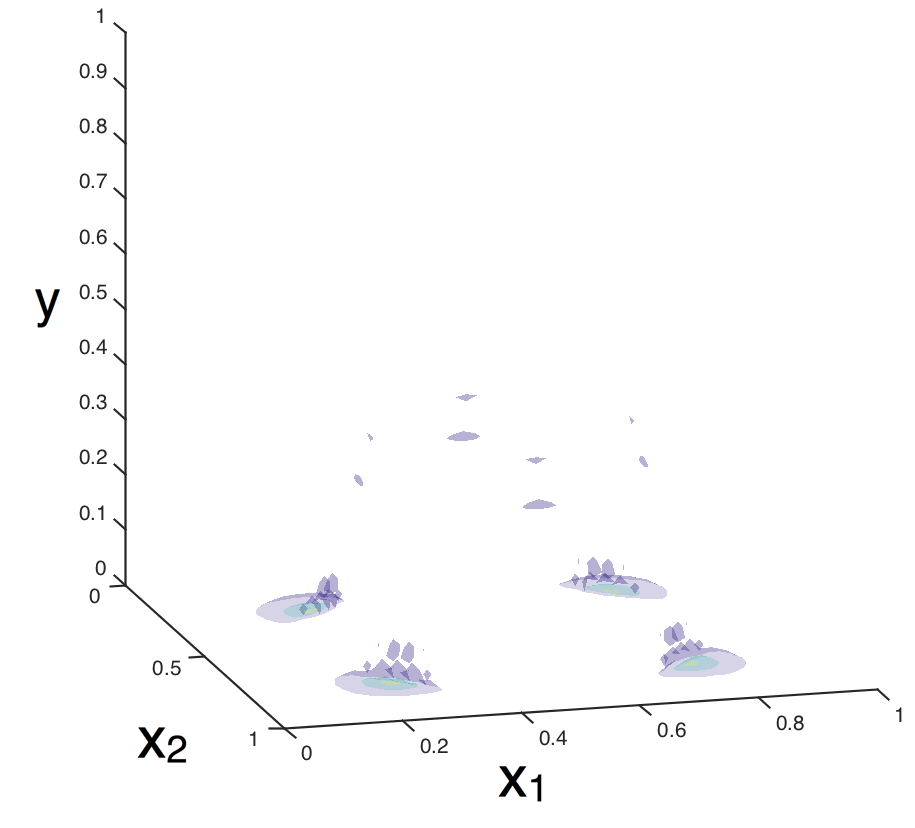}
	\\ $\begin{array}{c} \\[-2.5cm] m_1 \end{array}$ &\hspace{-2em}
		\includegraphics[width=.3\textwidth]{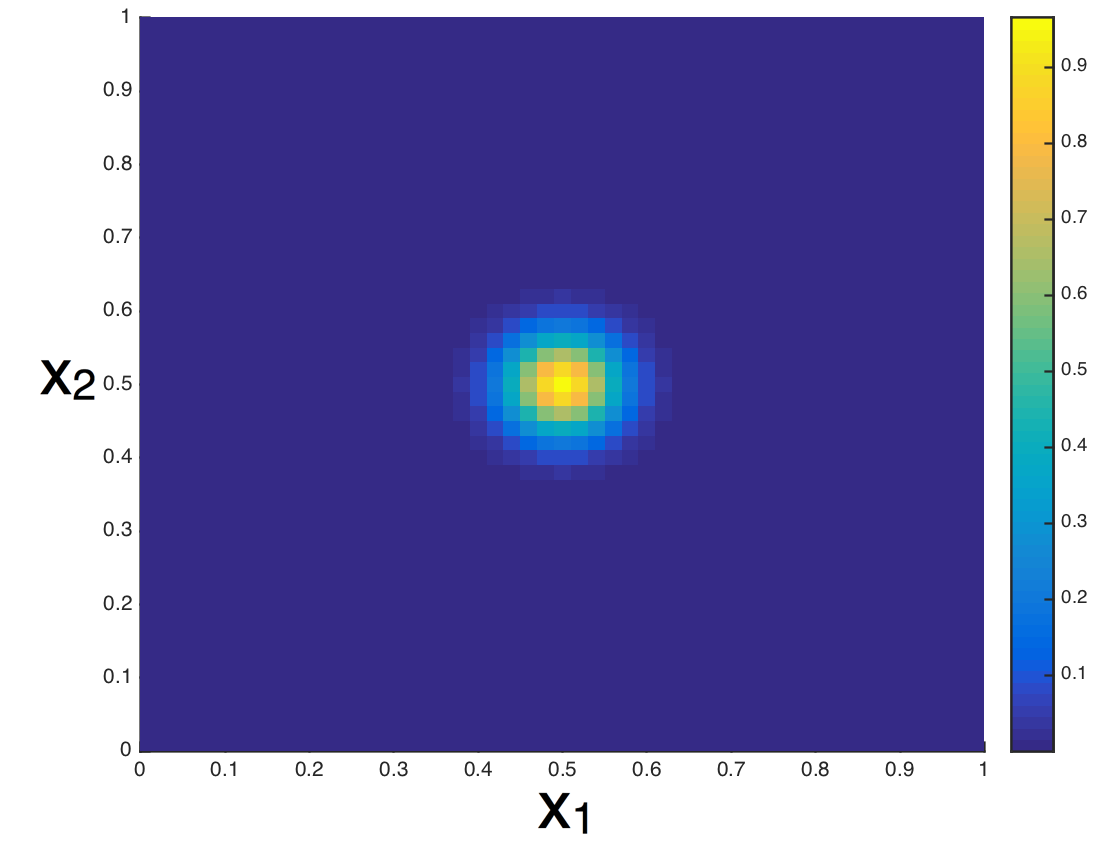} &\hspace{-2em}
		\includegraphics[width=.3\textwidth]{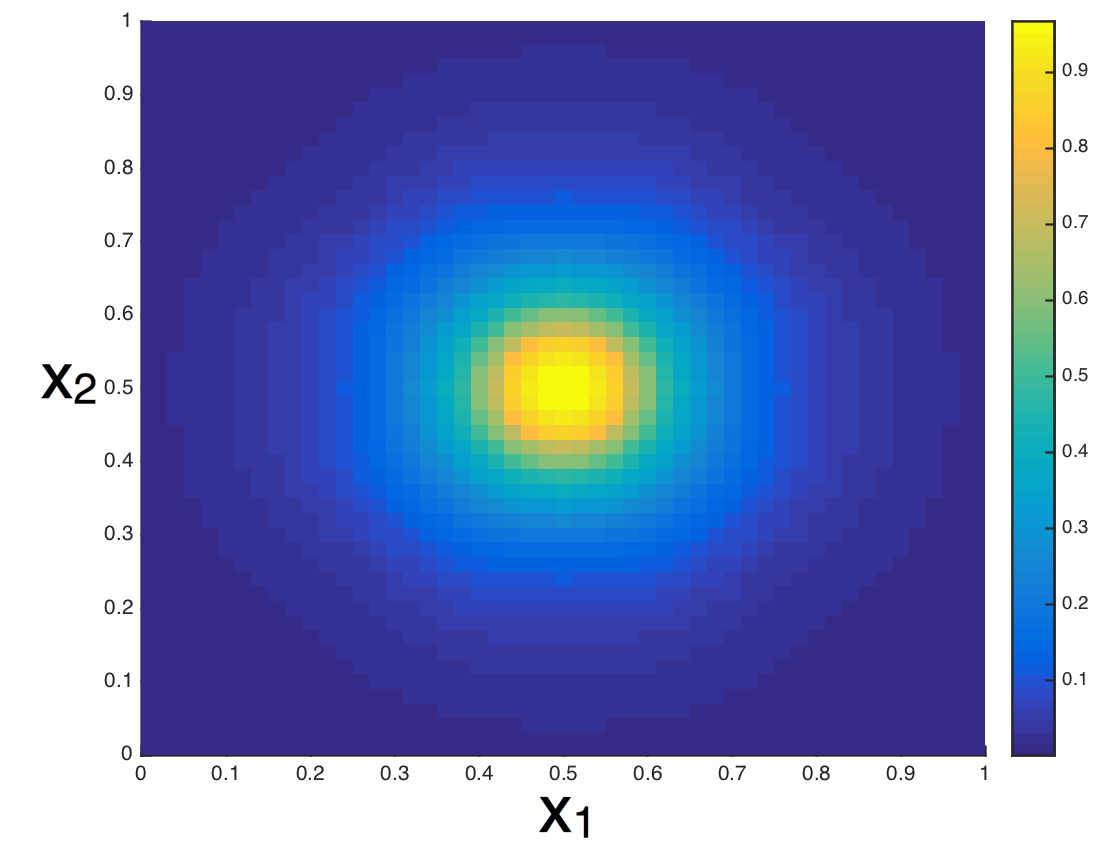} &\hspace{-2em}
		\includegraphics[width=.3\textwidth]{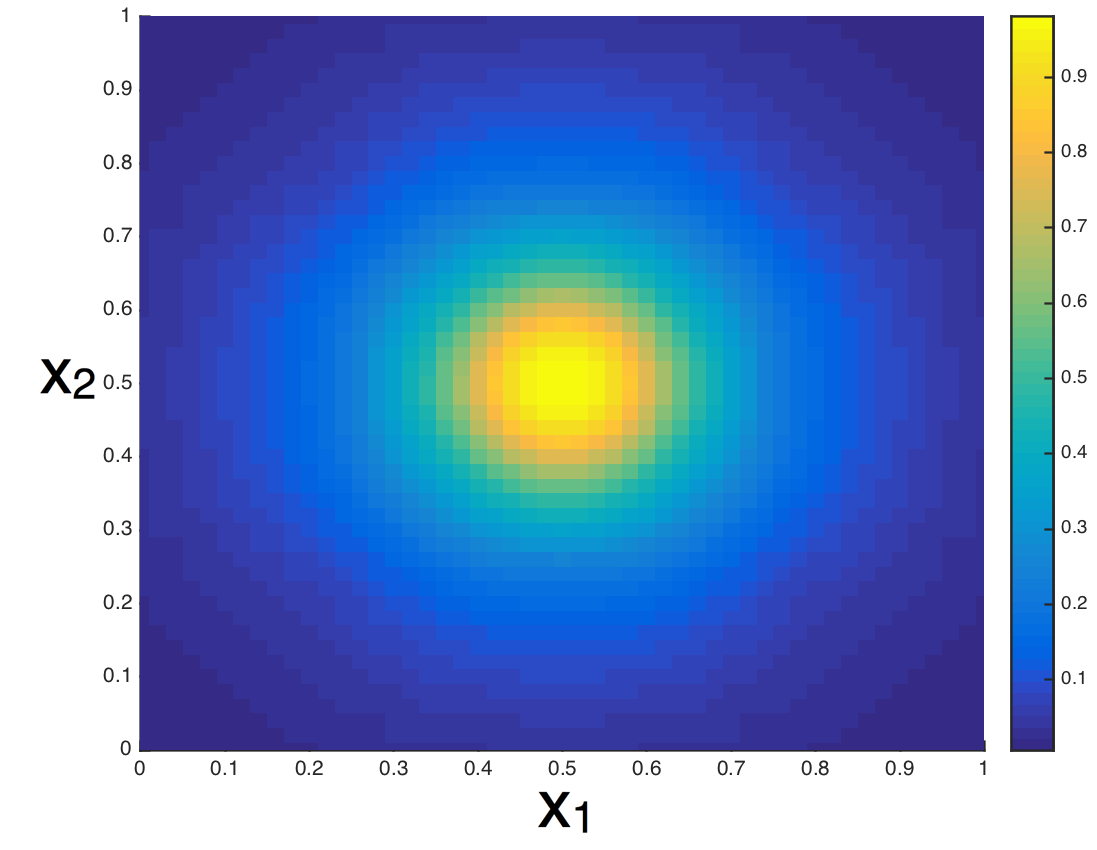}
	\\	&\hspace{-2em} $t=0$ &\hspace{-2em} $t=15$ &\hspace{-2em} $t=30$
	\end{tabular}
	\caption{Multi-cluster results from simulation of model (\ref{eq:IFNmodel_th}) for low affinity ($\lambda=0.01$) are given for $c(t,x,y,\alpha)$ in the spatio-metabolic domain (\textit{$1^{st}$ row}, $c_\alpha$), with $x$ on the horizontal plane and $\alpha$ on the vertical axis; in the spatio-binding domain (\textit{$2^{nd}$ row}, $c_y$), with $x$ on the horizontal plane and $y$ on the vertical axis; and for $m(t,x)$ in space (\textit{$3^{rd}$ row}), for $t\in\{0,15,30\}$ respectively.} \label{fig:Results_theta_1,1}
\end{figure}
\begin{figure}[t!] \centering \begin{tabular}{cccc}
	$\begin{array}{c} \\[-2.5cm] c_\alpha \end{array}$ &\hspace{-2em}
		\includegraphics[width=.3\textwidth]{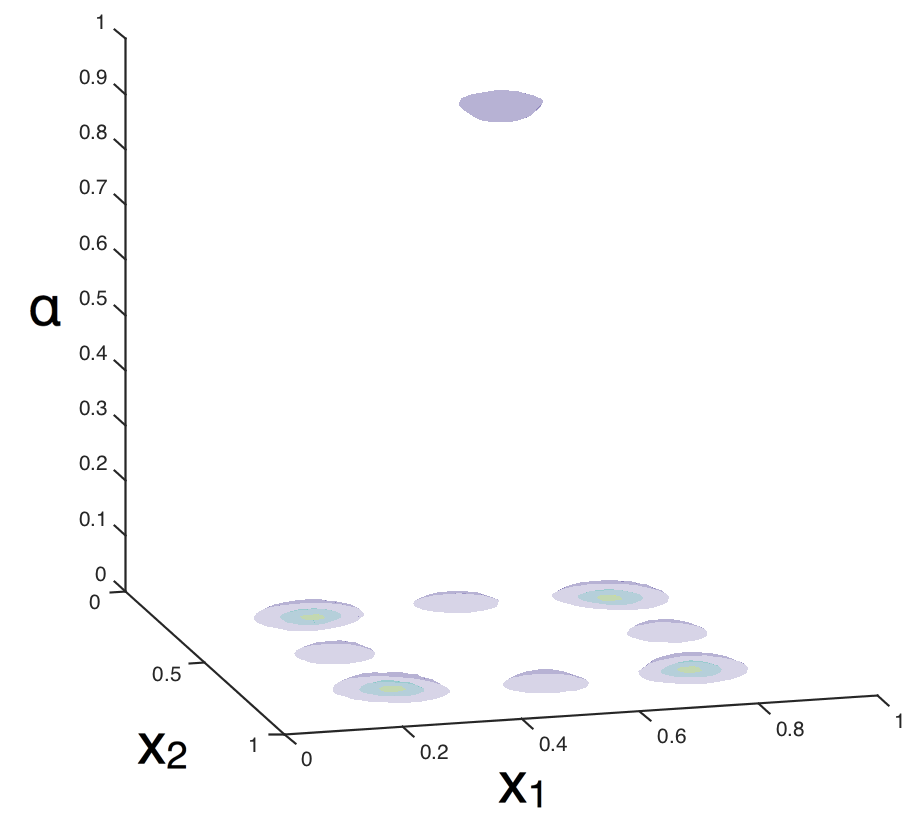} &\hspace{-2em}
		\includegraphics[width=.3\textwidth]{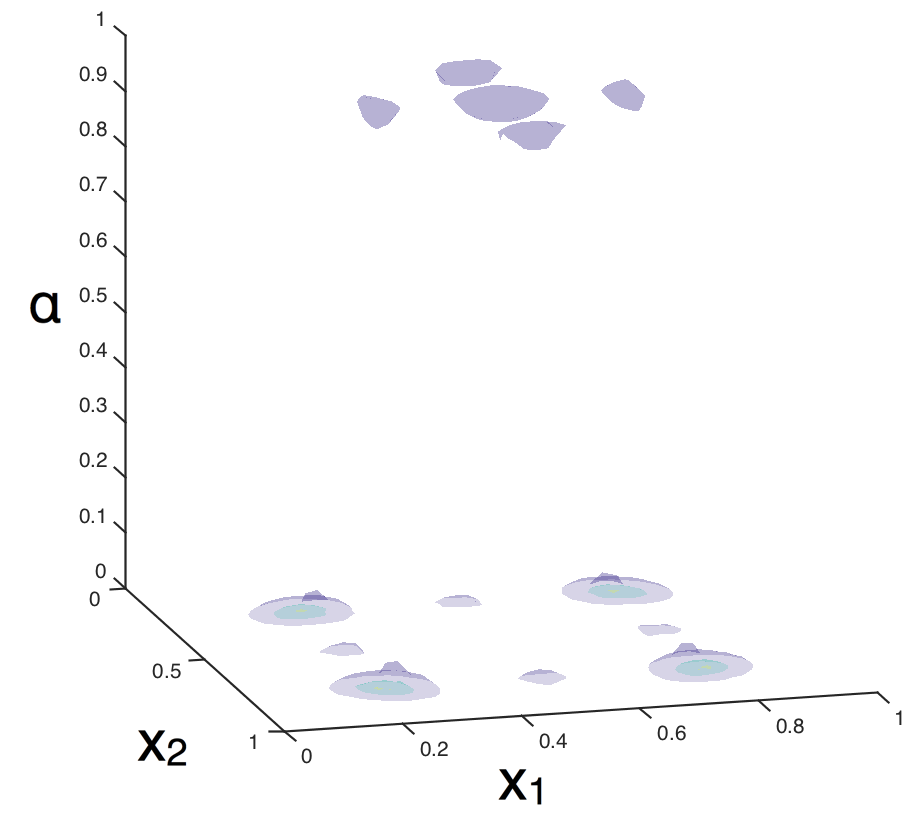} &\hspace{-2em}
		\includegraphics[width=.3\textwidth]{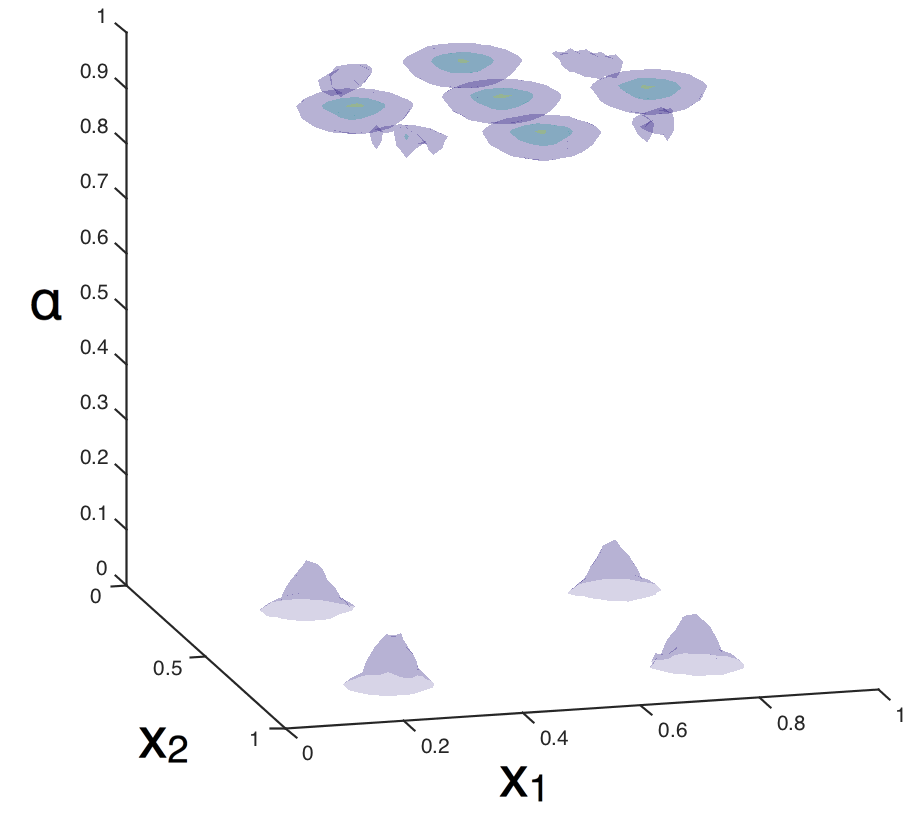}
	\\ $\begin{array}{c} \\[-2.5cm] c_y \end{array}$ &\hspace{-2em}
		\includegraphics[width=.3\textwidth]{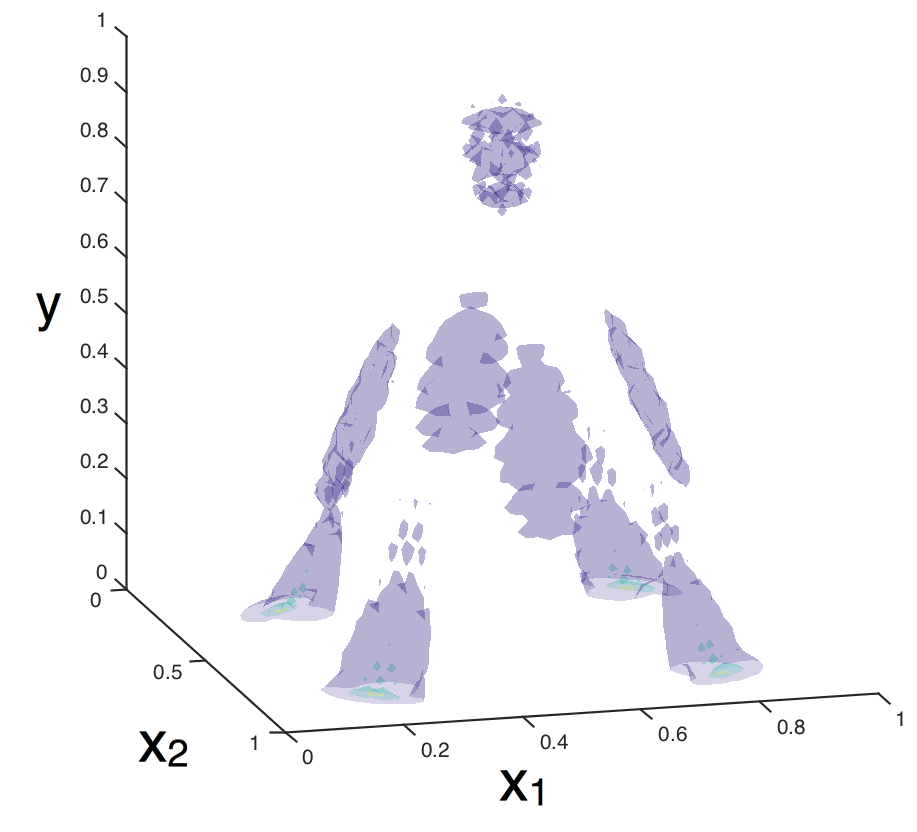} &\hspace{-2em}
		\includegraphics[width=.3\textwidth]{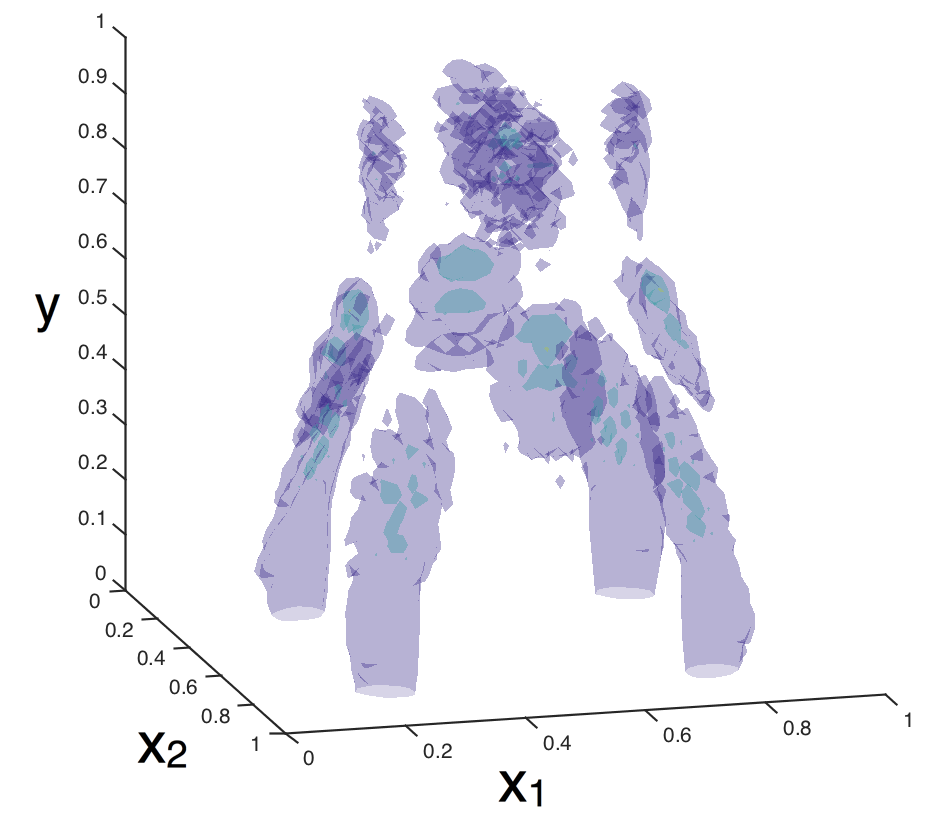} &\hspace{-2em}
		\includegraphics[width=.3\textwidth]{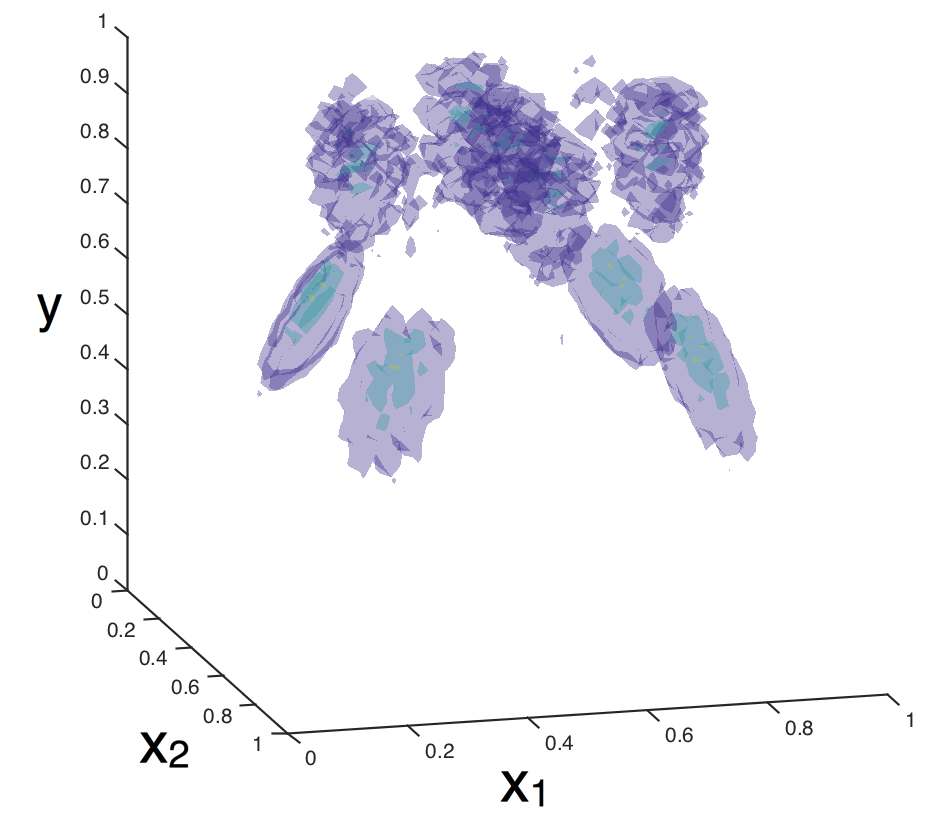}
	\\ $\begin{array}{c} \\[-2.5cm] m_1 \end{array}$ &\hspace{-2em}
		\includegraphics[width=.3\textwidth]{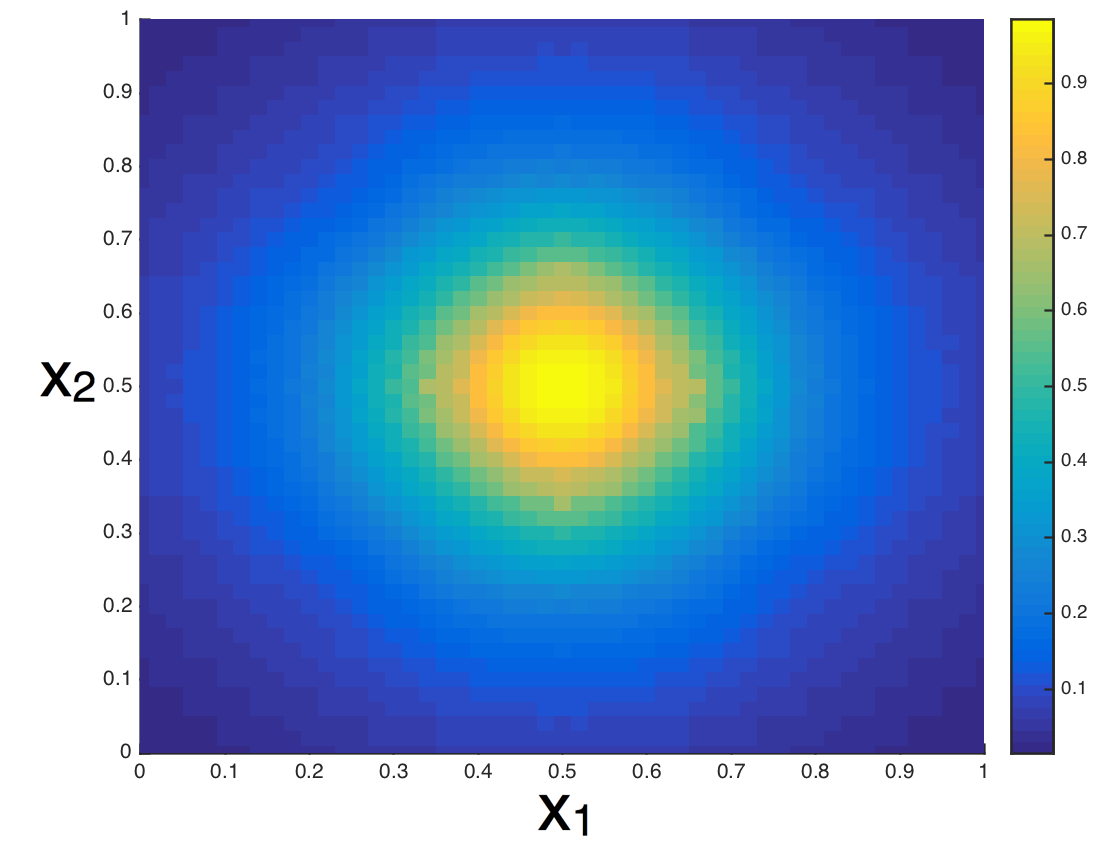} &\hspace{-2em}
		\includegraphics[width=.3\textwidth]{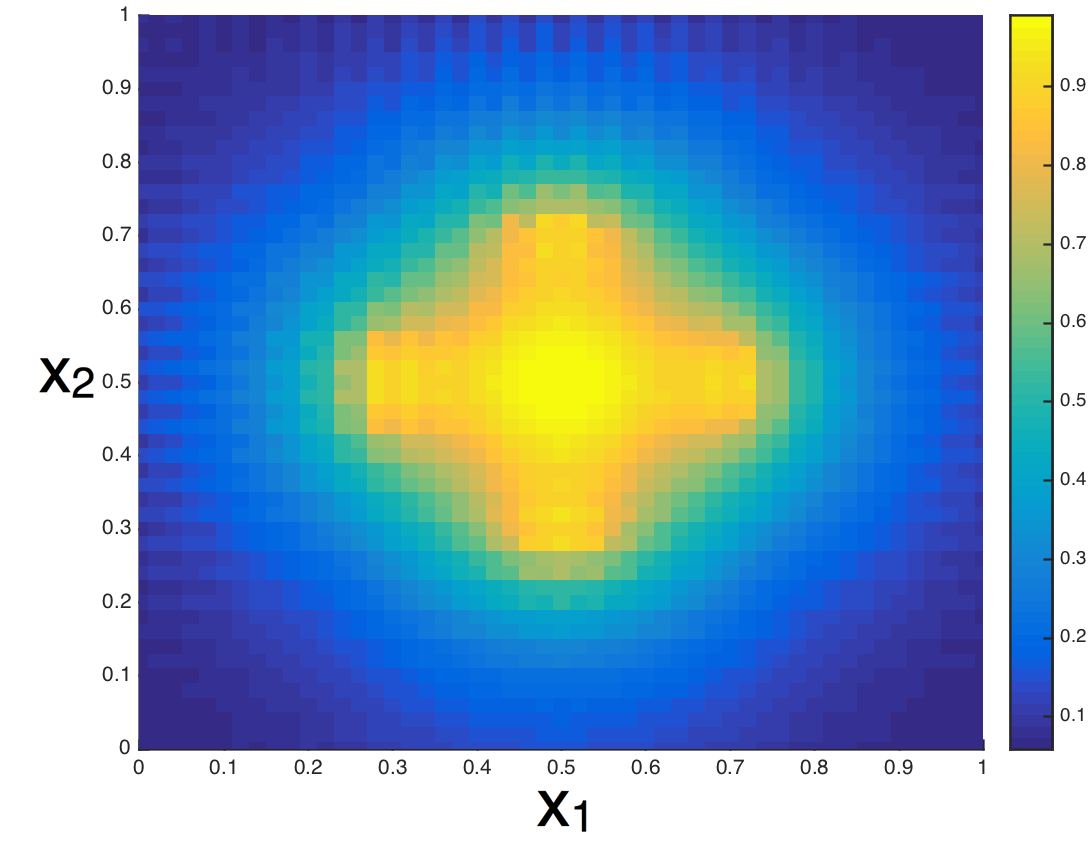} &\hspace{-2em}
		\includegraphics[width=.3\textwidth]{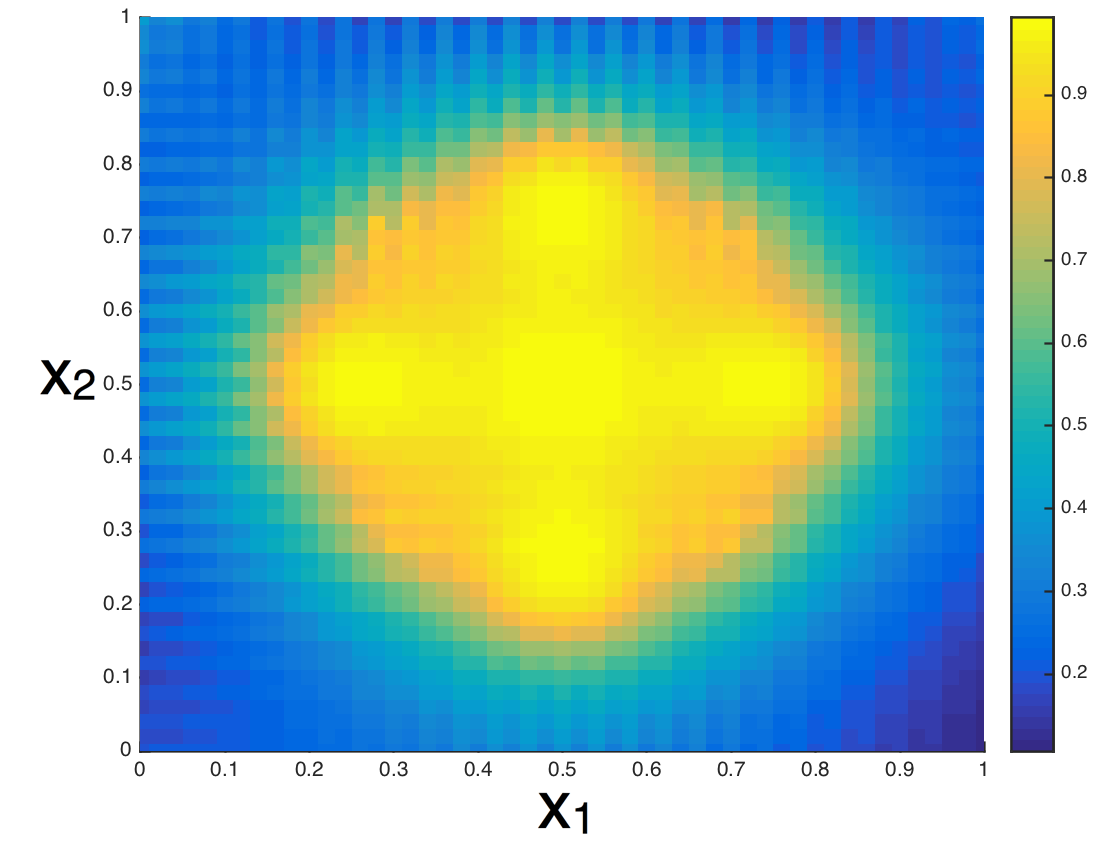}
	\\	&\hspace{-2em} $t=45$ &\hspace{-2em} $t=60$ &\hspace{-2em} $t=75$
	\end{tabular}
	\caption{Multi-cluster results from simulation of model (\ref{eq:IFNmodel_th}) for low affinity ($\lambda=0.01$) are given for $c(t,x,y,\alpha)$ in the spatio-metabolic domain (\textit{$1^{st}$ row}, $c_\alpha$), with $x$ on the horizontal plane and $\alpha$ on the vertical axis; in the spatio-binding domain (\textit{$2^{nd}$ row}, $c_y$), with $x$ on the horizontal plane and $y$ on the vertical axis; and for $m(t,x)$ in space (\textit{$3^{rd}$ row}), for $t\in\{45,60,75\}$ respectively.} \label{fig:Results_theta_1,2}
\end{figure}

\begin{figure}[t!] \centering \begin{tabular}{cccc}
	$\begin{array}{c} \\[-2.5cm] c_\alpha \end{array}$ &\hspace{-2em}
		\includegraphics[width=.3\textwidth]{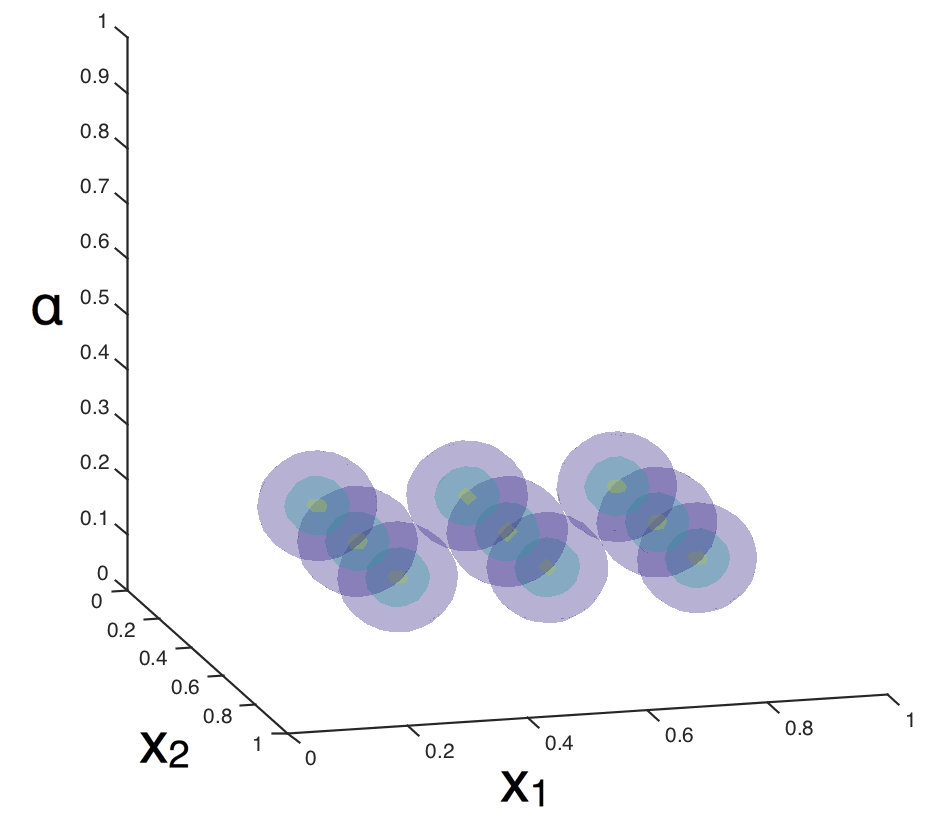} &\hspace{-2em}
		\includegraphics[width=.3\textwidth]{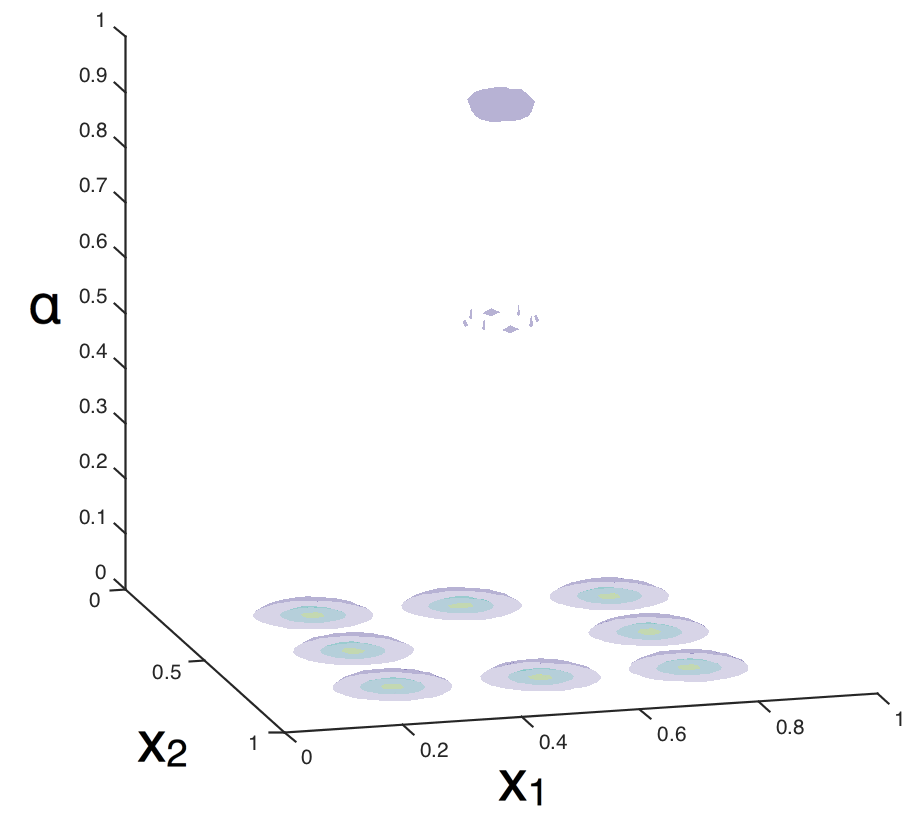} &\hspace{-2em}
		\includegraphics[width=.3\textwidth]{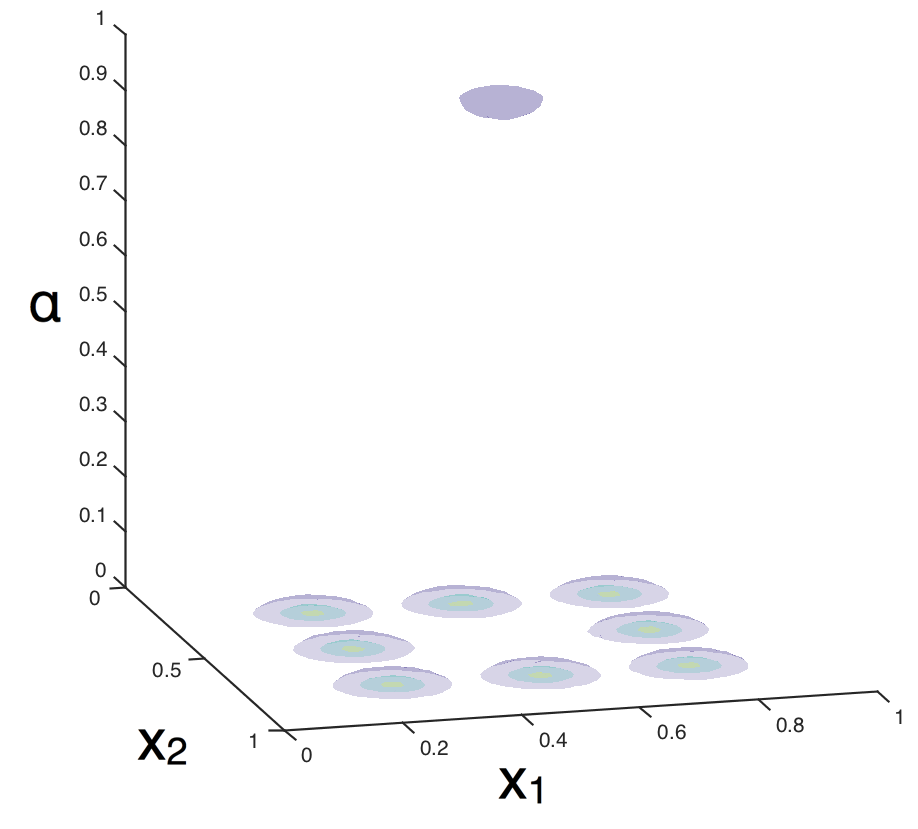}
	\\ $\begin{array}{c} \\[-2.5cm] c_y \end{array}$ &\hspace{-2em}
		\includegraphics[width=.3\textwidth]{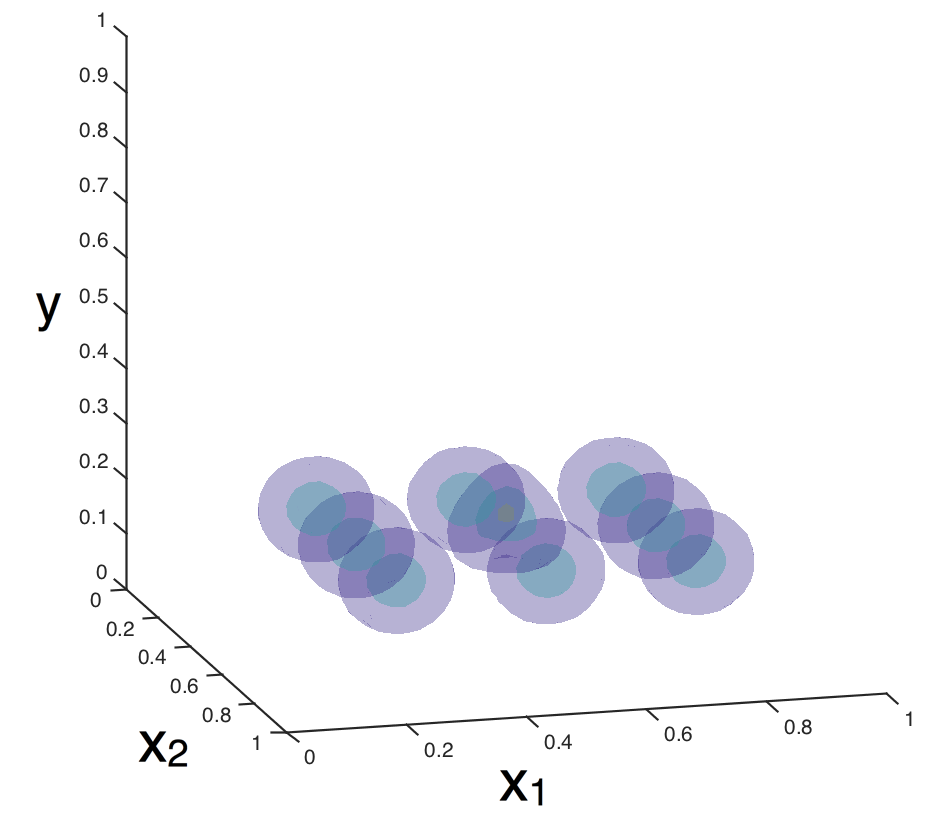} &\hspace{-2em}
		\includegraphics[width=.3\textwidth]{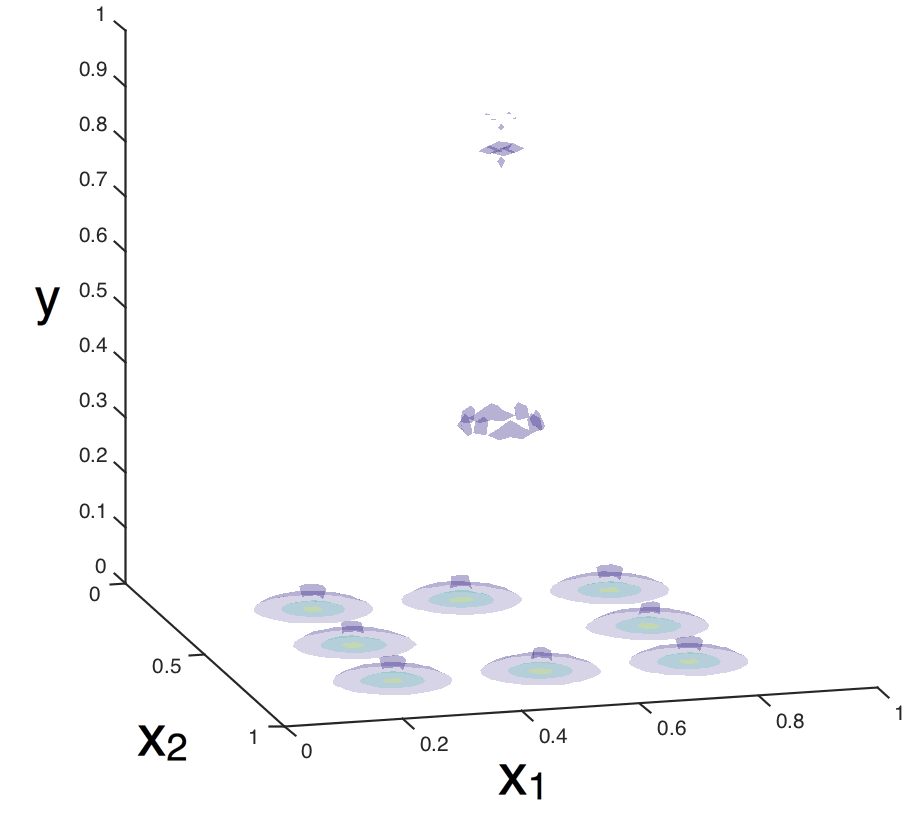} &\hspace{-2em}
		\includegraphics[width=.3\textwidth]{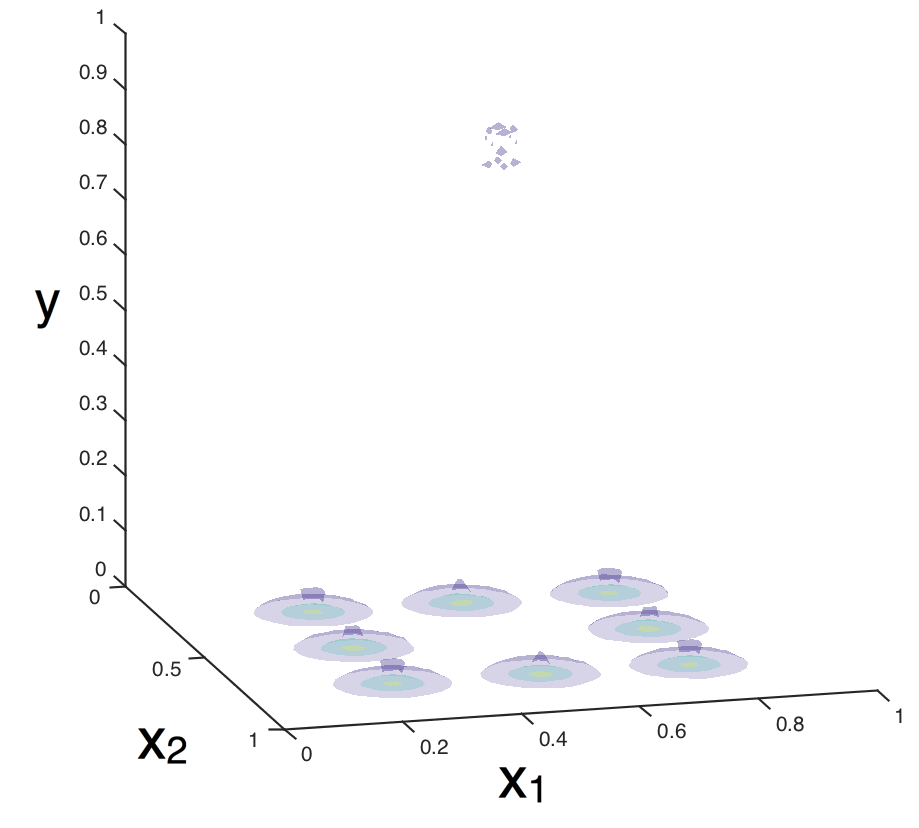}
	\\ $\begin{array}{c} \\[-2.5cm] m_1 \end{array}$ &\hspace{-2em}
		\includegraphics[width=.3\textwidth]{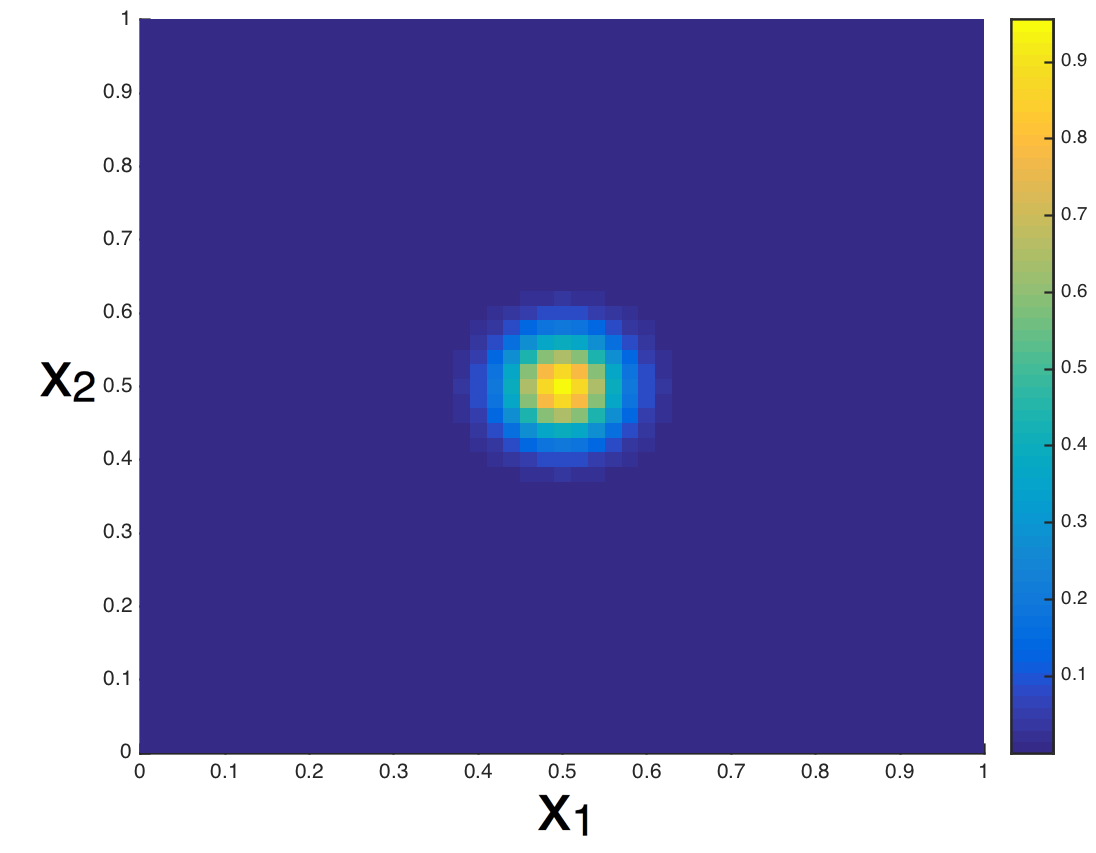} &\hspace{-2em}
		\includegraphics[width=.3\textwidth]{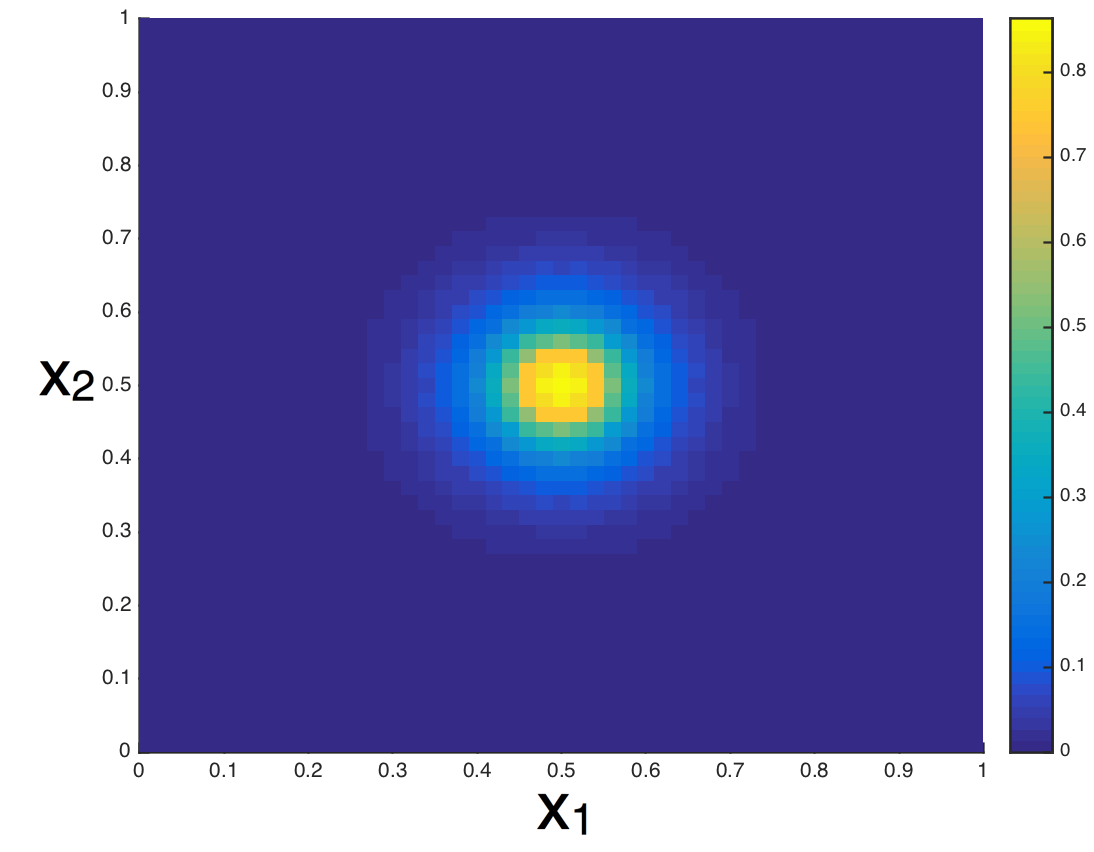} &\hspace{-2em}
		\includegraphics[width=.3\textwidth]{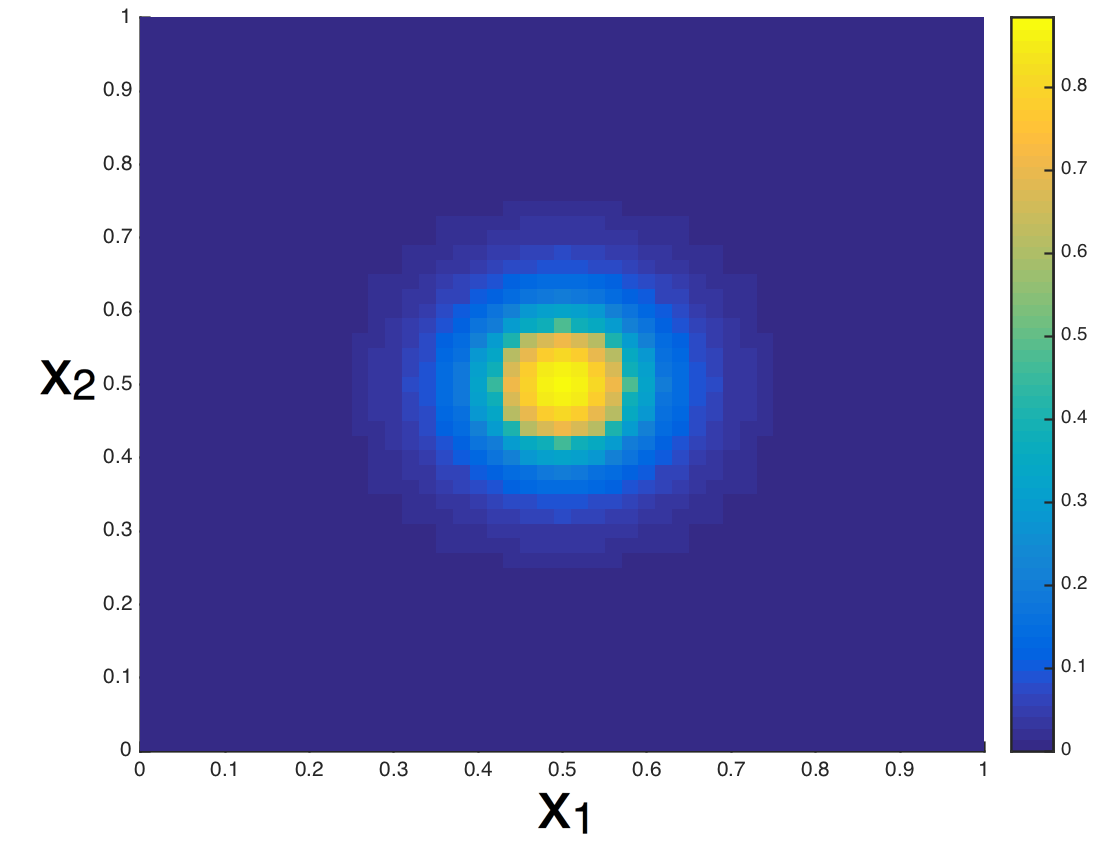}
	\\	&\hspace{-2em} $t=0$ &\hspace{-2em} $t=15$ &\hspace{-2em} $t=30$
	\end{tabular}
	\caption{Multi-cluster results from simulation of model (\ref{eq:IFNmodel_th}) for high affinity ($\lambda=0.5$) are given for $c(t,x,y,\alpha)$ in the spatio-metabolic domain (\textit{$1^{st}$ row}, $c_\alpha$), with $x$ on the horizontal plane and $\alpha$ on the vertical axis; in the spatio-binding domain (\textit{$2^{nd}$ row}, $c_y$), with $x$ on the horizontal plane and $y$ on the vertical axis; and for $m(t,x)$ in space (\textit{$3^{rd}$ row}), for $t\in\{0,15,30\}$ respectively.} \label{fig:Results_theta_50,1}
\end{figure}
\begin{figure}[t!] \centering \begin{tabular}{cccc}
	$\begin{array}{c} \\[-2.5cm] c_\alpha \end{array}$ &\hspace{-2em}
		\includegraphics[width=.3\textwidth]{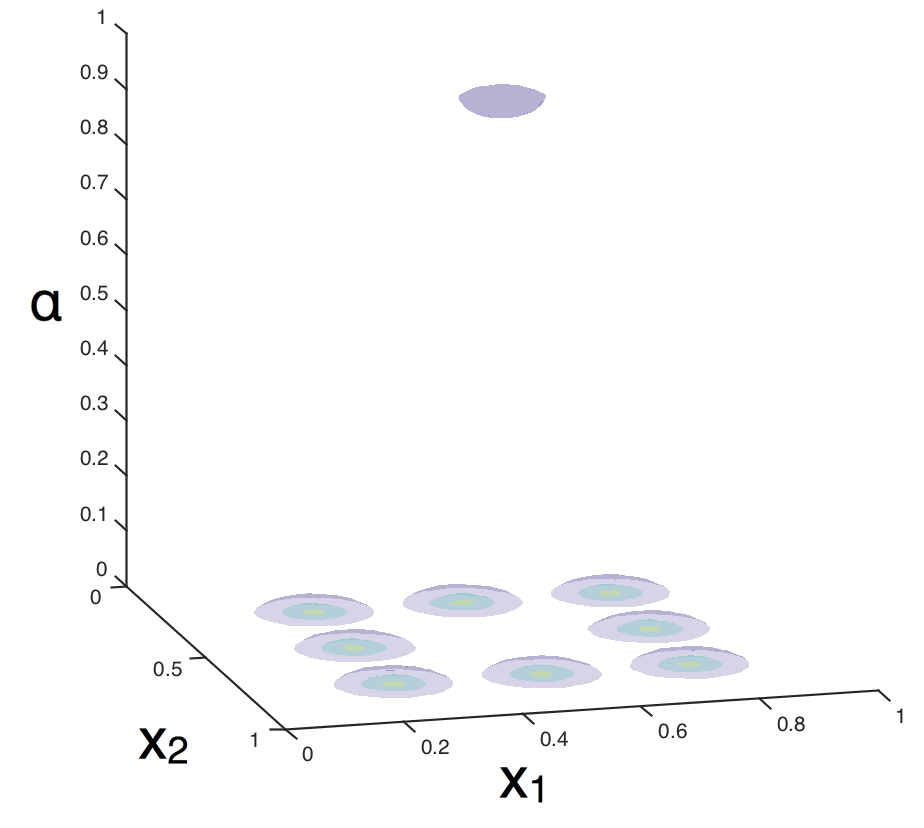} &\hspace{-2em}
		\includegraphics[width=.3\textwidth]{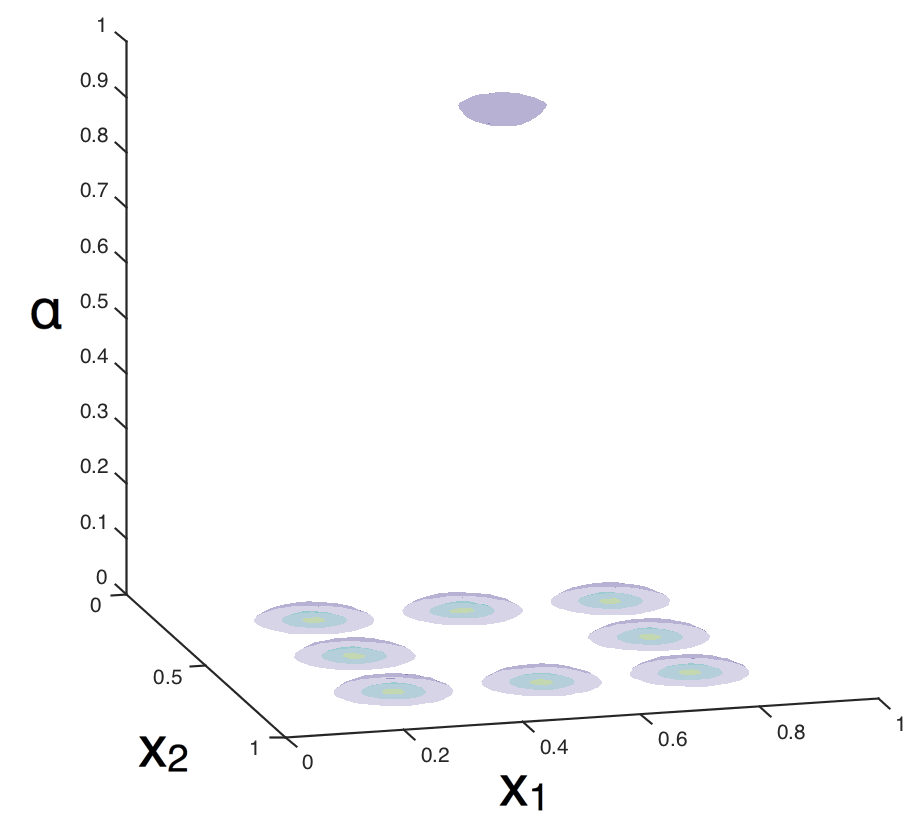} &\hspace{-2em}
		\includegraphics[width=.3\textwidth]{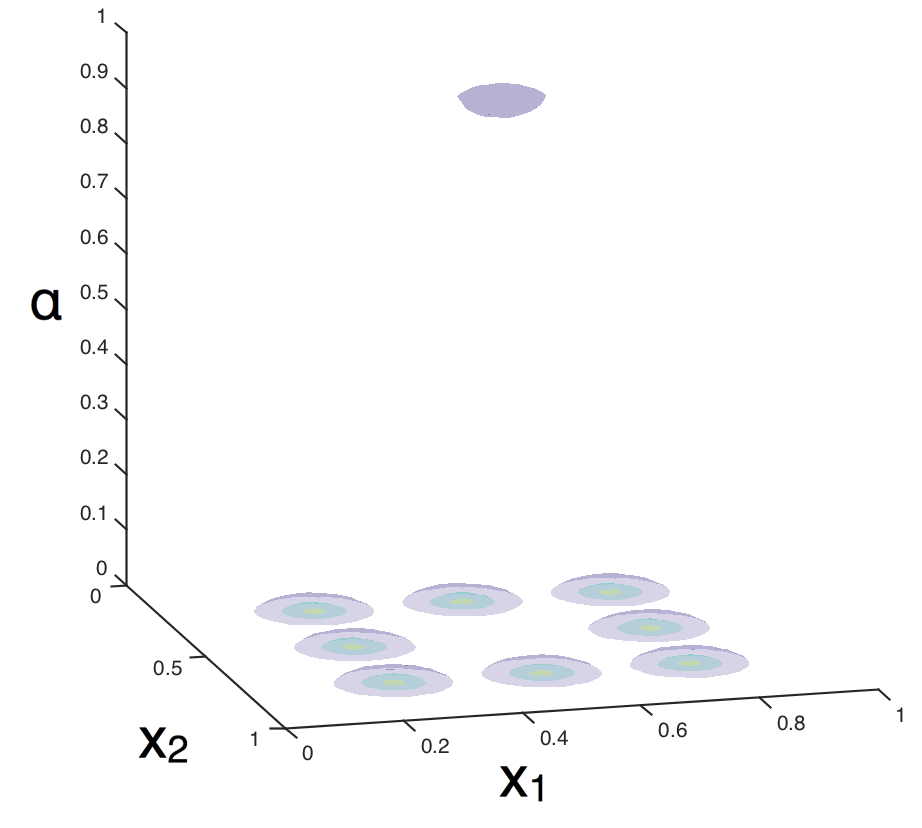}
	\\ $\begin{array}{c} \\[-2.5cm] c_y \end{array}$ &\hspace{-2em}
		\includegraphics[width=.3\textwidth]{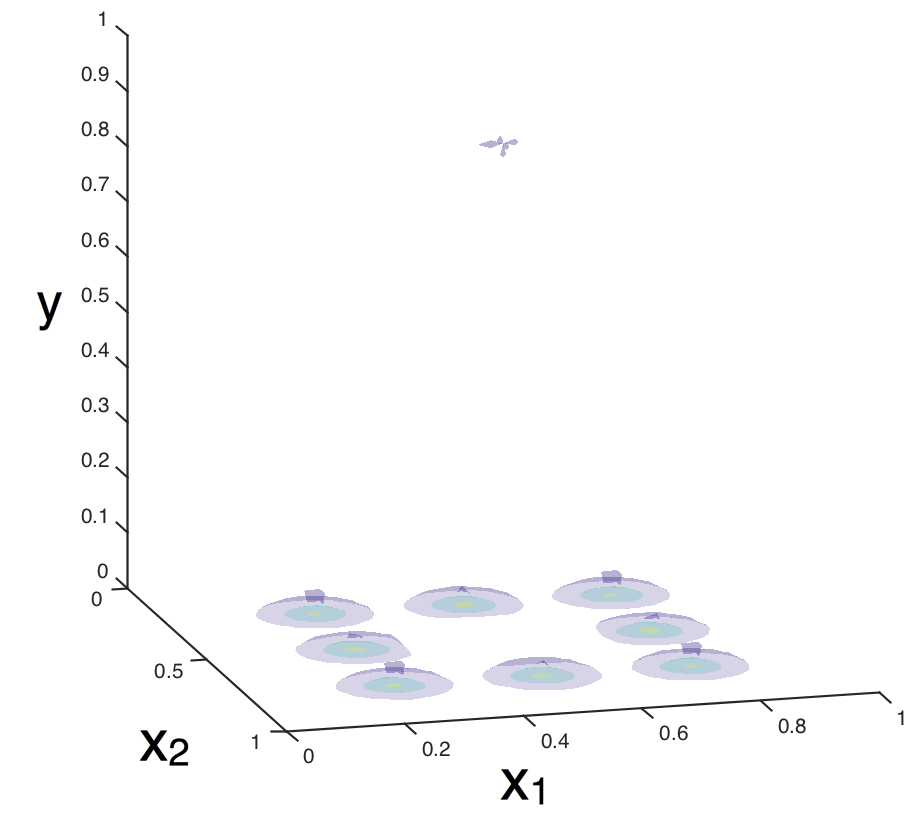} &\hspace{-2em}
		\includegraphics[width=.3\textwidth]{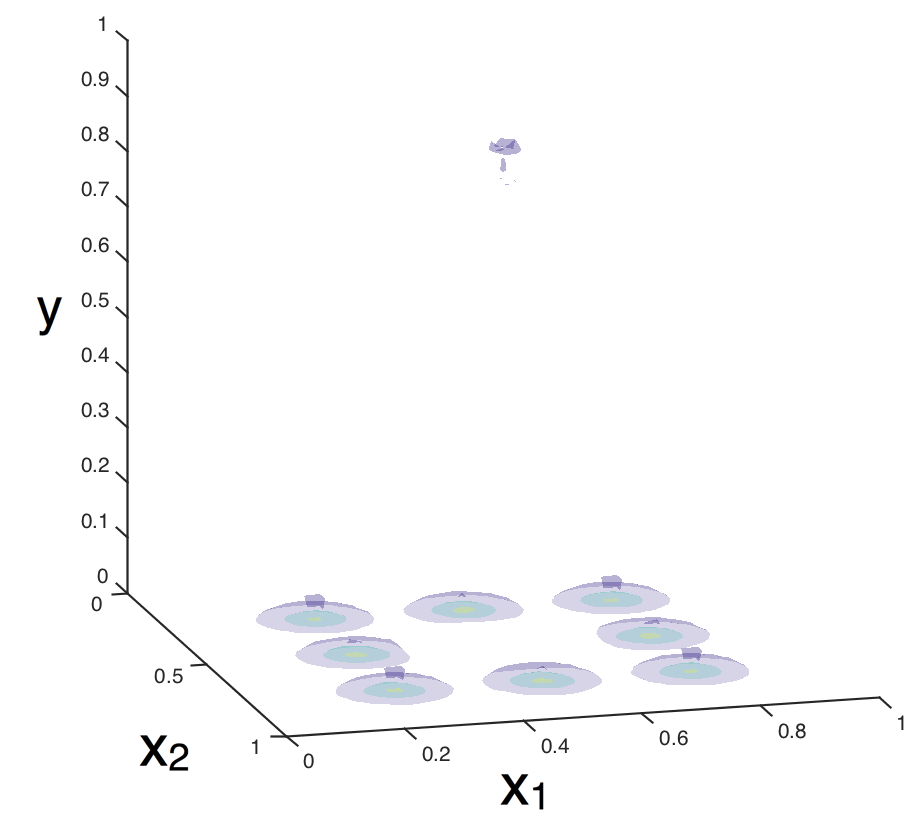} &\hspace{-2em}
		\includegraphics[width=.3\textwidth]{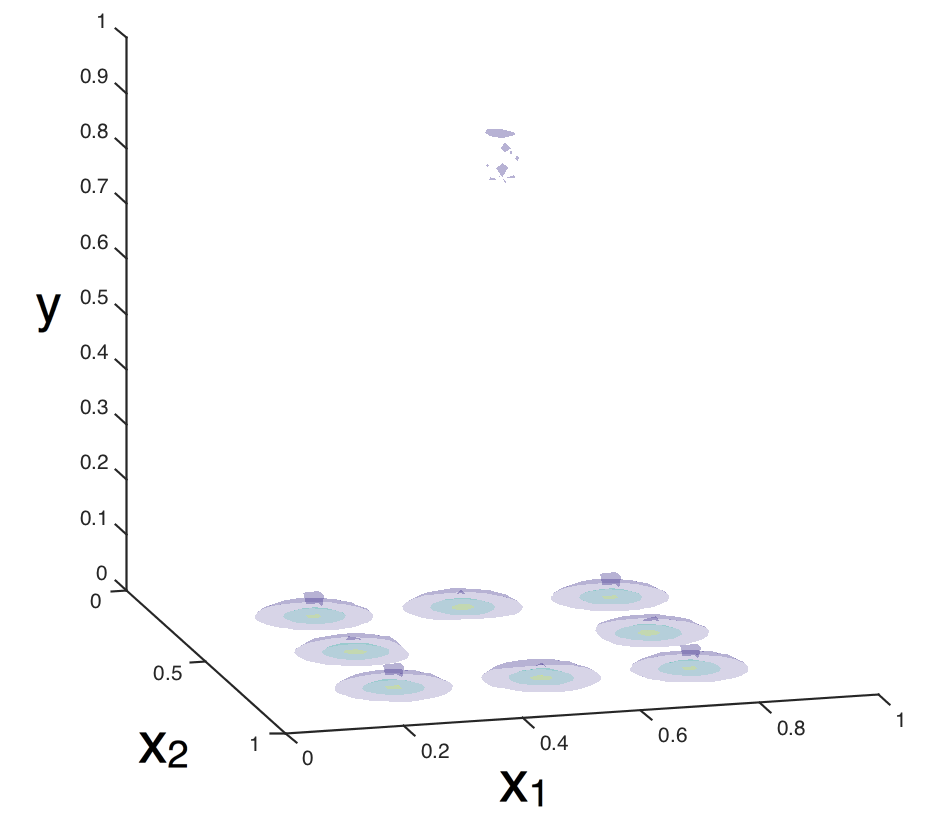}
	\\ $\begin{array}{c} \\[-2.5cm] m_1 \end{array}$ &\hspace{-2em}
		\includegraphics[width=.3\textwidth]{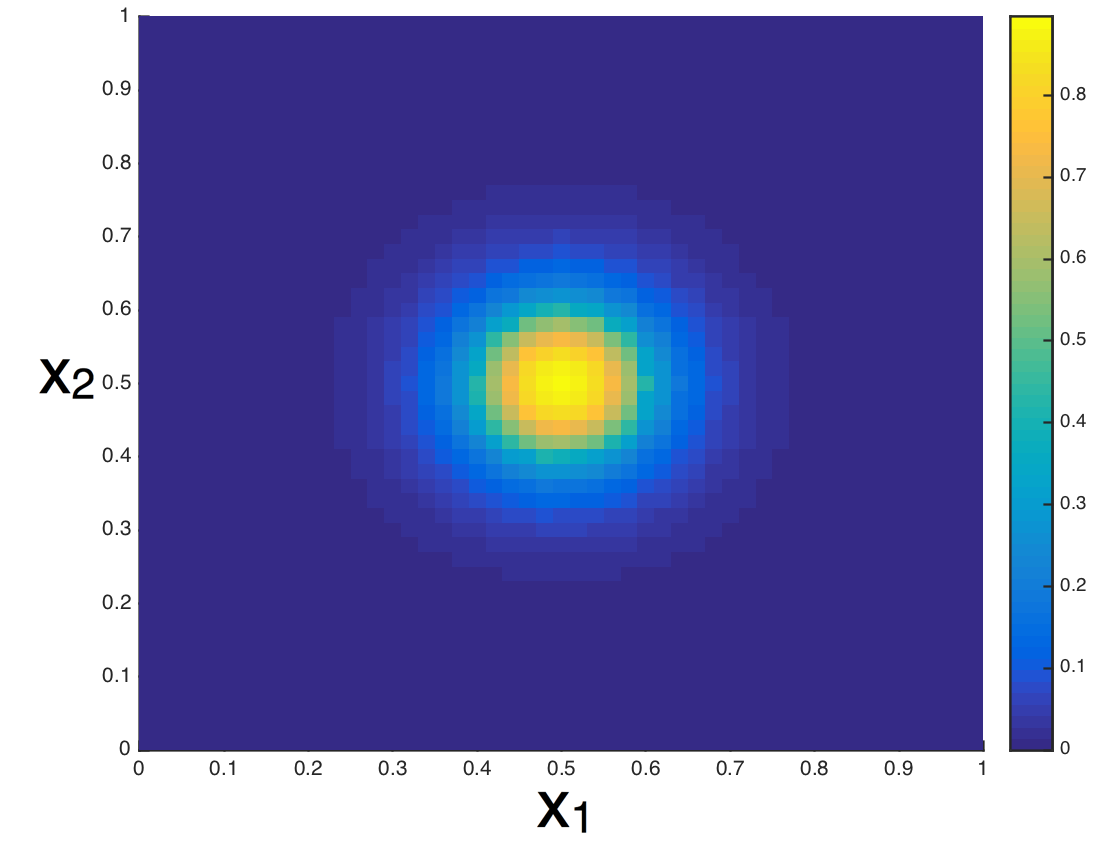} &\hspace{-2em}
		\includegraphics[width=.3\textwidth]{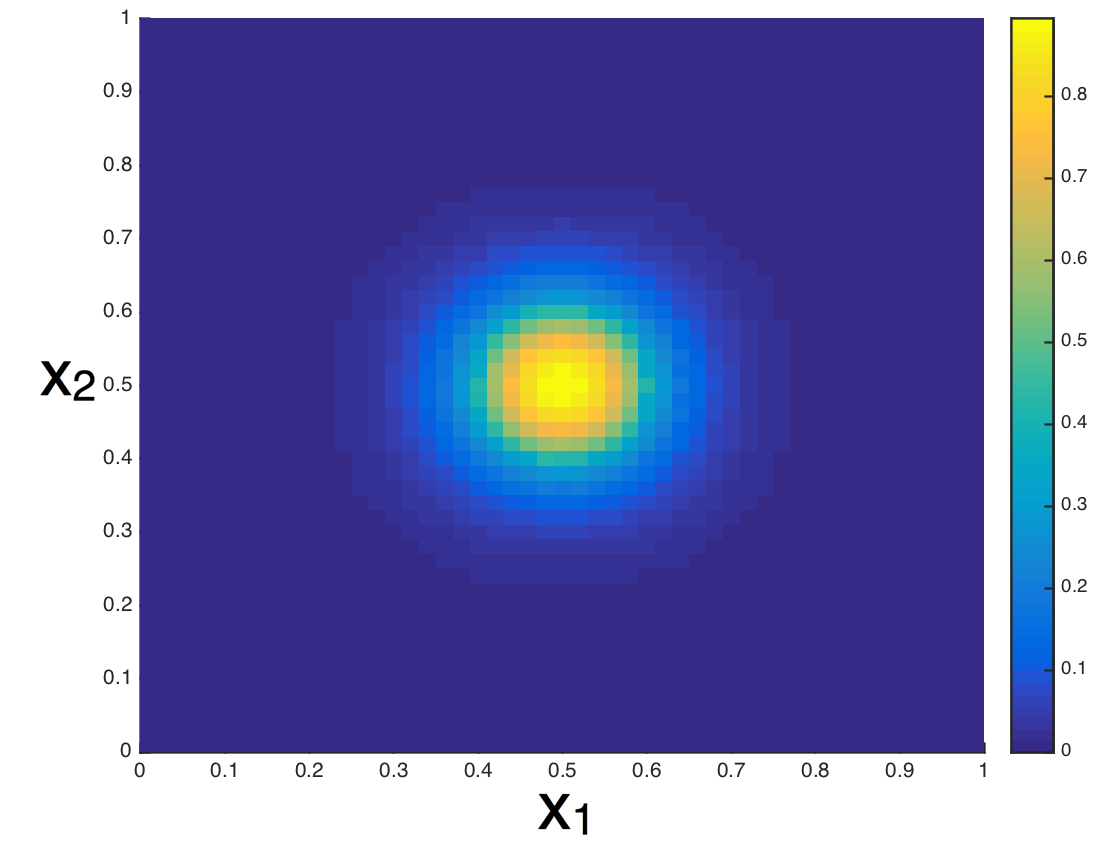} &\hspace{-2em}
		\includegraphics[width=.3\textwidth]{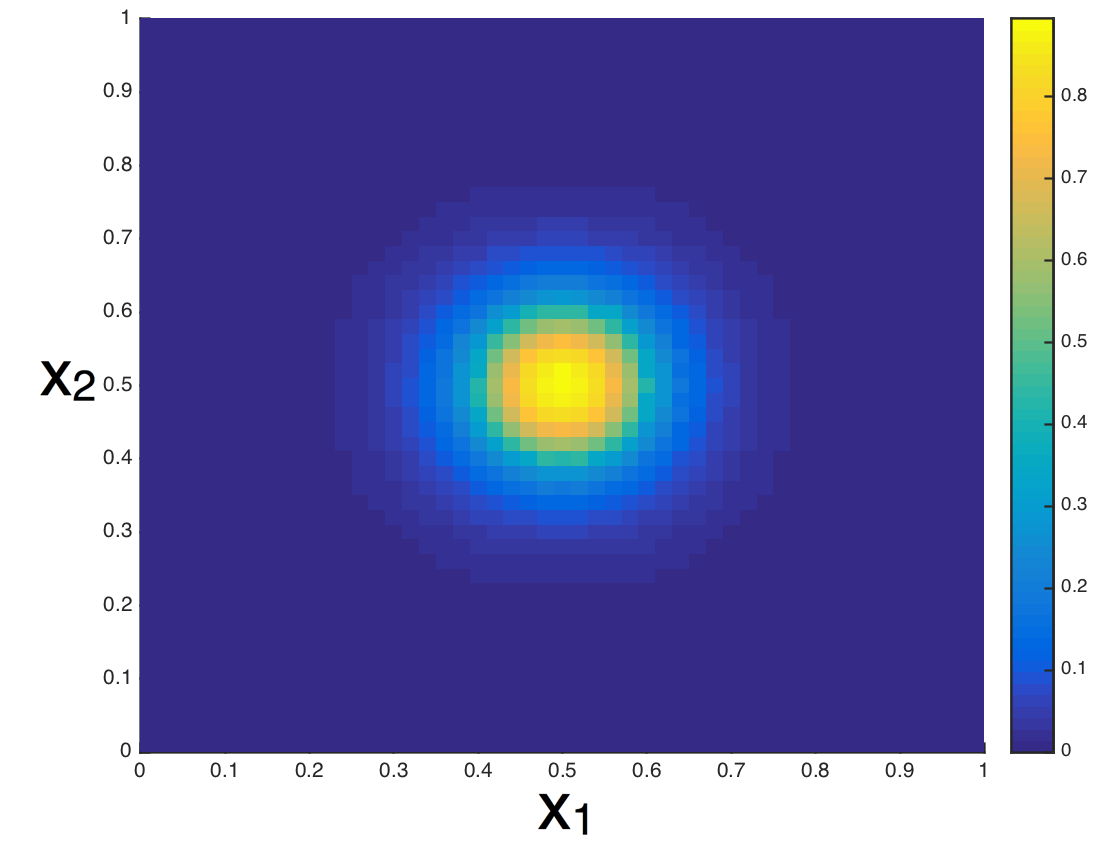}
	\\	&\hspace{-2em} $t=45$ &\hspace{-2em} $t=60$ &\hspace{-2em} $t=75$
	\end{tabular}
	\caption{Multi-cluster results from simulation of model (\ref{eq:IFNmodel_th}) for high affinity ($\lambda=0.5$) are given for $c(t,x,y,\alpha)$ in the spatio-metabolic domain (\textit{$1^{st}$ row}, $c_\alpha$), with $x$ on the horizontal plane and $\alpha$ on the vertical axis; in the spatio-binding domain (\textit{$2^{nd}$ row}, $c_y$), with $x$ on the horizontal plane and $y$ on the vertical axis; and for $m(t,x)$ in space (\textit{$3^{rd}$ row}), for $t\in\{45,60,75\}$ respectively.} \label{fig:Results_theta_50,2}
\end{figure}

Spatially static, single-cluster results were generated by simulating (\ref{eq:IFNmodel}), whilst multi-cluster results generate by simulating (\ref{eq:IFNmodel_th}). Spatially-dynamics results were generated by simulating (\ref{eq:IFNmodel_ch}). A full description of numerical techniques used, methods, and parameters for simulating this system of equations is given in \ref{app:NumMethods}, where parameters were used as appropriate for the simulated model.

\corr{In the following we will refer to two types of numerical simulations that differ by the type of initial
condition. Single cluster simulations start with a localized cell distribution having a single maximum.
Multiple cluster simulations start with an initial cell distribution having several maxima periodically
positioned in space.    }

\subsection{Spatially-static, single-cluster simulations}

Single-cluster results (Fig. \ref{fig:Single_cell,1} \& \ref{fig:Single_cell,2}) demonstrate an initial rise in average binding position, $c_y$, of the cellular population with a concurrent rise in average metabolic position, $c_\alpha$.
In $\breve{c}$ we also observe the rise in metabolo-binding state with a  focus developing at approximately $(y,\alpha)\approx(0.45,0.55)$, with a negatively graduated non-linear ridge, and a tail between the focus and $(y,\alpha)=(0,0)$.

Beyond $t=20$, the average distribution in the binding space remains largely static, whilst the population continues to redistribute itself into a teardrop geometry, around the average position. This indicates, firstly, that the cell is capable of sustaining its own binding state, through production, upon initial stimulation with IFN. The formation of this geometry could be as a result of the maximal concentration of producer cells being central, and thusly producing greater levels of IFN which can be bound by the population, itself.

The distribution in the metabolic space exhibits oscillation, around its average position, for all time points $t\geq15$ (Fig. \ref{fig:Single_cell,2} $c_\alpha$). This oscillation is both transverse and longitudinal, and is likely to occur as a result of the SAR-cycling between the metabolic and binding states of these cells. This demonstrates the importance of the establishment of heterogeneity within the cellular population as it acts to regulate the IFN output of the system, whilst concurrently maximising metabolic expedition from the available and bound IFN supplies. Interferon producer cells do not act in unison and, indeed, use heterogeneity to co-regulate cells within such a cluster.

The final observation that one wishes to make in the results for the single-cluster case is the visible SAR-cycle displayed within the metabolo-binding space (Fig. \ref{fig:Single_cell,1} \& \ref{fig:Single_cell,2} $\breve{c}$). Regions of the solution for the cellular population appear to increase their binding state of IFN; before concurrently increasing their metabolic state and slightly decreasing their binding state; subsequently decreasing their metabolic and binding states, together; and beginning this cycle, once more. Whilst the majority of the population maintains its position within the bulk of this distribution, there exist cells (or subpopulations of the cellular population) that are affected by this feedback cycle.

\subsection{Spatially-static, multi-cluster simulations}

In the multi-cluster results (Fig. \ref{fig:Results_theta_1,1}--\ref{fig:Results_theta_50,2}) we observe a significant difference in the behaviour of the metabolic and binding spaces, in comparison to those of the single cluster. One observes the appearance of stable regions within the metabolic space, at high values for $\alpha$; a phenomenon that we term `metabolic trapping'. In the low affinity case, where the focal point for metabolo-binding dynamics would be lower in value, this effect is likely due to the feeding back of IFN proteins between clusters that lead the internal feedback mechanism to be ineffective at downregulating the metabolic state of the cell. In the high affinity case, this is likely to be due to the high binding and retention rates, in comparison to the unbinding rate, which causes the internalisation rate to remain high.

The binding state (Fig. \ref{fig:Results_theta_1,1}--\ref{fig:Results_theta_50,2} $c_y$), on the other hand, demonstrate oscillatory dynamics which were before characteristic of the metabolic state. Upon the establishment of stable metabolic dynamics, at high values for $\alpha$, one expects that the conflict between the high rates of binding (caused by high rates of production and subsequent values for free chemical concentrations) and the feedback mechanism of the metabolic gene circuitry would cause such a behaviour. Cells will attempt to bind the high levels of IFN whilst the feedback mechanism continually acts to diminish the affinity of producer cells for IFN.

One should also notice that in the low affinity case (Fig. \ref{fig:Results_theta_1,1} \& \ref{fig:Results_theta_1,2}), as opposed to the high affinity case (Fig. \ref{fig:Results_theta_50,1} \& \ref{fig:Results_theta_50,2}), one observes that the signal is conveyed to the neighbouring cells. This can only be achieved through the implementation of a threshold in the binding dynamics for $c(t,x,y,\alpha)$ and this same threshold mediates the distance at which the signal can be conveyed.

Moreover, a simple comparative between the high affinity multi-cluster (Fig. \ref{fig:Results_theta_50,1} \& \ref{fig:Results_theta_50,2}), low affinity multi-cluster (Fig. \ref{fig:Results_theta_1,1} \& \ref{fig:Results_theta_1,2}), and single-cluster (Fig. \ref{fig:Single_cell,1} \& \ref{fig:Single_cell,2}) results will show that the concentrations of IFN produced by the low affinity multi-cluster system were far in excess of those in the other two cases. This is likely as a result of the cumulative production but also as a result of the production of the two, or more, clusters feeding back the IFN to one another, causing a metabolic trapping effect. This metabolic trapping is manifest as an emergence of the population at the upper boundary of the metabolic space and retention of this position. This effect is opposed to that of the metabolo-binding SAR-cycling that one observes in the single cluster case and is as a direct result of inter-cluster heterogeneity, where the promotion of the primed state in one cluster will facilitate the priming of the second, and so on.

\subsection{Spatially-dynamic, multi-cluster simulations}

\begin{figure}[t!] \centering \begin{tabular}{cccc}
	$\begin{array}{c} \\[-2.5cm] c_\alpha \end{array}$ &\hspace{-2em}
		\includegraphics[width=.3\textwidth]{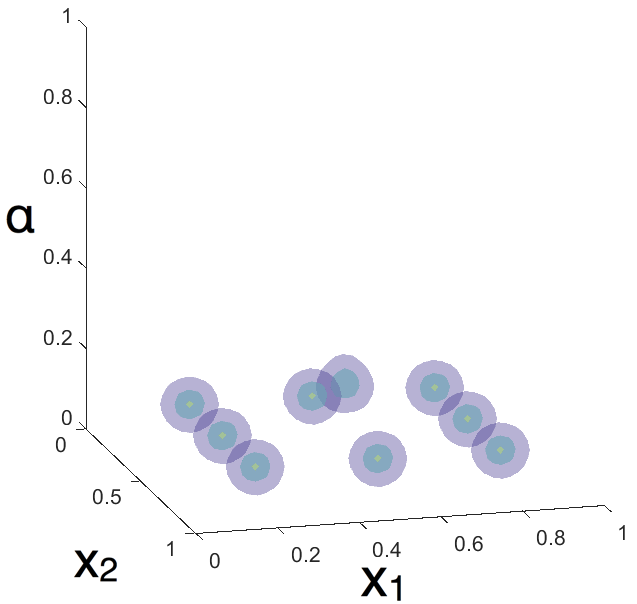} &\hspace{-2em}
		\includegraphics[width=.3\textwidth]{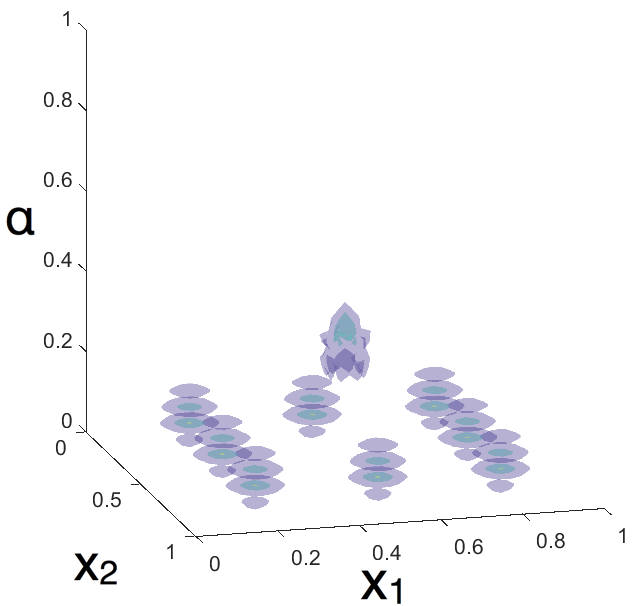} &\hspace{-2em}
		\includegraphics[width=.3\textwidth]{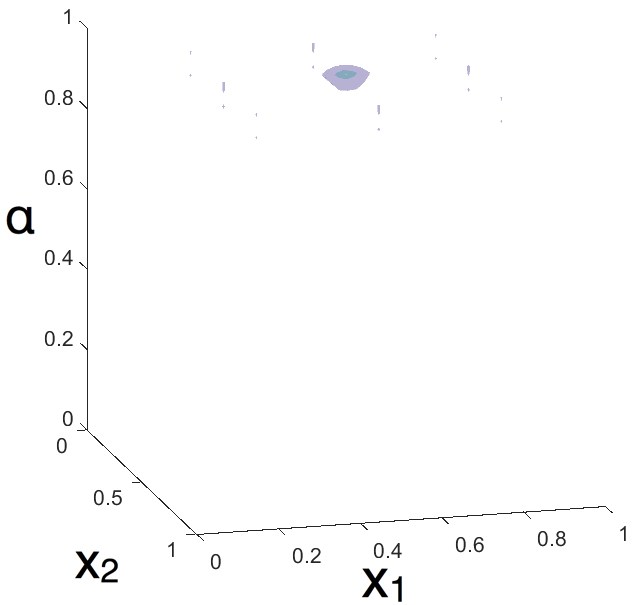}
	\\ $\begin{array}{c} \\[-2.5cm] c_y \end{array}$ &\hspace{-2em}
		\includegraphics[width=.3\textwidth]{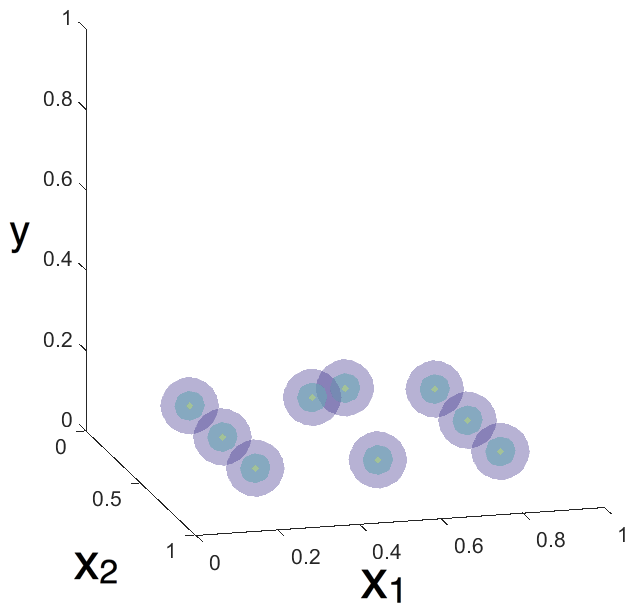} &\hspace{-2em}
		\includegraphics[width=.3\textwidth]{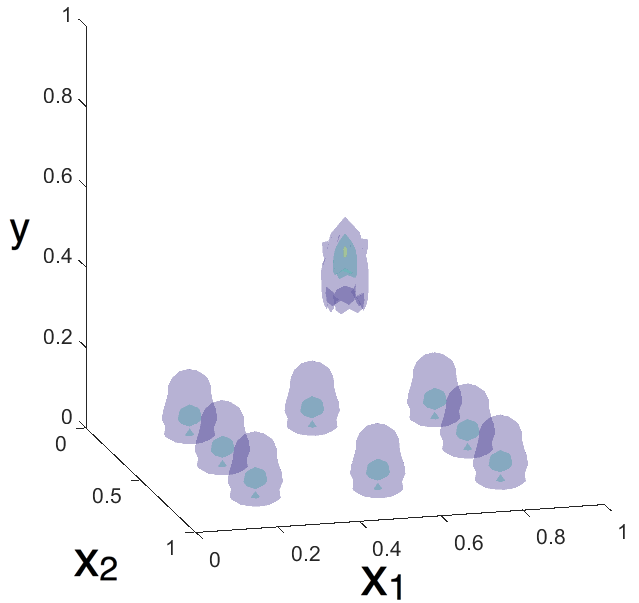} &\hspace{-2em}
		\includegraphics[width=.3\textwidth]{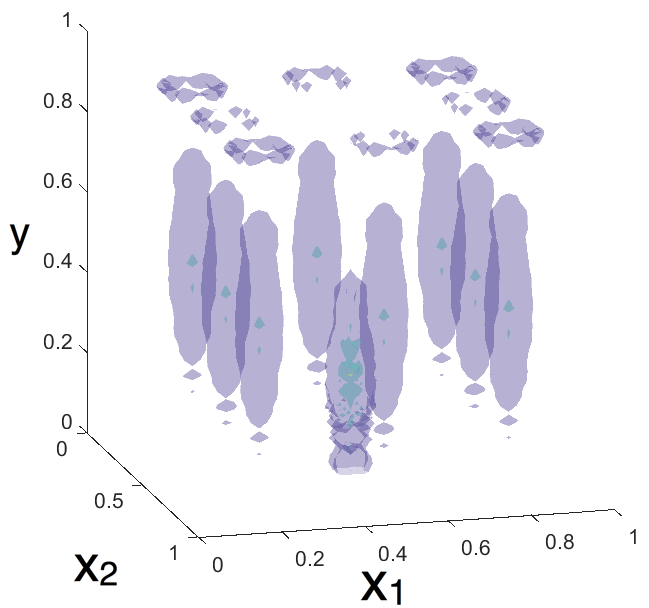}
	\\ $\begin{array}{c} \\[-2.5cm] m_1 \end{array}$ &\hspace{-2em}
		\includegraphics[width=.3\textwidth]{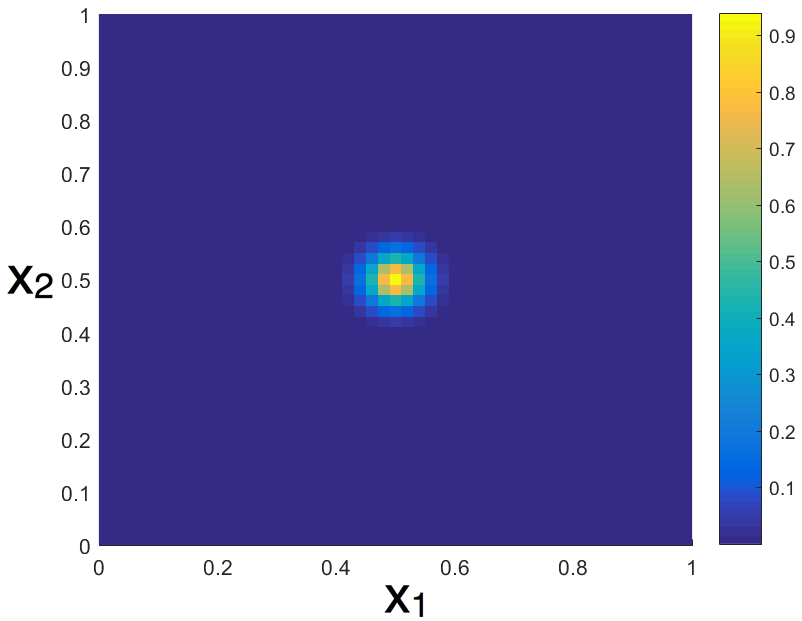} &\hspace{-2em}
		\includegraphics[width=.3\textwidth]{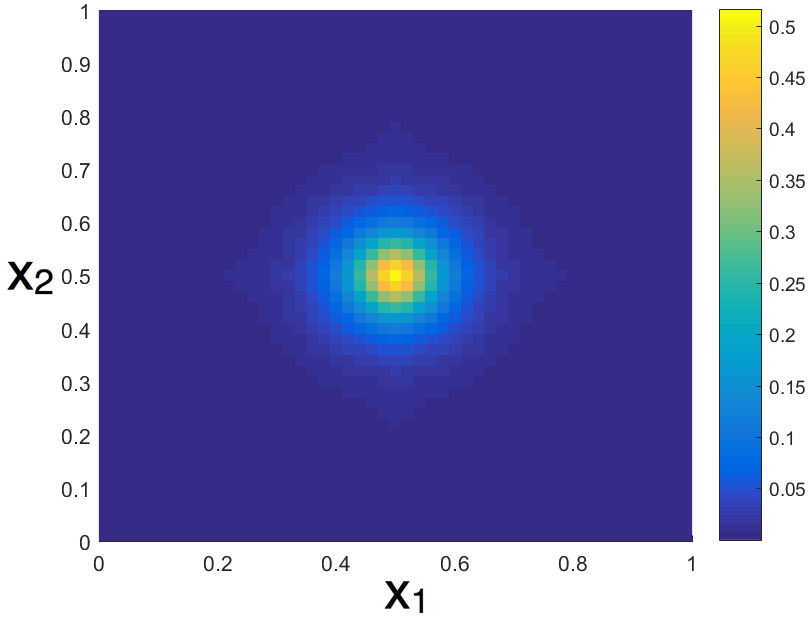} &\hspace{-2em}
		\includegraphics[width=.3\textwidth]{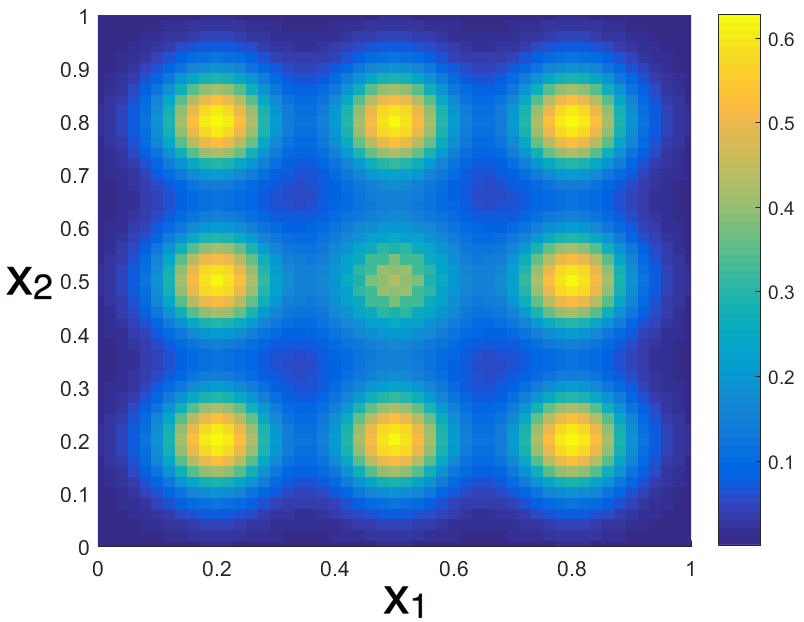}
	\\	&\hspace{-2em} $t=0$ &\hspace{-2em} $t=10$ &\hspace{-2em} $t=20$
	\end{tabular}
	\caption{Multi-cluster results from simulation of model (\ref{eq:IFNmodel_ch}) for high affinity ($\lambda=0.5$) are given for $c(t,x,y,\alpha)$ in the spatio-metabolic domain (\textit{$1^{st}$ row}, $c_\alpha$), with $x$ on the horizontal plane and $\alpha$ on the vertical axis; in the spatio-binding domain (\textit{$2^{nd}$ row}, $c_y$), with $x$ on the horizontal plane and $y$ on the vertical axis; and for $m(t,x)$ in space (\textit{$3^{rd}$ row}), for $t\in\{0,10,20\}$ respectively.} \label{fig:Results_chem_50,1}
\end{figure}
\begin{figure}[t!] \centering \begin{tabular}{cccc}
	$\begin{array}{c} \\[-2.5cm] c_\alpha \end{array}$ &\hspace{-2em}
		\includegraphics[width=.3\textwidth]{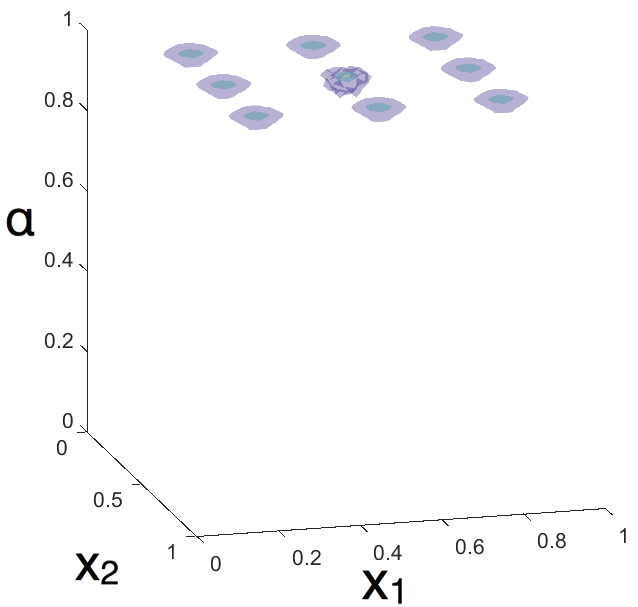} &\hspace{-2em}
		\includegraphics[width=.3\textwidth]{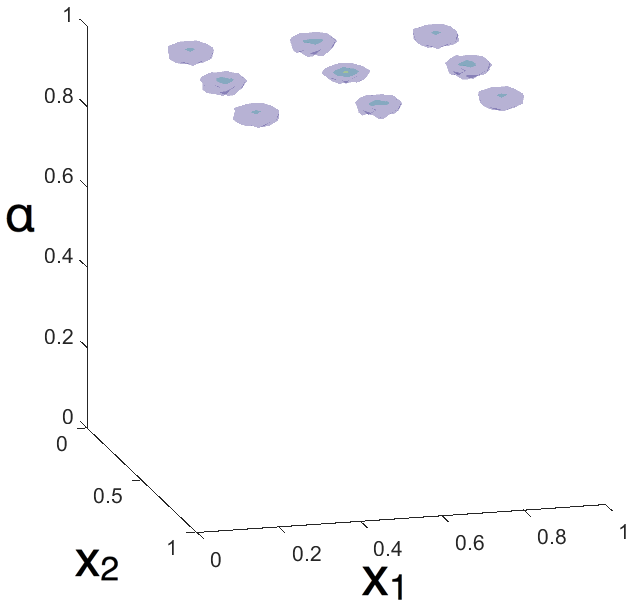} &\hspace{-2em}
		\includegraphics[width=.3\textwidth]{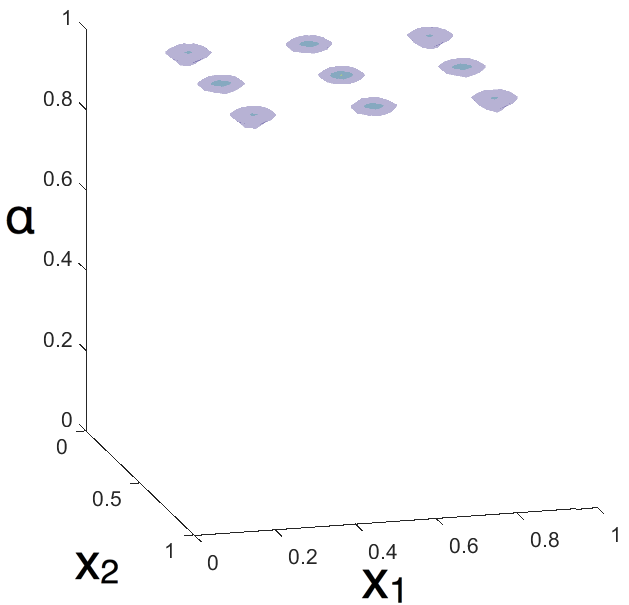}
	\\ $\begin{array}{c} \\[-2.5cm] c_y \end{array}$ &\hspace{-2em}
		\includegraphics[width=.3\textwidth]{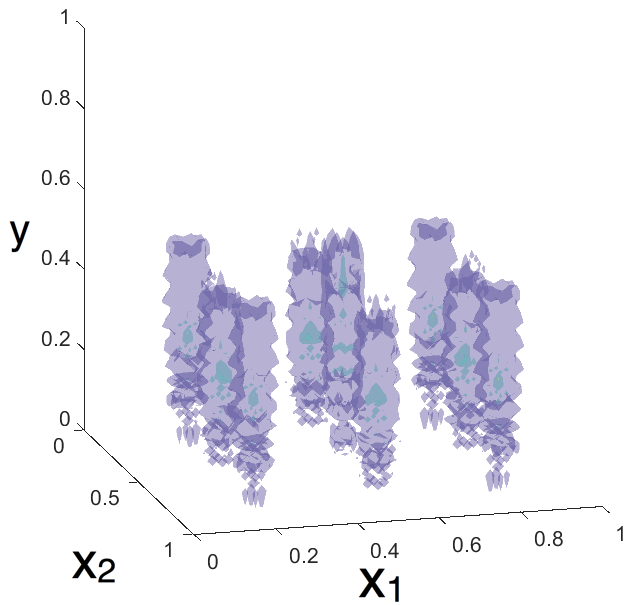} &\hspace{-2em}
		\includegraphics[width=.3\textwidth]{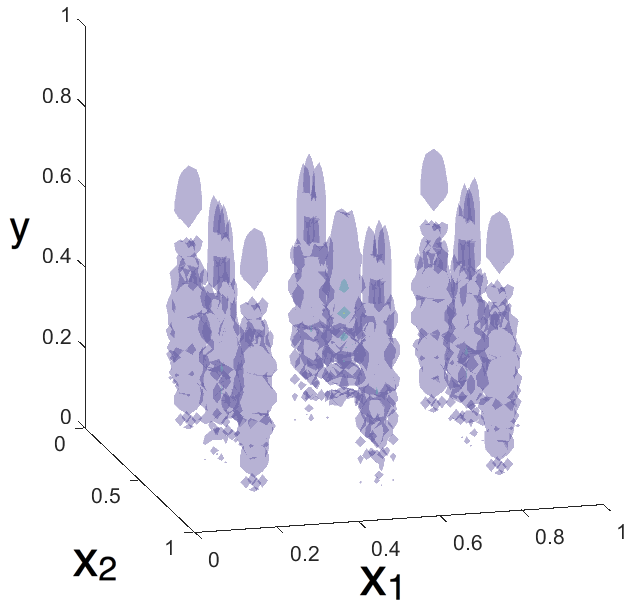} &\hspace{-2em}
		\includegraphics[width=.3\textwidth]{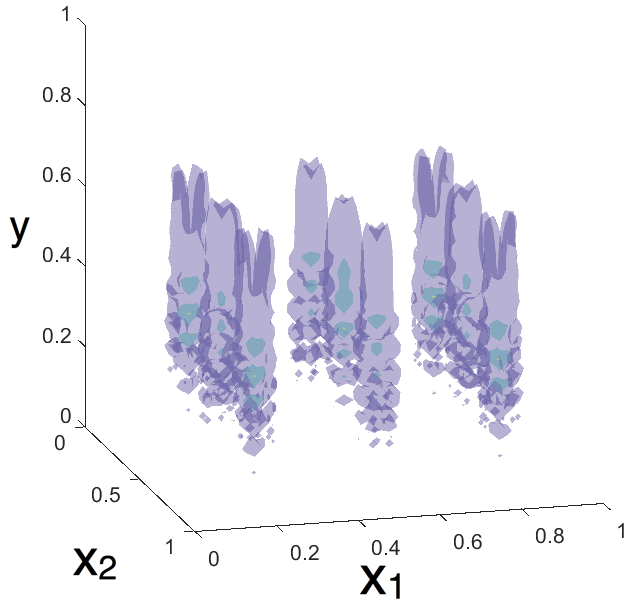}
	\\ $\begin{array}{c} \\[-2.5cm] m_1 \end{array}$ &\hspace{-2em}
		\includegraphics[width=.3\textwidth]{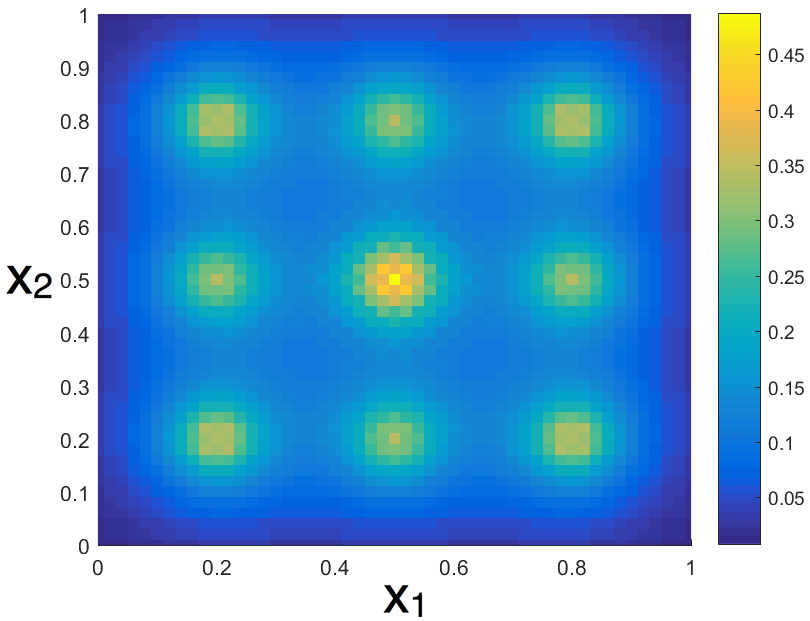} &\hspace{-2em}
		\includegraphics[width=.3\textwidth]{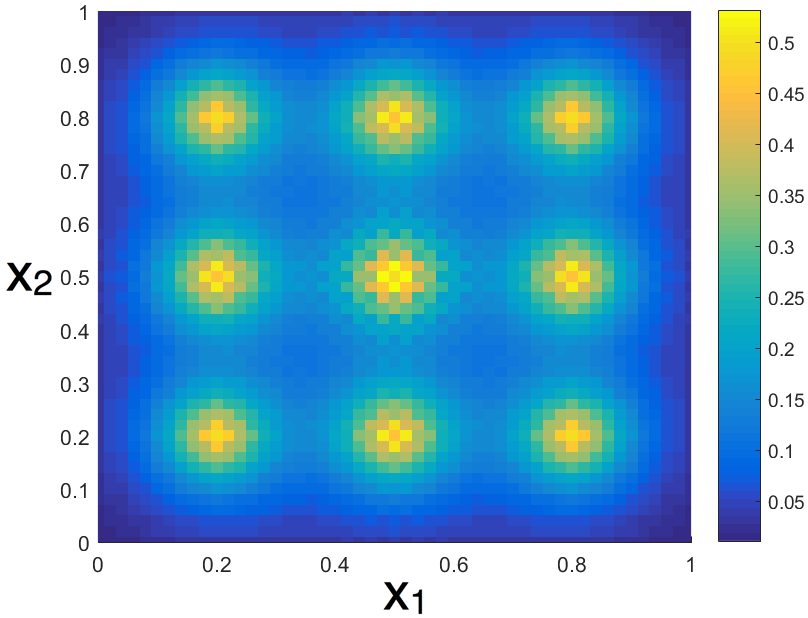} &\hspace{-2em}
		\includegraphics[width=.3\textwidth]{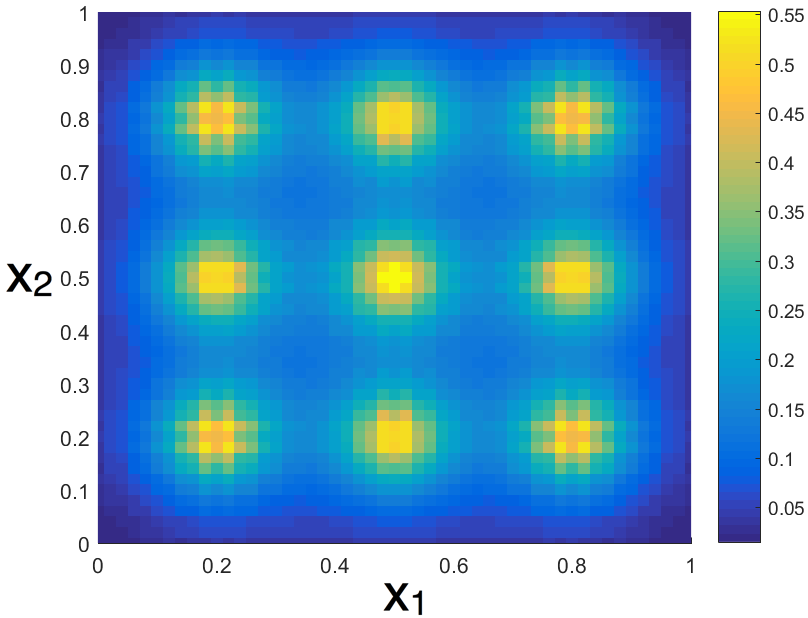}
	\\	&\hspace{-2em} $t=30$ &\hspace{-2em} $t=40$ &\hspace{-2em} $t=50$
	\end{tabular}
	\caption{Multi-cluster results from simulation of model (\ref{eq:IFNmodel_ch}) for high affinity ($\lambda=0.5$) are given for $c(t,x,y,\alpha)$ in the spatio-metabolic domain (\textit{$1^{st}$ row}, $c_\alpha$), with $x$ on the horizontal plane and $\alpha$ on the vertical axis; in the spatio-binding domain (\textit{$2^{nd}$ row}, $c_y$), with $x$ on the horizontal plane and $y$ on the vertical axis; and for $m(t,x)$ in space (\textit{$3^{rd}$ row}), for $t\in\{30,40,50\}$ respectively.} \label{fig:Results_chem_50,2}
\end{figure}

Consider, now, the numerically generated results for the system (\ref{eq:IFNmodel_ch}), with parameters given as in above sections (\ref{app:NumMethods}). We give the simulated solutions for the high-affinity, multi-cluster IFN case (Fig. \ref{fig:Results_chem_50,1} \& \ref{fig:Results_chem_50,2}), only, as the spatio-metabolo-binding dynamics are similar at both high and low affinities. One immediately observes the dissolution of the discrepancy between the two species in terms of their communicative capability. The high affinity SARs are able to communicate with one another under a spatially-dynamic, chemotactic regime.

In order to best understand these dynamics, one must observe them in the passage of time. The chosen initial conditions impose a stimulus on the central cluster of cells, whilst peripheral clusters are in a state of metabolic relaxation (Fig. \ref{fig:Results_chem_50,1}). The spatial dynamics of the central cluster, at early time-points, will be mainly balanced between diffusive processes and chemotactic auto-aggregation. In the peripheral clusters, however, the absence of IFN means that the spatial dynamics are mainly dictated by diffusive processes.

This diffusion in the cellular population allows some small subpopulation of cells to migrate sufficiently towards the central cluster so as to overcome the thresholding in the metabolic dynamics. Coupling this subpopulation with high affinity molecules, one achieves a fast dynamics in the binding and metabolic spaces on the perimeter of the peripheral clusters (Fig. \ref{fig:Results_chem_50,1}, $t=20$). Once these peripheral subpopulations have been potentiated to the point where they are capable of producing high affinity IFN, the cluster attains an intra-cluster supply and is capable of maintaining its own levels of IFN (Fig. \ref{fig:Results_chem_50,1}, $m(20,x)$), resulting in initially peaked levels of IFN concentration at peripheral sites.

In the chemotactic simulations, one can more clearly see the elements of inter-cluster oscillation as an illustration of similar intra-cluster events. One observes an initially raised production dynamics in the central clusters (Fig. \ref{fig:Results_chem_50,1}, $t=10$); followed by fast metabolic dynamics within, and a concurrent raising of the local concentrations around, the peripheral clusters (Fig. \ref{fig:Results_chem_50,1}, $t=20$); a subsequent response from the central cluster as the peripheral clusters feedback IFN to elevate binding rates (Fig. \ref{fig:Results_chem_50,2}, $t=30$); and the resolution of this oscillatory behaviour in the establishment of a quasi-equilibrium (Fig. \ref{fig:Results_chem_50,2}, $t\geq40$), where intra-cluster dynamics prevail but result in little macroscopic change. The initial inter-cluster heterogeneity is a necessary precursive state for the establishment of this uniformity in behavioural dynamics.

Moreover, the establishment of this synchronicity between the clusters leads to another effect stemming from the chemotactic dynamic. Not only are cells capable of communicating in the chemotactic paradigm but they also self-attenuate their diffusion and auto-aggregate upon the establishment of intra-cluster activation. This may have profound implications for immunity: If, as one might intuitively predict, cells who are inclined to utilise chemotactic dynamics were attracted to the first cluster, and activated at some gradualistic pace, then the infection of the organism by a pathogen would result in the accumulation of IFN excreting cells. The decay of the spatial diversity in the cells would then lead the body to become more vulnerable to infection at novel sites, as there would no longer be IFN SARs present. If, however and as predicted by our model, we have a slight diffusive process which allows the signal to be passed but followed by auto-aggregation, then the cells would remain mostly \textit{in situ} and would propogate the signal without compromising their position in the event of a further wave of infection.

\section{Discussion}


The model and framework that we have herein developed is also sufficiently general so as to be useful in cases that extend beyond the IFN system and even beyond the more general category of SARs. Generality is achieved through the biologically global forms of the binding and unbinding functions as well as the particularly general form chosen for the metabolic flow function, which describes a whole metabolic pathway in a reasoned but condensed single ODE form.

\medskip
The single-cluster model demonstrates a qualitative biological SAR-cycling between binding and metabolic dynamics of a SAR (Fig. \ref{fig:Single_cell,1} \& \ref{fig:Single_cell,2}). More basic, or simplistic, models may be capable of producing quantitatively similar results but could not capture the mechanistic heterogeneity within biological systems which cause them to function as they do. Alone, this illustrates the potential for SST systems to differentially mimic biological systems to a far greater accuracy than can current modelling techniques.

In terms of the biology, this model makes two important realisations: That low affinity molecules may be necessary, for the functioning of the system, in order that the concentrations of such molecules, at long range, are sufficiently high so as to activate distant clusters of producer cells. Also, the biological system actually has two important functions of heterogeneity internally, in order to self regulate clusters and maintain sensible levels of IFN, and externally between clusters, so as to convey the activation signal of one cluster by firstly priming an initially excited cluster at a distance.

The internal heterogeneity established by clusters informs one that the ability for a cluster of SARs (specifically those for IFN) to maintain optimal levels of metabolism and reciprocal output, it is necessary for some subpopulation of cells to sacrificially reduce their levels of binding. This appears to be as a consequence of the feedback between metabolism, $\alpha$, and binding, $y$, such that as one subpopulation rapidly increases its metabolism it will feel and subsequent inhibition of its ability to bind and will sacrifice itself such that another subpopulations may rapidly increase its binding and metabolism, due to the increased availability of local IFN. This is an important effect of intra-popular heterogeneity which we term `subpopular quiescence', and may explain several of the inter-cellular, intra-popular oscillatory events in biology.

The latter of these two realisations recognises the importance of heterogeneity to the biological system. We demonstrate that in order that a primary cluster be primed, upon excitation, it must be allowed to be internally heterogeneous such that more active cells serve to activate less active cells whilst down regulating their own activity. This is essential for maintenance of activity levels and eventually for switching the system off. We further show that this ability for one cluster to self-activate and autoregulate is essential to maintain the long range signal and activate further clusters, at a distance. This nuancing is not possible within the simple spatial model (\ref{eq:simple_sys}).

\medskip
One phenomenon, observed within the multi-dimensional model, which cannot be recreated within more simple mathematical models is that of `metabolic trapping', and therefore, production in the presence of inter-cluster cooperativity. In the simple models, one has a mechanism of feedback wherein a cluster will create IFN in the presence of IFN, amplifying a given local signal. This return, however, always achieves a maximal concentration and the rate is dependent only on local IFN concentration. In the SST context, one observes that the inter-cluster supply of IFN protein between clusters actually increases the metabolic state of all involved clusters causing the productions rates to increase, concurrently. This is a qualitative result which makes a qualitative difference to the final resting state of IFN concentration.

We recognise, also, that the conveyance of the signal in the low affinity cases (Fig. \ref{fig:Results_theta_1,1} \& \ref{fig:Results_theta_1,2}) is dependent on some thresholding parameter in the binding space, and can be justified through the biological realisation that sufficiently low quantities of chemical are insufficient to bind the receptor for long enough durations so as to cause co-phosphorylation of the internal proteins. This is a further major difference between this and previous modelling techniques, since previous modelling techniques make no comment on this phenomenon. A demonstrable advantage of this modelling framework is the ability to flag up novel biological problems, not necessarily perceptible to simpler state-variable frameworks.

\medskip
Spatially dynamic results demonstrated a breakdown in the different abilities of high and low affinity IFN to affect inter-cluster cell communication. This demonstrates that communication can be achieved either by means of reducing the barriers to the travelling molecule (affinity to consumer cells) or by cellular migration, reducing the distance between SARs themselves. In biology these dynamics may occur in environments which have more freedom for the cells to migrate and may not be achievable in many instances. In cases where migration is not possible it may be advantageous to increase production of lower affinity IFN, where high affinity IFN may be advantageous otherwise, due to the resultant increase in dynamic rate.

The biological significance of these processes are underscored by the intricate intra- and inter-cluster spatial and metabolo-binding dynamics. The major features are an intra-cluster oscillatory dynamic and a intra-cluster, post-potentiation auto-aggregation which may be immunologically advantageous (depending on the paradigm considered). In the paradigm where cells are capable of migration, however, one will immediately notice that any given signal is much harder to contain or confine to a local spatial domain. This may be important in organs, such as he brain, where the body wishes to localise inflammatory response and antiviral behaviour as far as is possible. Therefore, local biological considerations may effect the evolutionary choice of method for communication chosen.

\medskip
Finally, this framework is far more approachable for the biological community, in terms of understanding. The internal and inter-cluster heterogeneity described by the SST framework is relatable to biologists in a way that is conducive to dialogue. In line with this a further explanation proffered to the thresholding problem, however, could be that there are two such IFN molecules involved in this process; one of high and one of low affinity. The high affinity molecules may serve to perpetuate the activation of the considered cell, or cluster, whilst the low affinity molecule may serve to convey this signal to other producer cells. This is a theme that the authors intend to explore in a further publication.

\appendix

\section{Derivation of the Spatio-Structural-Temporal Model with Receptor and Metabolic Spaces}
\label{app:Formalism}

Following the same form as the derivation given in Domschke \textit{et al.} \cite{Domschke2016}, we derive of a spatio-structural-temporal (SST) model presented in (\ref{eq:GenFlux}).

Let $\D\subset\mathbb{R}^d$ with $d\in\{1,2,3\}$ be a bounded spatial domain, $\mathcal{I}=[0,T]\subset\mathbb{R}$, with $T>0$ be an arbitrary time interval. Further, let $\Y\subset\mathbb{R}^{\upsilon}$ with $\upsilon\in\mathbb{N}$ characterise the available binding sites, and corresponding binding space, for given receptor $\xi_i$ for $i\in\{1,...,\upsilon\}$, which may differ in structure dependent on the molecules capable of binding each $\xi_i$ and let $\P\subset\mathbb{R}^{p}$ characterise the binding space for the biological complexes bound to these available sites. Finally, let $\Gamma\subset\mathbb{R}^{\gamma}$ with $\gamma\in\mathbb{N}$ characterise the metabolic subspace of each $\Y$ whose boundary is given by the corresponding extreme currents for the effected metabolic gene network. Herein, the space defined by $\Y\times\P$ shall be referred to as the elementary-state ($e$-state) space and will give a characterisation of the total structure of an individual element's state.

Now, let the variables $x\in\mathcal{D}\subset\mathbb{R}^d$ represent space; $\xi\in\Y\subset\mathbb{R}^{\upsilon}$ represent receptoral state; $y\in\P\subset\mathbb{R}^{p}$ represent the binding state of these receptors; and $\alpha\in\Gamma\subset\mathbb{R}^{\gamma}$ represent the metabolic state of these cells. Therefore, we have also that $(\xi,y)\in\Y\cross\P$ gives the receptoro-binding state of the population, at any given point in the spatial domain, $\D$.

Further, let $U$, $V$, $W$ be rectangles in $\mathcal{D}$, $\Y\cross\P$, and $\Gamma$ respectively (i.e. $U\cross V\cross W \subseteq \D\cross\Y\cross\P\cross\Gamma$). Then the total amount of cells at a given time $t$ is given by
\ahh{\begin{equation} \begin{array}{rll}
	\hat{c}(t)	& =	& \int\limits_{W} \int\limits_{V} \int\limits_{U} c(t,x,(\xi,y),\alpha)\,dx\,d(\xi,y)\,d\alpha
\end{array} \end{equation}
the change in $\bar{c}:=\hat{c}(t,x,(\xi,y),\alpha)$  per unit time in the spatio-metabolo-receptoro-binding region
 $U\cross V\cross W$ is given by
\begin{equation} \begin{array}{rll}
	\frac{d\bar{c}(t)}{dt}	& =	& \int\limits_{W} \int\limits_{V} \int\limits_{U} \hat{S}(t,x,(\xi,y),\alpha)\,dx\,d(\xi,y)\,d\alpha\,
\\[1.2em]	&&	- \int\limits_{W} \int\limits_{V} \int\limits_{\partial U} \hat{F}(t,x,(\xi,y),\alpha)\cdot\mathfrak{n}(x)\,d\sigma_{d-1}(x)\,d(\xi,y)\,d\alpha\,
\\[1.2em]	&&	- \int\limits_{W} \int\limits_{U} \int\limits_{\partial V} [ \hat{G}(t,x,(\xi,y),\alpha),\hat{H}(t,x,(\xi,y),\alpha) ]^T \cdot\mathfrak{n}(\xi,y)\,d\sigma_{\upsilon+p-1}(\xi,y)\,dx\,d\alpha\,
\\[1.2em]	&&	- \int\limits_{V} \int\limits_{U} \int\limits_{\partial W} \hat{K}(t,x,(\xi,y),\alpha)\cdot\mathfrak{n}(\alpha)\,d\sigma_{\gamma-1}(y)\,dx\,d(\xi,y)\,d\alpha\,
\end{array} \end{equation} }
\ahh{where $\sigma_{d-1}$, $\sigma_{2r-1}$, and $\sigma_{\gamma-1}$ are surface measures on $\partial\mathcal{D}$, $\partial\mathcal{P}$, and $\partial\Gamma$, respectively. Supposing, now, that $F$, $G$, $H$, and $J$, are in the class of continuously differentiable vector fields, $\mathcal{C}^1$, one can use Stokes' Theorem to write
\begin{equation} \begin{array}{rll}
	\frac{d\bar{c}(t)}{dt}	& =	& \int\limits_{W} \int\limits_{V} \int\limits_{U} \hat{S}(t,x,(\xi,y),\alpha)\,dx\,d(\xi,y)\,d\alpha\,
\\[1.2em]	&&	- \int\limits_{W} \int\limits_{V} \int\limits_{U} \nabla_x\cdot \hat{F}(t,x,(\xi,y),\alpha)\,dx\,d(\xi,y)\,d\alpha\,
\\[1.2em]	&&	- \int\limits_{W} \int\limits_{U} \int\limits_{V} \nabla_{(\xi,y)}\cdot [ \hat{G}(t,x,(\xi,y),\alpha) , \hat{H}(t,x,(\xi,y),\alpha) ]^T \,d(\xi,y)\,dx\,d\alpha\,
\\[1.2em]	&&	- \int\limits_{U} \int\limits_{V} \int\limits_{W} \nabla_{\alpha}\cdot \hat{K}(t,x,(\xi,y),\alpha)\,d\alpha\,d(\xi,y)\,dx\,
\end{array} \end{equation} }
\ahh{and using Lebesgue's Dominated Convergence Theorem, one can move the time derivative within the integral for $\hat{c}$
\begin{equation} \begin{array}{rll}
	\int\limits_{W}\int\limits_{V}\int\limits_{U} \frac{\partial \hat{c}}{\partial t}\,dx\,d(\xi,y)\,d\alpha\,	& =	& \int\limits_{W} \int\limits_{V} \int\limits_{U} \hat{S}(t,x,(\xi,y),\alpha)\,dx\,d(\xi,y)\,d\alpha\,
\\[1.2em]	&&	- \int\limits_{W} \int\limits_{V} \int\limits_{U} \nabla_x\cdot \hat{F}(t,x,(\xi,y),\alpha)\,dx\,d(\xi,y)\,d\alpha\,
\\[.6em]	&&	- \int\limits_{V} \int\limits_{U} \int\limits_{W} \nabla_{(\xi,y)}\cdot \left(\begin{array}{c} \hat{G}(t,x,(\xi,y),\alpha) \\ \hat{H}(t,x,(\xi,y),\alpha) \end{array}\right) \,d(\xi,y)\,dx\,d\alpha\,
\\[1.6em]	&&	- \int\limits_{U} \int\limits_{V} \int\limits_{W} \nabla_{\alpha}\cdot \hat{K}(t,x,(\xi,y),\alpha)\,d\alpha\,d(\xi,y)\,dx\,,
\end{array} \end{equation} }
\ahh{which can be written
\begin{equation} \begin{array}{rll}
	\int\limits_{\mathbb{R}^{d+\upsilon+p+\gamma}} [ \frac{\partial \hat{c}}{\partial t}]\mathbf{1}_{U\cross V\cross W}(x,(\xi,y),\alpha)\,dx\,d(\xi,y)\,d\alpha\,	\hspace{-17em}&
\\[1.2em]	 &=&
			\int\limits_{\mathbb{R}^{d+\upsilon+p+\gamma}} [ \hat{S}(t,x,(\xi,y),\alpha)]\mathbf{1}_{U\cross V\cross W}(x,(\xi,y),\alpha)\,dx\,d(\xi,y)\,d\alpha
\\[1.2em]	&&	- \int\limits_{\mathbb{R}^{d+\upsilon+p+\gamma}} [ \nabla_x\cdot \hat{F}(t,x,(\xi,y),\alpha)]\mathbf{1}_{U\cross V\cross W}(x,(\xi,y),\alpha)\,dx\,d(\xi,y)\,d\alpha\,
\\[.6em]	&&	- \int\limits_{\mathbb{R}^{d+\upsilon+p+\gamma}} \left[ \nabla_{(\xi,y)}\cdot \left(\begin{array}{c} \hat{G}(t,x,(\xi,y),\alpha) \\ \hat{H}(t,x,(\xi,y),\alpha) \end{array}\right) \right]\mathbf{1}_{U\cross V\cross W}(x,(\xi,y),\alpha)\,dx\,d(\xi,y)\,\alpha\,
\\[1.6em]	&&	- \int\limits_{\mathbb{R}^{d+\upsilon+p+\gamma}} [ \nabla_{\alpha}\cdot \hat{K}(t,x,(\xi,y),\alpha)]\mathbf{1}_{U\cross V\cross W}(x,(\xi,y),\alpha)\,dx\,d(\xi,y)\,d\alpha\,.
\end{array} \end{equation} }
\ahh{There, since we have that
\begin{equation*} \begin{array}{c} \left\{ U\cross V\cross W ~\ | ~\ U,V,W \text{ - compact with piecewise smooth boundaries} \right\} \end{array} \end{equation*}
is a family of generators for the Borelian $\sigma$-algebra on $U \cross V\cross W$ we can denote $\mathbf{1}_A$ as the indicator function for any arbitrary $A\subseteq\D\cross\Y\cross\P\cross\Gamma$ and write
\begin{equation} \begin{array}{rll}
	\int\limits_{\mathbb{R}^{d+\upsilon+p+\gamma}} [ \frac{\partial \hat{c}}{\partial t}]\mathbf{1}_A(x,(\xi,y),\alpha)\,dx\,d(\xi,y)\,d\alpha\,	\hspace{-12.5em}&
\\	&=&		\int\limits_{\mathbb{R}^{d+\upsilon+p+\gamma}} [ \hat{S}(t,x,(\xi,y),\alpha)]\mathbf{1}_A(x,(\xi,y),\alpha)\,dx\,d(\xi,y)\,d\alpha\,
\\[1.2em]	&&	- \int\limits_{\mathbb{R}^{d+\upsilon+p+\gamma}} [ \nabla_x\cdot \hat{F}(t,x,(\xi,y),\alpha)]\mathbf{1}_A(x,(\xi,y),\alpha)\,dx\,d(\xi,y)\,d\alpha\
\\[1.2em]	&&	- \int\limits_{\mathbb{R}^{d+\upsilon+p+\gamma}} \left[ \nabla_{(\xi,y)}\cdot \left(\begin{array}{c} \hat{G}(t,x,(\xi,y),\alpha) \\ \hat{H}(t,x,(\xi,y),\alpha) \end{array}\right) \right]\mathbf{1}_A(x,(\xi,y),\alpha)\,dx\,d(\xi,y)\,d\alpha\,
\\[1.2em]	&&	- \int\limits_{\mathbb{R}^{d+\upsilon+p+\gamma}} [ \nabla_{\alpha}\cdot \hat{K}(t,x,(\xi,y),\alpha)]\mathbf{1}_A(x,(\xi,y),\alpha)\,dx\,d(\xi,y)\,d\alpha\,
\end{array} \end{equation}
for any arbitrary Borelian set $A$ in the $\sigma$-algebra on $\D\cross\Y\cross\P\cross\Gamma$. }\ahh{Then we can replace $\mathbf{1}_A$ with any simple function, as so
\begin{equation} \begin{array}{r} \begin{array}{rll}
	\int\limits_{\mathbb{R}^{d+2r+\gamma}} [ \frac{\partial \hat{c}}{\partial t}]\nu(x,(\xi,y),\alpha)\,dx\,d(\xi,y)\,d\alpha\,	\hspace{-11em}&
\\	&=&		\int\limits_{\mathbb{R}^{d+\upsilon+p+\gamma}} [ \hat{S}(t,x,(\xi,y),\alpha)]\nu(x,(\xi,y),\alpha)\,dx\,d(\xi,y)\,d\alpha\,
\\[1.2em]	&&	- \int\limits_{\mathbb{R}^{d+\upsilon+p+\gamma}} [ \nabla_x\cdot \hat{F}(t,x,(\xi,y),\alpha)]\nu(x,(\xi,y),\alpha) \,dx\,d(\xi,y)\,d\alpha\,
\\[.6em]	&&	- \int\limits_{\mathbb{R}^{d+\upsilon+p+\gamma}} \left[ \nabla_{(\xi,y)}\cdot \left(\begin{array}{c} \hat{G}(t,x,(\xi,y),\alpha) \\ \hat{H}(t,x,(\xi,y),\alpha) \end{array}\right) \right]\nu(x,(\xi,y),\alpha) \,dx\,d(\xi,y)\,d\alpha\,
\\[1.6em]	&&	- \int\limits_{\mathbb{R}^{d+\upsilon+p+\gamma}} [ \nabla_{\alpha}\cdot \hat{K}(t,x,(\xi,y),\alpha)]\nu(x,(\xi,y),\alpha) \,dx\,d(\xi,y)\,d\alpha\,
\end{array} \\
\forall \nu\in\mathcal{C}_{0}^{\infty}(\D\cross\P\cross\Gamma)\,.
\end{array} \end{equation}}

\ahh{Then, since this relation holds for any $\mathcal{C}^\infty$ test function, $\nu(x,(\xi,y),\alpha)$, we obtain the equation
\begin{equation} \begin{array}{rll}
	\frac{\partial \hat{c}}{\partial t}	 &=& \hat{S}(t,x,(\xi,y),\alpha) - \nabla_x\cdot \hat{F}(t,x,(\xi,y),\alpha) \\[.7em]&& - \nabla_{(\xi,y)}\cdot [ \hat{G}(t,x,(\xi,y),\alpha) , \hat{H}(t,x,(\xi,y),\alpha) ]^T - \nabla_{\alpha}\cdot \hat{K}(t,x,(\xi,y),\alpha) \,,
\end{array} \end{equation}
where the functions on the right-hand side describe fluxes in the cellular population density.}

\subsection{Derivation of a Structural Source Term}


The source term accounts for cell multiplication by division. It is clear that for the source term, therefore, one must consider the full, continuous transition from mother-cell to 2 daughter-cell and, finally, back to 2 second generation mother-cell. We achieve this by considering \ahh{mitosis as time dependent process that occurs on a} normalised micro-temporal scale, $\tau\in[0,1)$.

\ahh{Therefore, assuming uniform splitting of the receptors on the cell surface during cell differentiation, at a given spatio-temporal  node $(t,x)$, the amount of cells whose binding structure reside within an arbitrary rectangle $W\in\P$ is given by the difference between the cells that arrived within $W$ due to mitosis and those that leave $W$ through mitosis, and we re-cast this mathematically as}

\begin{equation} \begin{array}{rll}
	\int\limits_{W} \hat{S}(t,x,y)\,dy	& =	& 2\int\limits_{[0,1)}\int\limits_{(2-\tau)W} \phi(\tilde{y},c,v)\hat{c}(t,x,\tilde{y})\,d\tilde{y}\,d\tau
\\[14pt]	&& - \int\limits_W \phi(y,c,v)\hat{c}(t,x,y)\,dy
\end{array} \end{equation}
\ahh{Using} the change of variable
\begin{equation} \begin{array}{rll}
	\tilde{y}(y)	& =	& (2-\tau)y
\\	d\tilde{y}	& =	& (2-\tau)\,dy
\end{array} \end{equation}
we can \ahh{obtain}
\begin{equation} \begin{array}{rll}
	\int\limits_W \hat{S}(t,x,y)\,dy	& =	& 2\int\limits_{[0,1)}(2-\tau)^p \int\limits_{W}\phi((2-\tau)y,c,v)\hat{c}(t,x,(2-\tau)y)\,dy\,d\tau
\\[14pt]	&&	- \int\limits_W\phi(y,c,v)\hat{c}(t,x,y)\,dy
\\[18pt]	&=&	\int\limits_{W}2\int\limits_{[0,1)}(2-\tau)^p\phi((2-\tau)y,c,v)\hat{c}(t,x,(2-\tau)y)\,d\tau\,dy
\\[14pt]	&&	- \int\limits_W\phi(y,c,v)\hat{c}(t,x,y)\,dy\,.
\end{array} \end{equation}
\ahh{Thus, as this equality holds true for any rectangle $W$, via a standard measure theoretical argument, we obtain that }
\begin{equation}
\hat{S}(t,x,y)=2\int\limits_{[0,1)}(2-\tau)^p\phi((2-\tau)y,c,v)\hat{c}(t,x,(2-\tau)y)\,d\tau	- \phi(y,c,v)\hat{c}(t,x,y).
\end{equation}

\subsection{Derivation of a Structural Source Term for Systems with Receptors}

\ahh{We proceed similarly to derive the source term in the case when the dynamics of the receptors is also accounted for.} Consider again that \ahh{mitosis is a time dependent process that occurs on a} normalised micro-temporal scale, $\tau\in[0,1)$ and that we have uniform splitting of the receptors on the cell surface during cell differentiation, at a given spatio-temporal  node $(t,x)$. Then the amount of cells whose receptoral-binding structure reside within an arbitrary rectangle $V\cross W\in\Y\cross\P$ is given by the difference between the cells that arrived within $v\cross W$ due to mitosis and those that leave $V\cross W$ through mitosis, which can be expressed as
\begin{equation} \begin{array}{rll}
	\int\limits_{V\cross W} \hat{S}(t,x,\xi,y,\alpha)\,dy	& =	& 2\int\limits_{[0,1)}\int\limits_{(2-\tau)V\cross W} \phi((\tilde{\xi},\tilde{y}),c,v)\hat{c}(t,x,\xi,\tilde{y},\alpha)\,d(\tilde{\xi},\tilde{y})\,d\tau
\\[14pt]	&& - \int\limits_{V\cross W} \phi((\xi,y),c,v)\hat{c}(t,x,\xi,y,\alpha)\,d(\xi,y)
\end{array} \end{equation}
and using the change of variable
\begin{equation} \begin{array}{rll}
	(\tilde{\xi},\tilde{y})(\xi,y)	& =	& (2-\tau)(\xi,y)
\\	d(\tilde{\xi},\tilde{y})	& =	& (2-\tau)\,d(\xi,y)
\end{array} \end{equation}
we obtain
\begin{equation} \begin{array}{rll}
	\int\limits_{V\cross W} \hat{S}(t,x,\xi,y,\alpha)\,d(\xi,y)	\hspace{-9.5em}&\\[1em]
	&=\!\!\!	& 2\!\!\int\limits_{[0,1)}(2\!-\!\tau)^{(p+\gamma)} \!\!\int\limits_{V\cross W}\phi((2\!-\!\tau)(\xi,y),c,v)\hat{c}(t,x,(2\!-\!\tau)(\xi,y),\alpha)\,d(\xi,y)\,d\tau
\\[14pt]	&&	- \int\limits_{V\cross W}\phi((\xi,y),c,v)\hat{c}(t,x,\xi,y,\alpha)\,d(\xi,y)
\\[18pt]	&=\!\!\!	& \!\!\int\limits_{V\cross W}2 \!\!\int\limits_{[0,1)}(2\!-\!\tau)^{(p+\gamma)}\phi((2\!-\!\tau)(\xi,y),c,v)\hat{c}(t,x,(2\!-\!\tau)(\xi,y),\alpha)\,d\tau\,d(\xi,y)
\\[14pt]	&&	- \int\limits_{V\cross W}\phi((\xi,y),c,v)\hat{c}(t,x,\xi,y,\alpha)\,d(\xi,y)\,.
\end{array} \end{equation}
Since this relation holds for any rectangle $V\cross W$, then using the standard measure theory density argument as in the 2 preceding appendix sections, we arrive at our final expression of source flux for the total population as
\begin{equation} \begin{array}{rll} \label{eq:Source}
	S(t,x,\xi,y,\alpha)	\hspace{-2em}&\\[1em]
	&= 	& 2\int\limits_{[0,1)}(2-\tau)^{(p+\gamma)}\phi((2-\tau)(\xi,y),c,v)\hat{c}(t,x,(2-\tau)(\xi,y),\alpha)\,d\tau	\\[1em] &&
	-\phi((\xi,y),c,v)\hat{c}(t,x,\xi,y,\alpha) \,.
\end{array} \end{equation}


\section{Numerical Methods \& Parameters}
\label{app:NumMethods}

\subsection{Numerical methods}

We use the 4\th  order Runge-Kutta predictor for this system, given by
\[ \bar{c}_1^{\tau+1} := c_1^\tau + \frac{d\tau}{6}\left( F(k_{c_1,1}^\tau) + 2F(k_{c_1,2}^\tau) +2F(k_{c_1,3}^\tau) + F(k_{c_1,4}^\tau) \right) \,, \]
with
\begin{equation*} \begin{aligned}
	k_{c_1,1}^\tau := c_1^\tau, ~\
	k_{c_1,2}^\tau := c_1^\tau + \frac{h}{2}k_{c_1,1}^\tau, ~\ \\
	k_{c_1,3}^\tau := c_1^\tau + \frac{h}{2}k_{c_1,2}^\tau, ~\
	k_{c_1,4}^\tau := c_1^\tau + d\tau k_{c_1,3}^\tau \,,
\end{aligned} \end{equation*}
where $F(c_1^\tau):=F(c_1^{\tau},m^{\tau})$ are given by the local central difference approximation of the spatio-structural dynamics for $c_1^{\tau}:=c_1(t^{\tau},x,y,\alpha)$ at the given time point $t^{\tau}$. We then use a MacCormack corrector, of the form
\[ \hat c_1^{\tau+1} := \frac{c_1^\tau + \bar{c}_1^{\tau+1}}{2} + \frac{d\tau}{2}F(\bar{c}_1^{\tau+1}) \,. \]
Likewise, these formulae are used for the calculation of the solution for the IFN molecular species, $m(t,x)$.

We further apply the population-based constraint
\begin{equation} \label{eq:pop_constr}
	c_1^{\tau+1} := \hat c_1^{\tau+1}\frac{\int\limits_{\mathcal{P}}\int\limits_{\Gamma} c_1^0 \,d\alpha \,dy}{\int\limits_{\mathcal{P}}\int\limits_{\Gamma} \hat c_1^{\tau+1} \,d\alpha \,dy} \,,
\end{equation}
in order to constrain growth in the population due to the advective term under condition $c(t,x,\xi,y,\alpha)\geq0$. We can write this in the particular case give since $S(t,x,\xi,y,\alpha)=0$ and therefore we have that there is no overall change in population. Otherwise, however, this can be achieved by stepwise accumulation and conformity.

In order to compute accurate solutions to the multi-cluster distribution arrays, we denote one individual cluster as $c_{1,i}(t,x,y)$ for any $i\in\{1,\dots,k\}$, where $k$ is the total number of clusters or initial distributions. Then we have that the entire cellular population distribution is defined as \[ c_1(t,x,y,\alpha) := \sum\limits_{i=1}^k c_{1,i}(t,x,y,\alpha). \]
Observe that from the fundamental theorem of calculus we, therefore, have
\begin{equation} \begin{aligned} \label{eq:sum_div}
\nabla_x\cdot c_1(t,x,y,\alpha) \nabla_x m &= \nabla_x\cdot \left( \sum\limits_{i=1}^k c_{1,i}(t,x,y,\alpha) \right)\nabla_x m \\
		&= \sum\limits_{i=1}^k \nabla_x\cdot c_{1,i}(t,x,y,\alpha)\nabla_x m
\end{aligned} \end{equation}
Since we have that the overall population does not change with respect to changes time ($S(t,x,y,\alpha)=0$), we can use that (\ref{eq:pop_constr}) and (\ref{eq:sum_div}) imply that the population constraint holds on each individual cluster of IFN producer cells
\begin{equation}
c_{1,i}^{\tau+1} := \hat c_{1,i}^{\tau+1}\frac{\int\limits_{\mathcal{P}}\int\limits_{\Gamma} c_{1,i}^0 \,d\alpha \,dy}{\int\limits_{\mathcal{P}}\int\limits_{\Gamma} \hat c_{1,i}^{\tau+1} \,d\alpha \,dy}, ~\ ~\ \forall i.
\end{equation}
and then the total population changes with
\begin{equation}
c_{1}^{\tau+1} := \hat c_{1}^{\tau+1}\frac{\int\limits_{\mathcal{P}}\int\limits_{\Gamma} \sum\limits_{i=1}^kc_{1,i}^0 \,d\alpha \,dy}{\int\limits_{\mathcal{P}}\int\limits_{\Gamma} \hat c_{1}^{\tau+1} \,d\alpha \,dy}, ~\ ~\ \forall i.
\end{equation}
These constraints should either leave the population $c(t,x,y,\alpha)$ unaltered or correct for any small instabilities arising from the long-term cumulation of $\mathcal{O}(\delta^2)$ spatial advective errors, which are not adequately dealt with by the predictor-corrector methodology.

We also introduce the notations
\begin{equation*} \begin{aligned}
	c_\alpha &:= \int\limits_\P c(t,x,y,\alpha) \,dy
		~\ ~\ \text{and} ~\ ~\ &
	c_y &:= \int\limits_\Gamma c(t,x,y,\alpha) \,d\alpha
\end{aligned} \end{equation*}
as quantifying the spatio-metabolic and spatio-binding distributions, respectively, and
\begin{equation*}
	\breve{c} := \iint\limits_\D c(t,x,y,\alpha) \,dx
\end{equation*}
as quantifying the non-spatial metabolo-binding distribution of the cellular population $c(t,x,y,\alpha)$.

\subsection{Parameters}

Here we give the table of parameters for the complete, SST system:
\begin{table}[h!] \centering \begin{tabular}{c | c | c c c c c}
	dependent	&	independent	&	parameters	\\
	variable		&	variable		\\[.25em] \hline &&&&& \\[-.75em]
	$\hat{c}$	&	$x$	&	$D_c=10^{-5}$	&&	$\chi_m=10^{-4}$	\\[.1em]
			&	$y$	&	$\beta=2\lambda$	&&	$\upsilon=10^{-1}$	&&	$\theta_m=10^{-1}$ \\[.1em]
			&	$\alpha$	&	$d=\frac{1}{4}\beta$	&&	$\mu_0=10^{-1}$ \\[.25em] \hline &&&&& \\[-.75em]
	$m$	&	$x$	&	$D_m=4\cross10^{-3}$ \\[.1em]
		&		&	$\varepsilon=10^{-2}$	&&	$\theta_\alpha=10^{-1}$	&&	$\phi=1$
\end{tabular}
\caption{Table of parameters} \end{table}

\section*{Acknowledgements}
The first author would like to acknowledge the funding provided by the \'Ecole Doctorale de l'Universit\'e de Montpellier, without which this project would have been impossible.

\bibliographystyle{spmpsci}
\bibliography{uPA_txxiy}

\end{document}